\documentclass[a4paper]{article}
\usepackage[top=3cm, bottom=3cm, left=3cm, right=3cm, headheight=13pt, marginparwidth=2.6cm, marginparsep=0.3cm]{geometry}
\usepackage{amsmath,amssymb,amsthm,mathtools,bm,amsfonts}
\usepackage[natbibapa]{apacite}
\usepackage{microtype}
\usepackage[appendix=append,bibliography=common]{apxproof} 
\numberwithin{equation}{section}
\theoremstyle{plain} 
\newtheoremrep{theorem}{Theorem}[section]
\newtheoremrep{proposition}[theorem]{Proposition}
\newtheoremrep{corollary}[theorem]{Corollary}
\newtheoremrep{lemma}[theorem]{Lemma}

\newcommand{\ProofSeeAppendix}[1]{%
  \noindent\emph{Proof.} See the \hyperref[#1]{Appendix}.%
  \unskip\nobreak\hfill\mbox{\ensuremath{\qedsymbol}}\par
}

\usepackage{multirow}
\usepackage{xcolor}
\usepackage{indentfirst} 
\usepackage{authblk}

\usepackage{float}
\usepackage{booktabs}
\usepackage{graphicx}   
\usepackage{subcaption} 
\usepackage{hyperref}
\usepackage{accents}

\newcommand{\lowtilde}[1]{\accentset{\smash{\lower2.15ex\hbox{$\widetilde{\phantom{#1}}$}}}{#1}}
\newcommand{\lowtildeO}[1]{\accentset{\smash{\lower1.9ex\hbox{$\widetilde{\phantom{#1}}$}}}{#1}}

\newcommand{\Vega}{\mathrm{Vega}}
\newcommand{\Vanna}{\mathrm{Vanna}}
\newcommand{\Volga}{\mathrm{Volga}}

\providecommand{\keywords}[1]{\par\smallskip\noindent\textbf{Keywords: }#1\par}

\providecommand{\MSC}[2][]{%
  \par\smallskip\noindent\textbf{MSC #1:}~#2\par\smallskip
}
\providecommand{\keywords}[1]{%
  \par\smallskip\noindent\textbf{Keywords:}~#1\par\smallskip
}

\title{Curved Greeks: A Geometric Layer for Option P\&L Adjustments}

\author[1,2]{Pedro Pablo P\'erez Velasco\thanks{Corresponding author: \href{mailto:ppperez@comillas.edu}{ppperez@comillas.edu}}}
\author[3]{Mengjue Lu}
\author[1,2]{Daniel Arrieta\thanks{\href{mailto:darrieta@comillas.edu}{darrieta@comillas.edu}; \href{mailto:daniel.arrieta@gruposantander.com}{daniel.arrieta@gruposantander.com}}}

\affil[1]{Department of Quantitative Methods, ICADE (Universidad Comillas), Madrid, Spain\\ \href{mailto:ppperez@comillas.edu}{ppperez@comillas.edu}}
\affil[2]{Model Risk, Banco Santander, Madrid, Spain\\ \href{mailto:pedpperez@gruposantander.com}{pedpperez@gruposantander.com}}
\affil[3]{Instituto Complutense de An\'alisis Econ\'omico (ICAE), Complutense University of Madrid, Madrid, Spain\\ \href{mailto:mengjuel@ucm.es}{mengjuel@ucm.es}}
\date{\today}

\begin{document}
\maketitle

\begin{abstract}
Greek-based risk management provides a local description of option profit and loss.
Standard Taylor P\&L is written in a chosen set of risk-factor coordinates.
By treating coordinate increments as the primitive shocks and ordinary Hessians as second-order Greeks, it fixes a flat affine convention that is usually left implicit.
This paper makes that convention explicit and introduces a geometric layer for short-horizon option books.
The valuation model and first-order hedge are preserved, but the quadratic P\&L term is written as a covariant Hessian associated with a connection.
This separates two ways in which desk conventions and trading frictions can enter a local P\&L representation: Through curvature overlays, which change the second-order P\&L attributed to the value function, and through metric or penalty overlays, which assign absolute costs or risk weights to factor moves and hedge trades.
The construction is developed at second order, the first level at which the choice of geometric convention affects Greeks beyond the ordinary differential.
It can be calibrated either from target quadratic P\&L coefficients or from cost and risk specifications, depending on the economic object being represented.
Vanna--Volga smile targets, liquidity and execution-cost penalties, and book-level hedge netting are used as representative applications, illustrating the two channels without exhausting the range of possible local quadratic overlays.
\end{abstract}


\MSC[2020]{Primary 91G20; Secondary 91G60, 53B20, 62P05.}


\section*{Acknowledgments}
The authors would like to thank Juan Ballesteros, Gabriela Godoy, Luis Herraiz, Alejandro Olmos, Gonzalo Rodr\'iguez and Ignacio Vellido for valuable discussions and feedback that improved the exposition. All content has been reviewed and edited by the authors, who assume full responsibility for the manuscript and for any remaining errors or omissions.



\section{Introduction}
\label{sec:introduction}
\nosectionappendix

Greek-based risk management is the working local language of option trading. 
On a desk, Greeks are not only derivatives of a pricing formula. 
They are also the quantities used to hedge, explain short-horizon profit and loss (P\&L), set limits, aggregate positions and decide whether rebalancing is worthwhile. 
In most systems this language is implemented through a Taylor expansion around the current market state. 
Small moves in spot, volatility, rates, curves or smile variables are translated into local P\&L through Delta, Gamma, Vega, Vanna, Volga and related sensitivities.

This standard construction is simple and useful but it leaves an important convention implicit. 
Once a set of risk factors has been chosen, the usual Taylor expansion treats coordinate increments as the elementary short-horizon shocks and ordinary second derivatives as the relevant second-order Greeks. 
This amounts to a flat affine convention on the local risk-factor space. 
At first order the issue is usually not visible because the first differential gives the desk's hedge sensitivities. 
At second order, however, the convention matters. 
Gamma, Vanna, Volga and other quadratic terms depend on how shocks are represented, how smile variables are transported and how marking conventions are imposed. 
Thus the quadratic P\&L layer may depend on choices that are operationally natural but often left outside the formal notation.

In practice, desks already modify this local structure. 
A raw Black--Scholes (BS) delta--gamma approximation may be supplemented with smile adjustments,\footnote{Through explicit smile models or local rules such as Vanna--Volga,~\cite{Wystup2006Book,CastagnaMercurio2007}.} marking conventions, execution costs, liquidity assumptions, market-impact estimates, bucketing rules and book-level netting. 
These additions are often useful and sometimes unavoidable. 
They are also usually introduced through separate modules, local rules and aggregation conventions. 
As a result, it can become unclear what is being changed. 
Sometimes the desk wants to modify the second-order P\&L attributed to the value function. 
At other times the desk wants to attach a cost, risk weight or penalty to a factor move or hedge trade. 
These are different economic objects and they should not be represented in the same way.

This paper proposes a local geometric overlay for short-horizon option P\&L and hedging. 
The overlay is designed to sit on top of an existing pricing model, not to replace it. 
It preserves the marginal valuation model and the first-order hedge convention used by the desk. 
Its role is to make the local second-order convention explicit and to provide a systematic way either to modify the quadratic P\&L layer or to attach local quadratic costs and penalties when desk conventions, smile rules, liquidity inputs or other local features require it. 
The construction is developed at second order because this is the first level at which the choice of geometric convention affects Greeks beyond the first differential. 
Higher-order extensions are possible in principle but the second-order layer is the one most directly connected to standard desk practice through Gamma, Vanna, Volga and quadratic cost approximations.

The framework separates two channels. 
The first channel concerns \emph{curvature overlays}. 
These specify target quadratic coefficients in the local P\&L expansion. 
The key device is a covariant Hessian built from an affine connection. 
The connection changes the quadratic term used for local P\&L attribution while leaving the first-order sensitivities unchanged. 
In the smile application below, the connection is calibrated so that the adjusted quadratic predictor matches a Vanna--Volga target. 
More generally, a target quadratic coefficient may come from an alternative model Hessian, a desk marking convention such as sticky-delta or sticky-strike, a curve roll-down rule, a tenor-bucket transport rule, a cross-gamma adjustment or a local fit to realized short-horizon residuals when enough data are available. 
The point is not that all such effects are the same. 
Rather, they enter the local P\&L representation through the same type of object: a prescribed second-order attribution.

The second channel concerns \emph{cost} and \emph{risk overlays}. 
These do not primarily modify the curvature of the valuation function. 
Instead, they assign an absolute cost, charge or penalty to a factor move or hedge trade. 
Liquidity costs, execution costs, bucketed market impact, risk normalisations, scenario penalties and capital add-ons have this character. 
At a local quadratic level they are represented by a positive semidefinite quadratic form or by a metric on the effective subspace when the form is non-degenerate. 
This distinction is important. 
A curvature overlay tells us how to adjust second-order P\&L attribution. 
A cost or risk overlay tells us how expensive, risky or penalized a move is. 
The scale of the penalty is part of the object being modelled and therefore cannot be absorbed into a change of second-order Greeks.

The same local interface can therefore be used for two different purposes, provided the economic role of the object is kept clear. 
For curvature overlays, calibration is to target quadratic P\&L coefficients. 
For cost and risk overlays, calibration is to quoted widths, representative clips, liquidity buckets, impact inputs, risk weights or other specifications of local cost or risk. 
When a hedge map is combined with a quadratic cost on hedge trades, it induces a quadratic penalty on factor moves. 
This induced penalty can be used for cost-aware rebalancing or for local risk measurement. 
If it is smooth and non-degenerate on the appropriate subspace, it may also be used to induce adjusted second-order sensitivities. 
The primary object in that case, however, remains the cost or risk penalty itself.

The relevant risk factors are the variables through which the desk explains and manages local P\&L. 
They may be simple market coordinates, such as spot and volatility, or a richer set of quoting, hedging and marking variables. 
The choice of these variables is itself part of the local convention. 
For this reason, the framework treats the state description explicitly rather than assuming that one particular coordinate system is canonical. 
This is important at second order, where ordinary Hessians depend on the chosen representation. 
A geometric formulation makes clear which part of the local P\&L predictor is attached to the economic overlay and which part is a consequence of the coordinates used to describe the book. 
A useful consequence is comparability across common parameterizations: once the quadratic overlay is specified consistently, the same local predictor can be expressed in spot, forward, log-forward or other smooth coordinates without changing the economic adjustment being represented. 
This coordinate awareness is useful in practice, but it is a consequence of the construction rather than its main objective.

Book-level aggregation is a central motivation. 
Greeks are often computed instrument by instrument but hedging and execution happen at the level of the net book. 
For curvature overlays, book-level aggregation means that target quadratic P\&L coefficients and exposure vectors must be combined consistently. 
For execution costs, the implication is even more direct: costs apply to the net hedge trade. 
Netting, internal crossing or cost amplification should therefore arise from the same quadratic cost object, not from unrelated per-deal add-ons. 
This is one reason for keeping the curvature and cost channels separate. 
They interact through the hedge map and the book but they represent different things.

The paper makes three contributions. 
First, it gives a local geometric interpretation of second-order Greeks as convention-dependent P\&L objects. 
The usual second-order Taylor term is recovered when the curvature correction is set to zero. 
Second, it develops two channels through which additional local features can enter an existing option model without changing the base price or first-order hedge: curvature overlays, represented by a connection, and cost or risk overlays, represented by quadratic penalty forms. 
Third, it shows how these objects can be calibrated and aggregated at book level, with transparent identification conditions, controlled coordinate changes and explicit hedge netting.

The applications in the paper are chosen to illustrate the two channels in settings that are familiar to option desks. 
On the curvature side, we develop the construction for Vanna--Volga smile targets. 
On the cost side, we develop liquidity and execution-cost penalties built from quoted widths, representative clips and hedge responses. 
These examples are not meant to exhaust the possible uses of the framework. 
They are used because they are complex enough to show the difference between changing second-order P\&L attribution and assigning a local cost or risk penalty.

We illustrate the workflow on two foreign-exchange barrier case studies, \texttt{EURUSD} and \texttt{USDTRY}. 
The \texttt{EURUSD} case is a stability check: the smile correction is small, so the overlay should not distort an already adequate local predictor. 
The \texttt{USDTRY} case illustrates the cost side of the framework in a lower-liquidity setting, where execution costs can be economically material. 
The examples are diagnostic rather than a broad empirical validation. 
Their purpose is to show how the two channels of the framework can be implemented from market and desk inputs.

The framework is intentionally local. 
It is designed for short-horizon quadratic P\&L, hedging and attribution. 
It is not a replacement for a global pricing model, a full smile-dynamics model or a multi-period execution model. 
Effects such as path dependence, no-arbitrage dynamics of the state variables, optimal execution over time and full dynamic smile modelling require additional structure. 
This also clarifies the connection with no-arbitrage. 
As a local overlay, the method can sit on top of the desk's arbitrage-free marginal pricing model. 
If the overlay is instead embedded in a full dynamic model, then the usual martingale restrictions of the chosen pricing measure must be imposed. 
Liquidity is treated in the same spirit, as an execution-cost layer on hedge trades, not as a change in the marginal arbitrage-free price process.

Section~\ref{sec:related} reviews related work. 
Section~\ref{sec:notation} fixes notation and conventions. 
Section~\ref{sec:framework} introduces the general geometric overlay. 
Section~\ref{sec:vv-connection} specializes the curvature construction to Vanna--Volga smile targets. 
Section~\ref{sec:liquidity} develops the liquidity and execution-cost penalty. 
Section~\ref{sec:constLambda} studies the practically important case of bucketed, state-constant impact inputs. 
Section~\ref{sec:example} presents the \texttt{EURUSD} and \texttt{USDTRY} case studies. 
Section~\ref{sec:extensions} outlines further local applications. 
Section~\ref{sec:local_geometry_dynamics} discusses the boundary between the local overlay, pricing dynamics and no-arbitrage. 
Section~\ref{sec:conclusion} concludes. 
The appendices collect metric-reconstruction formulas, coordinate-change identities, component formulas for the liquidity penalty and deferred proofs.


\section{Related Work}
\label{sec:related}
\nosectionappendix

This paper relates to several strands of the literature, but its contribution is distinct from each of them. 
We do not propose a new diffusion model, a new smile model or a new execution model. 
The object of the paper is the local Greek-based P\&L layer itself: the convention by which short-horizon factor moves, first-order hedge sensitivities and second-order P\&L terms are represented. 
We develop a local framework that lets a desk add additional quadratic features to an existing option model without replacing the underlying pricing model or its first-order hedge convention. 
In this sense, the closest point of contact is option P\&L attribution and hedging practice. 
The smile, liquidity, transaction-cost and geometry literatures provide ingredients that we combine at the local quadratic level.

A first relevant strand studies local option P\&L and its decomposition into exposures to the underlying and implied volatility. 
For example, \cite{CarrWu2020} links option valuation to daily P\&L attribution through the Black--Scholes--Merton formula. 
Our paper differs in both objective and construction. 
We do not derive a new pricing relation from no-arbitrage conditions. 
Instead, we take the pricing model and first-order hedge convention as given and study how the second-order P\&L layer can be modified or supplemented. 
Put differently, attribution papers typically decompose P\&L under a chosen Greek system, whereas we focus on the local structure of that Greek system when additional desk features are introduced.

A second strand concerns smile corrections and short-horizon approximations. 
In practice, desks routinely move beyond Black--Scholes (BS) to account for smiles and skews, and in foreign exchange the Vanna--Volga (VV) recipe remains a widely used benchmark, cf.~\cite{CastagnaMercurio2007,Wystup2006Book}. 
On the model side, local volatility and stochastic volatility frameworks generate short-time implied-volatility expansions, including~\cite{Dupire1994}, the SABR asymptotics of~\cite{Hagan2002} and more general small-time expansions~\cite{BerestyckiBuscaFlorent2002,LorigPagliaraniPascucci2015}. 
Our contribution is different. 
We do not commit to a particular diffusion, smile dynamics or global volatility surface model. 
We show how a chosen local quadratic smile adjustment, such as VV, can be represented as a curvature overlay on the P\&L layer of an existing model while preserving the desk's first-order hedge convention.

A third strand studies liquidity, execution costs and market impact. 
Quadratic execution costs are standard in optimal execution and inventory models, beginning with~\cite{AlmgrenChriss2000}, and later work allows for transient impact and cross-impact, cf.~\cite{BouchaudFarmerLillo2009,GatheralSchied2013,BenzaquenMastromatteoBouchaud2017}. 
Related but distinct is the literature on option hedging and pricing with transaction costs, from volatility-adjusted replication to utility-based and no-transaction-region approaches, cf.~\cite{Leland1985,DavisPanasZariphopoulou1993,WhalleyWilmott1997}. 
We use these ideas in a different way. 
Rather than solving a dynamic execution or transaction-cost pricing problem, we use quoted widths, representative clips and a desk rebalancing rule to induce a local quadratic penalty on factor moves. 
The primary object is the cost of the net hedge trade. 
When the resulting penalty is smooth and non-degenerate on the relevant subspace, it can also be used as a metric for adjusted second-order sensitivities. 
Netting, internal crossing and cost amplification therefore arise from the same quadratic cost rather than from separate add-ons on a per-deal basis.

A fourth strand uses geometric structure in finance. 
Geometric and heat-kernel approaches typically start from a diffusion model and interpret its local covariance structure as a metric governing short-maturity behavior, cf.~\cite{Varadhan1967,HenryLabordere2008}. 
Our use of geometry is more modest and more local. 
It is not a new pricing geometry for the underlying market model. 
It is a way to make explicit the convention used to encode local quadratic P\&L overlays on top of an existing model. 
At second order, ordinary Hessians depend on the chosen risk-factor representation. 
A connection provides a controlled way to represent curvature overlays, while a metric or quadratic penalty represents costs, risks or constraints whose scale is part of the object being modelled.

Finally, covariance-based distances provide one practical source of additional quadratic penalties when direct execution data are limited. 
Choosing a metric from the inverse covariance of factor moves connects to the Mahalanobis distance,~\cite{Mahalanobis1936}, and, when a likelihood is available, to the Fisher-metric viewpoint in information geometry, cf.~\cite{Amari2016}. 
In the present paper, these covariance-based constructions are secondary. 
They matter because they show that the same local framework can accommodate cost, risk and constraint overlays beyond the smile and liquidity examples developed in detail.

The paper also clarifies how the local overlay relates to pricing dynamics and no-arbitrage. 
This is not a separate contribution to arbitrage-free pricing with transaction costs. 
The point is narrower: as a local P\&L and hedging layer, the overlay can sit on top of the desk's marginal pricing model. 
If the overlay is embedded in a full dynamic model, then the usual martingale restrictions of the chosen pricing measure must be imposed. 
Similarly, liquidity enters as an execution-cost layer on hedge trades, not as a replacement of the marginal arbitrage-free pricing process.

Taken together, these literatures motivate the components of our framework but they do not deliver the object studied here: a local geometric layer for Greek-based risk management that separates two channels for enriching short-horizon option P\&L. 
Curvature overlays modify second-order P\&L attribution through a connection. 
Cost and risk overlays assign quadratic penalties to factor moves or hedge trades. 
This separation allows smile corrections, liquidity costs and other local quadratic desk features to be added to an existing option model in a coherent and coordinate-aware way.


\section{Notation and conventions}
\label{sec:notation}
\nosectionappendix

This section fixes the notation and conventions used throughout the paper.
There are three basic objects: the local state variables used to describe market moves, the hedge-trade space in which rebalancing takes place, and the local quadratic overlays built on top of them.
Most examples are displayed in the spot/ATM-volatility coordinates $(S,\sigma)$.
When needed, however, the local state can be enlarged to include additional desk variables, such as risk reversals (RR), butterflies (BF), curve factors, liquidity states or empirical P\&L factors.
All objects are local, evaluated around the current market state, and dependence on the state is sometimes suppressed when no confusion can arise.

We use standard conventions for matrix and differential notation. 
All vectors are column vectors. 
For a symmetric matrix $A$, $A \succ 0$ means that $A$ is positive definite (SPD), and $A \succeq 0$ means that $A$ is positive semidefinite. 
The superscript ${}^\top$ denotes transpose. 
For an SPD matrix $W$, we write $\|v\|_W^2 = v^\top W v$. 
For time derivatives we may write $\dot{\gamma}(t)=\mathrm{d}\gamma/\mathrm{d}t$.

Let $X$ be a finite-dimensional local state space with coordinates $x=(x^1,\ldots,x^d)$. 
In many displays we single out the two-coordinate block $y = (S,\sigma)$, where $S$ denotes the underlying level in the chart currently in use and $\sigma=\sigma_{\mathrm{ATM}}$ denotes the ATM implied-volatility level. 
More generally, the state may contain additional local coordinates needed by the overlay under consideration. 
We write
\[
x=(y,\theta),
\qquad
y=(S,\sigma_{\mathrm{ATM}}),
\qquad
\theta=(\theta^1,\ldots,\theta^p).
\]
For smile applications, $\theta$ may contain smile-shape variables such as risk reversals and butterflies, for example $\theta=\chi=(RR,BF,\ldots)$. 
Other choices are possible depending on the feature being represented: curve or carry factors, basis variables, correlation factors, funding or liquidity-state variables, margin or capital drivers, or empirical factors used in a local P\&L fit. 
Discrete regimes, such as time-of-day or liquidity buckets, are treated as fixed labels within a local calculation unless a smooth interpolation is introduced.

Unless stated otherwise, volatility coordinates are measured in vol points, so $\Vega=V_\sigma$ is in premium currency per vol point.\footnote{If decimal volatility $\hat\sigma\in[0,1]$ is used instead, then $\sigma=100\,\hat\sigma$ and derivatives rescale accordingly. The same convention applies to other vol-like coordinates. When spot and forward are treated as alternative charts over a short horizon, e.g. $(S,\sigma)$ versus $(F,\sigma)$, we hold discount factors/carry fixed over the short horizon. If rates or carry are material, they should be added to the state vector.} 
When needed, we also write $F$ for the forward price and $z=\log F$.

State indices $i,j,k$ run over the full state $x$. 
Repeated generic state indices are summed over the full state unless stated otherwise. 
For displayed components we use coordinate labels directly, such as $S,\sigma,\mathrm{RR}$ or $\mathrm{BF}$. 
For example, $\widetilde H_{SS}$ denotes the spot--spot component of the adjusted Hessian. 
These coordinate labels are component labels, not summation indices. 
Thus, in an expression such as $C^k_{SS}V_k$, only the generic index $k$ is summed, and it is summed over all coordinates in $x$.

Instrument indices are $r,s\in\{1,\dots,m\}$. 
Repeated instrument indices are summed over the hedge instruments, consistently with usual matrix notation. 
For instance, component formulas for products such as $M^\top\Lambda \, M$ use the same convention. 
To avoid clashing with the instrument index $s$, we write $s^{\mathrm{price}}$ for price half-spreads and $s^{\mathrm{vol}}$ for volatility half-spreads, with optional qualifiers such as ``per unit'' or ``clip''. 
We reserve $\Delta$ for the Greek Delta and $\delta$ for small increments.

For each hedge instrument $r$ we fix a trade unit $u_r$, such as cash notional for spot, contracts for futures or vega units for options. 
Hedge positions and hedge trades are expressed in the same vector space $\mathcal{H}\simeq\mathbb{R}^m$. 
A hedge position is written $q=(q_1,\dots,q_m)^\top$, where $q_r$ is the signed position in instrument $r$ measured in units $u_r$. 
A rebalancing trade is an increment $\delta q$. 
The hedge price $P_r(x)$ is quoted in premium currency per unit $u_r$, so $q_rP_r(x)$ is a premium-currency amount.

Since execution costs act on rebalancing trades, a desk rebalancing rule is represented locally by
\begin{equation}
\label{eq:hedging_policy}
\delta q = M(x)\,\delta x,
\qquad
M(x)\in\mathbb{R}^{m\times d}.
\end{equation}
Equation~\eqref{eq:hedging_policy} encodes how a small factor move $\delta x$ over the hedging horizon translates into hedge trades $\delta q$ in trade units. 
The matrix $M$ may be obtained from a differentiable desired hedge position $q=q(x)$, in which case $M_{ri}(x)=\partial_{x^i}q_r(x)$. 
More generally, it can be viewed as the local linearization of the desk's rebalancing rule. 
When a calculation is restricted to a subset of state directions, such as the $(S,\sigma)$ block, $M$ denotes the corresponding column restriction.

Separately, the hedge-price sensitivities with respect to the state are
\begin{equation}
\label{eq:hedge_price_sens}
A_{ri}(x)=\partial_{x^i}P_r(x),
\qquad
A(x)\in\mathbb{R}^{m\times d},
\end{equation}
so $\delta P \approx A(x)\,\delta x$, with $P=(P_1,\dots,P_m)^\top$ and $\delta P=(\delta P_1,\dots,\delta P_m)^\top$. 
The matrices $A$ and $M$ play different roles: $A$ tells us how hedge prices move when factors move, while $M$ tells us what the desk trades when factors move under the chosen rebalancing rule.

Let $V(x)$ denote the baseline price. 
On the displayed spot/ATM-volatility block we use the standard notation
\[
V_S=\Delta,\qquad
V_{SS}=\Gamma,\qquad
V_\sigma=\Vega,\qquad
V_{S\sigma}=\Vanna,\qquad
V_{\sigma\sigma}=\Volga.
\]
If additional coordinates are included, their first-order sensitivities are written explicitly when useful, for example $V_{\mathrm{RR}}$ or $V_{\mathrm{BF}}$, or with desk mnemonics such as $RR01$ and $BF01$. 
The ordinary coordinate Hessian is $\overline H_{ij}=V_{ij}$. 
When referring to a particular instrument $r$ we append a subscript, e.g. $V_{ij,r}$. 
A superscript $\star$ denotes a generic target local quadratic form, for example $H^{\mathrm{VV}}$ for a Vanna--Volga target or $H^{\mathrm{emp}}$ for an empirical local P\&L target.

To avoid confusion with option Gamma, we denote the coefficients of an affine connection by $\smash{C^k_{ij}}$ rather than by the standard geometric notation $\smash{\Gamma^k_{ij}}$. 
Unless stated otherwise, we work with torsion-free connections, $\smash{C^k_{ij}}=\smash{C^k_{ji}}$, so that the covariant Hessian is symmetric and defines a quadratic form, cf.~\cite{do2016differential,Lee2018Riemannian}.

We use overlines for ordinary second-order objects and tildes for covariant or adjusted ones. 
The ordinary second-order Taylor predictor is
\begin{equation}
\label{eq:ordinary_predictor}
\overline{\delta V}
\approx
V_i\,\delta x^i+\tfrac12\,\overline H_{ij}\,\delta x^i\delta x^j,
\qquad
\overline H_{ij}=V_{ij}.
\end{equation}
The adjusted second-order object used throughout the paper is
\begin{equation}
\label{eq:curved_second_order}
\widetilde H_{ij}=V_{ij}-C^k_{ij}V_k,
\end{equation}
and the associated local quadratic predictor is
\begin{equation}
\label{eq:curved_predictor}
\widetilde{\delta V}
\approx
V_i\,\delta x^i+\tfrac12\,\widetilde H_{ij}\,\delta x^i\delta x^j.
\end{equation}
The key point is that the overlay acts only on the quadratic term: the baseline first-order hedge Greeks remain $V_i$.

When the reported P\&L predictor is evaluated on the displayed $(S,\sigma)$ block, we write the relevant components directly. 
For example,
\[
\widetilde H_{SS}
=
V_{SS}
-
C^k_{SS}V_k,
\qquad
\widetilde H_{S\sigma}
=
V_{S\sigma}
-
C^k_{S\sigma}V_k,
\qquad
\widetilde H_{\sigma\sigma}
=
V_{\sigma\sigma}
-
C^k_{\sigma\sigma}V_k.
\]
In these expressions, the repeated index $k$ is summed over the full state. 
Thus, if $x=(S,\sigma,\mathrm{RR},\mathrm{BF})$, then
\[
\widetilde H_{SS}
=
V_{SS}
-
C^S_{SS}V_S
-
C^\sigma_{SS}V_\sigma
-
C^{\mathrm{RR}}_{SS}V_{\mathrm{RR}}
-
C^{\mathrm{BF}}_{SS}V_{\mathrm{BF}}.
\]
Consequently, delta and ATM-vega neutrality alone need not eliminate a smile curvature overlay. 
What matters is first-order exposure to the full state included in the local model.

For the liquidity calibration we also use positive reference clips $Q=(Q_1,\dots,Q_m)^\top$. 
The clip $Q_r$ is measured in the same trade unit $u_r$ as $q_r$ and represents a typical executable size within a calibration bucket, for example a time-of-day or liquidity regime over which widths and clips are treated as approximately constant. 
These clips are inputs used to translate quoted widths into quadratic cost coefficients per unit. 
They are not hedge positions and they are not optimization variables.

With the trade units fixed above, the trade-cost matrix $\Lambda$ is a symmetric bilinear form on the hedge space $\mathcal{H}$. 
Unless otherwise stated, we shall take $\Lambda \succ 0$ so that $\tfrac12\,\delta q^\top\Lambda\,\delta q$ is a cost in premium currency for any mix of trade units, as $[\Lambda_{rs}] = \text{(premium-currency amount)} / (u_r u_s)$. 
Some applications allow $\Lambda\succeq0$, in which case the induced object is a seminorm and regularization is needed only when an inverse or non-degenerate metric is required.

If option widths are quoted in vol points, we convert them to price half-spreads per unit through
\begin{equation}
\label{eq:price-half-spread-per-unit}
s^{\mathrm{price}}_{r,\text{per unit}}
=
s^{\mathrm{vol}}_r\,\Vega^{\mathrm{quote}}_r(S,\sigma,T),
\end{equation}
where $\Vega^{\mathrm{quote}}_r$ is the positive quote-vega, expressed per trade unit $u_r$, used by the desk to translate a volatility width into a premium width. 
Signed vega exposure is still denoted by $V_\sigma$ or $\Vega$ in the P\&L expansion. 
For spot, and for instruments quoted directly in price units, we use the quoted price half-spread per unit.

Combining $\Lambda$ with the local hedge map yields a quadratic cost on factor moves,
\begin{equation}
\label{eq:lq_metric}
g_\ell(x)=M(x)^\top\Lambda\,M(x),
\end{equation}
so that $\tfrac12\,\delta x^\top g_\ell(x)\,\delta x$ is again a cost in premium currency. 
If $\Lambda\succ0$ and $M$ has full column rank, then $g_\ell$ is positive definite. 
Otherwise it is positive semidefinite. 
This is still sufficient for computing local execution costs. 
A non-degenerate or regularized version is needed only when $g_\ell$ is used as a genuine metric to define a connection. 
The paper later uses $g_\ell$ as the main example of a quadratic penalty induced by execution costs.

The construction is not specific to foreign exchange. 
What matters is unit consistency: $\Lambda$ must be expressed in premium currency and $M$ must be measured in the trade units used for hedging. 
With those conventions in place, the same $\Lambda\mapsto g_\ell$ mapping carries over to other asset classes.


\section{A geometric overlay for local quadratic features}
\label{sec:framework}
\nosectionappendix

Desks rarely run option books on a raw second-order Taylor expansion.
In practice they add smile corrections, marking conventions, liquidity adjustments, scenario terms, capital charges and other local features that matter for short-horizon P\&L and hedging, see Sec.~\ref{sec:extensions}.
Not every such feature is intrinsically quadratic, and some require additional modelling.
The object studied here is narrower: a local quadratic representation over the hedging horizon.
Within that scope, the difficulty is not that these features are unavailable, but that they are often added one by one, with different conventions and at different levels of aggregation.
This section introduces the general object used in the paper: a local geometric overlay that lets quadratic desk features be incorporated into an existing option model without replacing the underlying pricing model or its first-order hedge convention.
Table~\ref{tab:desk_to_framework} summarizes the translation between desk language and the formal objects used below.

\begin{table}[t]
\centering
\small
\begin{tabular}{p{0.40\textwidth} p{0.52\textwidth}}
\toprule
\textbf{Desk object} & \textbf{Representation in the framework} \\
\midrule
Risk factors &
State vector $x=(x^1,\ldots,x^d)$, e.g. $(S,\sigma)$,
$(\log F,\sigma)$ or $(S,\sigma,\theta)$ \\

Deal or book value &
Scalar value function $V(x)$ or portfolio value $V_\pi(x)$ \\

Small market move &
Factor increment $\delta x$ \\

First-order hedge convention &
Gradient $V_i$, kept unchanged by the overlay \\

Curvature overlay
(e.g.\ smile correction, marking conv., empirical P\&L target) &
Target quadratic form $H^\star$ \\

Adjusted second-order P\&L term &
$\tfrac12\,\widetilde H_{ij}\delta x^i\delta x^j$,
where $\widetilde H_{ij}=V_{ij}-C^k_{ij}V_k$ \\

Desk rebalancing rule &
Local hedge response $\delta q\approx M(x)\delta x$. $M_{ri}=\partial_{x^i}q_r$ if generated by a differentiable desired hedge position $q(x)$.  \\

Quadratic hedge-trade cost &
$L(\delta q)\approx \tfrac12\,\delta q^\top\Lambda\,\delta q$ \\

Induced execution-cost penalty in factor space &
$g_\ell(x) = M(x)^\top \Lambda \, M(x)$ \\

Other cost or risk overlays
(e.g. scenario terms, risk normalization, capital charges) &
Quadratic penalties $g_a \succeq 0$ combined into
$g_{\mathrm{eff}} = \sum_a \eta_a g_a$, with $\eta_a \ge 0$ \\

Book-level netting &
Cost applied to the net hedge trade
$\delta q_\pi=\sum_\nu w_\nu\,\delta q_\nu$ \\
\bottomrule
\end{tabular}
\caption{Main desk objects and their representation in the framework.}
\label{tab:desk_to_framework}
\end{table}

Let $\overline{\delta V} \approx V_i\,\delta x^i+\tfrac12\,V_{ij}\,\delta x^i\delta x^j$ be the baseline local predictor delivered by the underlying model.
The overlay does not replace the base model and it does not alter the baseline first-order hedge convention.
For curvature overlays, its role is to modify only the quadratic P\&L attribution.
The adjusted predictor, as introduced in Sec.~\ref{sec:notation}, Eqs.~\eqref{eq:curved_second_order} and~\eqref{eq:curved_predictor}, is
\begin{equation}
\label{eq:overlay_predictor}
\widetilde{\delta V}
\approx
V_i\,\delta x^i + \tfrac12\,\widetilde H_{ij}\,\delta x^i\delta x^j,
\qquad
\widetilde H_{ij}=V_{ij}-C^k_{ij}V_k.
\end{equation}
The connection $C^k_{ij}$ is not assumed a priori to come from a global pricing model or a diffusion.
It is chosen locally so that the quadratic term reflects the curvature feature the desk wants to represent.
Cost and risk overlays enter separately as explicit quadratic penalties, as described below.

The framework distinguishes two classes of local quadratic input. First, one may specify a target quadratic form $H^\star$ for the second-order P\&L term. This is the right object for curvature overlays: smile curvature corrections, alternative model Hessian targets, marking convention effects, cross-gamma overlays or empirical short-horizon residuals. The target $H^\star$ is a P\&L coefficient and need not be positive semidefinite. Second, one may specify one or more quadratic penalties on factor moves,
\begin{equation}
\label{eq:generic_penalty}
\tfrac12\,\delta x^{\!\top}  g_a(x)\,\delta x,
\qquad
g_a(x) \succeq 0,
\end{equation}
to represent costs, risk charges or constraints. Examples include execution costs, liquidity, risk normalization, scenario penalties and capital add-ons. When several such penalties matter, they can be combined into an effective penalty
\begin{equation}
\label{eq:effective_metric}
g_{\mathrm{eff}}(x)=\sum_{a=1}^A \eta_a\,g_a(x),
\end{equation}
where the coefficients $\eta_a$ convert the different terms into a common desk objective or P\&L scale.
When the penalties are expressed in common P\&L or desk-objective units, the same local description can be written as a cost-adjusted or penalty-adjusted local objective,
\begin{equation}
\label{eq:net_overlay_predictor}
\widetilde{\delta V}_{\!\!\mathrm{net}}
\approx
V_i\,\delta x^i
+\tfrac12\,\widetilde H_{ij}\,\delta x^i\delta x^j
-\tfrac12\,\delta x^\top g_{\mathrm{eff}}(x)\,\delta x.
\end{equation}
Depending on the application, the penalty may be recorded explicitly as a cost term, used to define cost-aware or constraint-aware rebalancing rules, or both.

A penalty need not represent execution cost. It may also represent a risk normalization. For example, let $\Xi(x)=\operatorname{Cov}[\delta x\mid x]$ be an empirical covariance matrix of factor moves over the hedging horizon, estimated from historical or regime-filtered data. After the usual shrinkage or eigenvalue regularization, the inverse covariance $g_{\mathrm{cov}}(x) = \Xi(x)^{-1}$ defines the local Mahalanobis penalty $\tfrac12\,\delta x^\top g_{\mathrm{cov}}(x)\,\delta x$. This term does not measure execution cost and does not by itself change the mark-to-market value. It measures the size of a move in risk units, accounting for the relative scale and correlation of the factors. It is useful when the desk wants rebalancing thresholds, scenario comparisons or model diagnostics to be expressed in statistically comparable units rather than raw spot and volatility units. This is the usual Mahalanobis normalization, cf.~\cite{Mahalanobis1936}.

It is useful to distinguish the two ways geometry enters the framework. For a curvature overlay, the primitive input is a target quadratic P\&L coefficient $H^\star$. A connection is the device used to represent that target through the covariant Hessian. No metric, distance or cost scale is required. For a cost or risk overlay, the primitive input is a positive quadratic penalty $g$, such as an execution cost, margin, capital or risk-normalization penalty. This penalty can be used directly to measure costs, distances or constraint violations. When it is smooth and non-degenerate on the relevant subspace, it may also be treated as a metric and its Levi--Civita connection $C(g)$ can be used in the covariant Hessian. These two routes coincide only when the chosen affine connection is the Levi--Civita connection of a metric.

The distinction matters for interpretation. A metric contains an absolute scale for costs and distances, while its Levi--Civita connection is unchanged by a constant rescaling $g\mapsto \alpha g$, $\alpha>0$. Thus a uniform rescaling of a quadratic penalty changes cost levels, distance thresholds and risk budgets, but not the second-order adjustment obtained from $C(g)$. Relative weights, cross-terms and state-dependence of the penalty are what change the induced connection.

The most direct use of the overlay is to replace the baseline quadratic term by a local target chosen by the desk. The next proposition states the idea in its most general form.

\begin{propositionrep}[Local matching of a quadratic target]
  \label{prop:local_matching}
  Fix a state $x$ and a baseline value function $V$. Let $H^\star(x)$ be a
  symmetric target quadratic form. Suppose that a torsion-free affine
  connection satisfies
  \begin{equation}
    \label{eq:local_matching_condition}
    C^k_{ij}(x)\,V_k(x)
    =
    V_{ij}(x)-H^\star_{ij}(x),
    \qquad
    i,j=1,\ldots,d.
  \end{equation}
  Then the adjusted predictor in Eq.~\eqref{eq:overlay_predictor} has the
  same first-order term as the baseline predictor and second-order term equal
  to the target:
  \begin{equation}
    \label{eq:target}
    \widetilde{\delta V}
    \approx
    V_i\,\delta x^i
    +
    \tfrac12\,H^\star_{ij}\,\delta x^i\delta x^j.
  \end{equation}
  If $H^\star_{ij}=V_{ij}$, then the contracted correction vanishes for $V$
  and the baseline quadratic predictor is recovered.
\end{propositionrep}

\ProofSeeAppendix{apx-prop:local_matching}

\begin{appendixproof}
  \phantomsection\label{apx-prop:local_matching}

  By definition of the covariant Hessian,
  \[
  \widetilde H_{ij}
  =
  V_{ij}
  -
  C^k_{ij}V_k.
  \]
  Using Eq.~\eqref{eq:local_matching_condition} gives
  \[
  \widetilde H_{ij}
  =
  V_{ij}
  -
  \big(V_{ij}-H^\star_{ij}\big)
  =
  H^\star_{ij}.
  \]
  The first-order term is unchanged because the covariant derivative of a
  scalar is its ordinary derivative, so
  $\nabla_i V=\partial_i V=V_i$. Substituting
  $\widetilde H_{ij}=H^\star_{ij}$ into
  Eq.~\eqref{eq:overlay_predictor} gives the stated predictor. If
  $H^\star_{ij}=V_{ij}$, then Eq.~\eqref{eq:local_matching_condition} gives
  $C^k_{ij}V_k=0$, so $\widetilde H_{ij}=V_{ij}$ for the value function being
  expanded.
\end{appendixproof}

Proposition~\ref{prop:local_matching} captures the main economic role of the curvature overlay. A desk can keep the baseline model and its first-order hedge convention while replacing only the quadratic term by a target that encodes the local feature of interest. The proposition is conditional: it states what happens once the contracted correction has been chosen. The identification problem is separate. In practice, the coefficients are recovered by applying Eq.~\eqref{eq:local_matching_condition} across calibration instruments. Stable recovery requires the relevant first-order gradients to span the state directions through which the correction is meant to act.

Only the contracted term $C^k_{ij}V_k$ needs to be identified for the local P\&L predictor. This makes calibration local and tractable. Suppose instrument $r$ has baseline gradient $V_{k,r}$ and target quadratic form $H^\star_{ij,r}$. Then, for each quadratic component $(ij)$,
\begin{equation}
\label{eq:generic_calibration_by_instruments}
C^k_{ij}\,V_{k,r}
=
V_{ij,r}-H^\star_{ij,r}.
\end{equation}
Stacking instruments at the same state gives a linear system. Define the Greek design matrix $\mathcal G$ by $\mathcal{G}_{rk} = V_{k,r}$, and define
\[
u_{ij}
=
\begin{bmatrix}
C^1_{ij}\\
\vdots\\
C^d_{ij}
\end{bmatrix},
\qquad
b_{ij}
=
\begin{bmatrix}
V_{ij,1}-H^\star_{ij,1}\\
\vdots\\
V_{ij,m}-H^\star_{ij,m}
\end{bmatrix}.
\]
The calibration problem is
\begin{equation}
  \label{eq:calibration_problem}
  \mathcal G\,u_{ij} = b_{ij}.
\end{equation}
In the two-factor case, the rows of $\mathcal G$ are simply $(\Delta_r,\Vega_r)$, and one solves the three systems corresponding to $(ij)=(SS),(S\sigma),(\sigma\sigma)$. In an enlarged state, the same rows may also contain sensitivities to smile or other local coordinates, for example
\[
(\Delta_r,\Vega_r,RR01_r,BF01_r,\ldots).
\]
Thus a spot/ATM-volatility curvature component such as $\widetilde H_{SS}$ may depend on first-order exposure to the full state, not only on delta and ATM vega. When the system is overidentified or poorly conditioned, weighted least squares or ridge regularization can be used. Section~\ref{sec:vv-connection} applies this construction to a Vanna--Volga target.

A connection-only curvature overlay has one important limitation. It changes the scalar Hessian through the contraction $C^k_{ij}V_k$. Therefore, if the value being expanded is first-order neutral with respect to every state coordinate included in $x$, the correction vanishes for that value. This is not a problem for a book that is merely delta neutral and ATM-vega neutral if the state also includes smile variables such as RR or BF coordinates. It does mean that any residual curvature of a book neutral to the entire included state must be represented either by enlarging the state or by adding an independent residual quadratic term.

A second use of the framework is to encode frictions or constraints through local quadratic penalties. Execution costs are the leading example. Starting from a quadratic trade-space penalty
\begin{equation}
  \label{eq:trade_space_penalty}
  L(\delta q)\approx \tfrac12 \, \delta q^{\!\top} \Lambda\,\delta q
\end{equation}
and the local hedge map $\delta q\approx M(x)\,\delta x$, we obtain the induced factor-space penalty
\begin{equation}
\label{eq:factor_space_penalty}
L(\delta x)\approx \tfrac12\,\delta x^\top g_\ell(x)\,\delta x,
\qquad
g_\ell(x)=M(x)^\top\Lambda\,M(x).
\end{equation}
The same construction applies to any other quadratic cost or constraint that can be expressed locally in factor space.

This cost/risk view matters for two reasons. First, it gives a local cost or penalty for moving in the state variables, which can be used to define liquidity-aware distances, least-cost directions and rebalancing schedules. Second, when one wants the same penalty to shape the geometry of the quadratic predictor itself, a natural choice is to take the connection in Eq.~\eqref{eq:overlay_predictor} to be the Levi--Civita connection of $g_{\mathrm{eff}}$, provided the effective penalty is smooth and non-degenerate on the relevant subspace. This induced connection is optional: the primary economic object on the cost side is the penalty itself.

The construction is naturally book-level. Desks do not rebalance deals in isolation: they rebalance net portfolio exposures and execution costs are paid on the net hedge trade. This is important because quadratic costs do not aggregate deal by deal. If portfolio $\pi$ has deals indexed by $\nu=1,\ldots,N_\pi$, weights $w_\nu$ and net hedge trade $\delta q_\pi = \sum_{\nu=1}^{N_\pi} w_\nu\,\delta q_\nu$, then the portfolio cost is
\begin{equation}
\label{eq:portfolio_cost_netting}
L_\pi
=
\tfrac12 \, \delta q_\pi^\top \Lambda \, \delta q_\pi
=
\sum_{\nu=1}^{N_\pi}
\tfrac12\,w_\nu^2 \, \delta q_\nu^\top \Lambda \, \delta q_\nu
+
\sum_{\nu<\mu}
w_\nu \, w_\mu \, \delta q_\nu^\top \Lambda \, \delta q_\mu.
\end{equation}
The cross terms in Eq.~\eqref{eq:portfolio_cost_netting} are the mechanical source of internal crossing when hedge trades offset and of cost amplification when they reinforce one another. Thus netting is not an extra adjustment layered on top of the framework. It appears directly once costs are applied to the book trade.

The same book-level view applies on the curvature side. Let $V_\pi = \sum_{\nu=1}^{N_\pi} w_\nu V_\nu$ be the portfolio value. Once the connection has been fixed at book level, the covariant Hessian is linear in the scalar value function:
\begin{equation}
\label{eq:portfolio_covariant_hessian}
\widetilde H_{ij,\pi}
=
\sum_{\nu=1}^{N_\pi} w_\nu\,\widetilde H_{ij,\nu}.
\end{equation}
Equivalently, one may apply the covariant Hessian directly to $V_\pi$. This equality assumes that the same book connection is used for every term. Hessians adjusted at deal level and computed with different deal-level connections need not aggregate consistently. This is one practical advantage of the overlay: it gives a common quadratic layer for the portfolio rather than a collection of add-ons per deal.

A useful consequence of the framework is coordinate awareness.
The components of $\widetilde H$ change when the desk writes factor moves in spot, forward, log-forward or other smooth coordinates, but the scalar quadratic predictor for the same local move does not change when the connection, target and increments are transformed consistently.\footnote{Under a smooth change of state coordinates $y^\alpha = y^\alpha(x)$, e.g.\ $y = (\log F,\sigma)$, increments transform as $\delta y^\alpha = \frac{\partial y^\alpha}{\partial x^i}\delta x^i$. The ordinary coordinate Hessian $V_{ij}$ is not tensorial: $V_{\alpha\beta} = \frac{\partial x^i}{\partial y^\alpha}\frac{\partial x^j}{\partial y^\beta}\,V_{ij} + \frac{\partial^2 x^k}{\partial y^\alpha\partial y^\beta}\,V_k$.}
This is practical because it makes quadratic attribution and backtests comparable across systems and risk-factor parameterizations.\footnote{Marking conventions such as sticky-strike or sticky-delta should not be confused with mere coordinate changes: they specify how the smile is transported and therefore enter through the target quadratic form or the connection. Once specified, however, the resulting local overlay can be expressed in any smooth coordinate system.}
Appendix~\ref{app:connection_transform} illustrates this point in a one-dimensional forward/log-forward example.

\begin{propositionrep}[Parameterization invariance]
\label{prop:parameterization_invariance}
For any affine connection $\nabla$, the adjusted second-order object
$\widetilde H=\nabla(\mathrm{d}V)$ is a $(0,2)$ tensor. In particular, under
a smooth change of coordinates $\bar x=\bar x(x)$,
\begin{equation}
\label{eq:covariant_hessian_transform}
\widetilde H_{ab}
=
\frac{\partial x^i}{\partial \bar x^a}
\frac{\partial x^j}{\partial \bar x^b}
\,\widetilde H_{ij},
\end{equation}
where $a,b$ denote components in the $\bar x$-coordinate system. If
$\delta x$ and $\delta\bar x$ are the components of the same tangent move,
then
\begin{equation}
\label{eq:quadratic_predictor_invariant}
V_i\,\delta x^i+\tfrac12\,\widetilde H_{ij}\,\delta x^i\delta x^j
=
V_a\,\delta \bar x^a+\tfrac12\,\widetilde H_{ab}\,\delta \bar x^a\delta \bar x^b.
\end{equation}
If the connection is torsion free, then $\widetilde H$ is symmetric and
defines a quadratic form.
\end{propositionrep}

\ProofSeeAppendix{apx-prop:parameterization-invariance}

\begin{appendixproof}
  \phantomsection\label{apx-prop:parameterization-invariance}
  Define $\widetilde H$ coordinate freely by, for vector fields $U,W$,
  \[
  \widetilde H(U,W)
  =
  U\big(W(V)\big)-(\nabla_U W)(V).
  \]
  This expression is $C^\infty(X)$-linear in each argument, hence defines a
  $(0,2)$ tensor. Evaluating at $U=\partial_i$, $W=\partial_j$ yields the
  stated coordinate expression. For the symmetry part, compute
  $\widetilde H(\partial_i,\partial_j)-\widetilde H(\partial_j,\partial_i)$,
  use $[\partial_i,\partial_j]=0$ in coordinates, and use
  \[
  T(\partial_i,\partial_j)
  =
  \nabla_{\partial_i}\partial_j
  -
  \nabla_{\partial_j}\partial_i
  -
  [\partial_i,\partial_j],
  \]
  which gives
  \[
  \widetilde H_{ij}-\widetilde H_{ji}
  =
  -\,T^k_{ij}\,V_k.
  \]
  Thus $\widetilde H$ is symmetric when the connection is torsion free. See
  \cite{do2016differential,Lee2018Riemannian}.
\end{appendixproof}

By contrast, the ordinary coordinate Hessian $V_{ij}$ is not tensorial.
The ordinary Hessian term therefore depends on the chosen chart unless it is interpreted as part of a full coordinate-specific Taylor expansion.
In the present framework, parameterization invariance is not the main objective but it is a useful consequence of writing the adjusted quadratic layer intrinsically.

Appendix~\ref{app:connection_transform} gives the explicit coordinate-change formulas used in the implementation, including the log-forward chart $z=\log F$. In particular, if a desk target $H^\star$ is specified in one quoting chart and the calculation is carried out in another, $H^\star$ is transported as a quadratic form rather than recomputed as an ordinary Hessian.

This section provides the general object used in the rest of the paper. Section~\ref{sec:vv-connection} develops the curvature overlay in detail for a Vanna--Volga target snd Sec.~\ref{sec:liquidity} develops the cost/risk overlay for liquidity and execution costs.


\section{Smile Corrections as a Local Quadratic Target}
\label{sec:vv-connection}
\nosectionappendix

This section specializes the curvature-overlay part of the framework to a familiar desk adjustment: Vanna--Volga (VV) smile corrections, cf.~\cite{Wystup2006Book,CastagnaMercurio2007}.
The objective is not to replace the baseline pricing model by a global VV model.
Rather, VV is used here as a source of a local quadratic target for the second-order P\&L layer (common in practice).
The baseline price and first-order hedge convention are kept fixed, while the quadratic term is adjusted to match the chosen VV-implied curvature target at the calibration state.
Once this target has been specified, the calibration is the same as for any other curvature overlay, such as an alternative-model Hessian, a marking convention or an empirical short-horizon P\&L target.

Let $x$ denote the state used by the desk in the local calculation. In the simplest implementation $x=(S,\sigma)$, where $S$ is the underlying level in the chosen quoting convention and $\sigma=\sigma_{\mathrm{ATM}}$ is the ATM implied-volatility level. More generally, the state may include additional coordinates,
\[
x=(S,\sigma,\theta),
\]
where $\theta$ may contain smile variables such as risk reversals and butterflies, or other local variables relevant to the overlay. In this section we display the spot/ATM-volatility block because this is the usual form of the desk P\&L expansion. Unless explicitly stated otherwise, derivatives of $V^{\mathrm{BS}}$ are derivatives of the baseline mark with respect to the chosen state coordinates. In the two-coordinate case these are the usual Black--Scholes (BS) Greeks,
\[
V_S=\Delta,\qquad
V_\sigma=\Vega,\qquad
V_{SS}=\Gamma,\qquad
V_{S\sigma}=\Vanna,\qquad
V_{\sigma\sigma}=\Volga.
\]
In component formulas in this section, derivatives without an explicit superscript are baseline derivatives. In an enlarged smile state, first-order derivatives such as $V_{\mathrm{RR}}$ or $V_{\mathrm{BF}}$ are the corresponding chain-rule sensitivities of the baseline mark to the desk's smile coordinates. These sensitivities may arise through the surface map or quoting convention used by the desk. If the baseline mark is truly independent of a coordinate at the calibration point, then that coordinate cannot contribute to the contracted correction for that mark.

For an instrument $r$, the adjusted quadratic object is
\begin{equation}
\label{eq:vv_cov_hessian}
\widetilde H_{ij,r}
=
V^{\mathrm{BS}}_{ij,r}
-
C^k_{ij}\,V^{\mathrm{BS}}_{k,r},
\end{equation}
where the displayed pair $(ij)$ may be, for example, $SS$, $S\sigma$ or $\sigma\sigma$, while the repeated index $k$ is summed over the full state included in $x$. The connection coefficients $C^k_{ij}$ are chosen locally, at the calibration state, so that $\widetilde H_{ij,r}$ matches a smile-implied target quadratic form. The first-order term remains the baseline hedge convention, so the local predictor for the instrument or book is
\begin{equation}
\label{eq:vv_local_predictor}
\widetilde{\delta V}
\approx
V_i^{\mathrm{BS}}\delta x^i
+
\tfrac12\,\widetilde H_{ij}\delta x^i\delta x^j.
\end{equation}
If a desk also wants to change the first-order hedge convention, one may replace $V_i^{\mathrm{BS}}$ by the VV first derivatives. That gives the ordinary local Taylor expansion of the VV-corrected price. The overlay studied here instead keeps the baseline first-order hedge convention fixed and changes only the quadratic layer.

For a two-coordinate state $x=(S,\sigma)$, writing the displayed components explicitly gives
\begin{align}
\widetilde{H}_{SS}        &= V_{SS}-C^{S}_{SS}V_S-C^{\sigma}_{SS}V_\sigma
= \Gamma-C^{S}_{SS}\,\Delta-C^{\sigma}_{SS}\,\Vega \qquad = \widetilde{\Gamma},\label{eq:vv_eff_gamma}\\
\widetilde{H}_{S\sigma}    &= V_{S\sigma}-C^{S}_{S\sigma}V_S-C^{\sigma}_{S\sigma}V_\sigma
= \Vanna-C^{S}_{S\sigma}\,\Delta-C^{\sigma}_{S\sigma}\,\Vega = \widetilde{\Vanna},\label{eq:vv_eff_vanna}\\
\widetilde{H}_{\sigma\sigma}&= V_{\sigma\sigma}-C^{S}_{\sigma\sigma}V_S-C^{\sigma}_{\sigma\sigma}V_\sigma
= \Volga-C^{S}_{\sigma\sigma}\,\Delta-C^{\sigma}_{\sigma\sigma}\,\Vega \,\,\, = \widetilde{\Volga}.\label{eq:vv_eff_volga}
\end{align}
In an enlarged state the same displayed components may depend on additional first-order exposures. For example, if $x=(S,\sigma,\mathrm{RR},\mathrm{BF})$, then
\[
\widetilde H_{SS}
=
V_{SS}
-
C^S_{SS}V_S
-
C^\sigma_{SS}V_\sigma
-
C^{\mathrm{RR}}_{SS}V_{\mathrm{RR}}
-
C^{\mathrm{BF}}_{SS}V_{\mathrm{BF}}.
\]
Thus delta neutrality and ATM-vega neutrality alone do not make the smile correction vanish when the book still has first-order exposure to the smile-shape coordinates included in the state.

For displayed spot/ATM-volatility moves, with the remaining state coordinates held fixed, the adjusted predictor is
\begin{equation}
\label{eq:vv_effective_pl}
\widetilde{\delta V}
\approx
\Delta\,\delta S
+
\Vega\,\delta\sigma
+
\tfrac12
\left[
\widetilde{\Gamma} \, (\delta S)^2
+
2 \, \widetilde{\Vanna} \, \delta S \, \delta\sigma
+
\widetilde{\Volga} \, (\delta\sigma)^2
\right],
\end{equation}
with the understanding that the adjusted coefficients may have been obtained from the full state through the contraction in Eq.~\eqref{eq:vv_cov_hessian}. This is the desk interpretation of the construction: the overlay replaces the baseline quadratic Greeks by smile-adjusted quadratic Greeks while leaving the first-order hedge ratios unchanged.

Fix the quoting and marking coordinates used to compute the VV adjustment.
Let $V^{\mathrm{VV}}_r$ denote the VV-corrected price of calibration instrument $r$.
In those coordinates, we define the VV target quadratic coefficient by differentiating $V^{\mathrm{VV}}_r$ with respect to the displayed directions:
\begin{equation}
\label{eq:vv_target_hessian}
H^{\mathrm{VV}}_{ij,r}
=
\frac{\partial^2 V^{\mathrm{VV}}_r}{\partial x^i\partial x^j},
\qquad
(ij)\in\{SS,S\sigma,\sigma\sigma\}.
\end{equation}
Equivalently, for the displayed spot/ATM-volatility block,
\[
\big( H^{\mathrm{VV}}_{ij,r} \big)_{i,j\in\{S,\sigma\}}
=
\begin{pmatrix}
\Gamma^{\mathrm{VV}}_r & \Vanna^{\mathrm{VV}}_r \\[2pt]
\Vanna^{\mathrm{VV}}_r & \Volga^{\mathrm{VV}}_r
\end{pmatrix}.
\]
This definition is conditional on the quoting and marking convention used to compute the VV adjustment.
Once specified, $H^{\mathrm{VV}}$ is treated as the desk's target quadratic form at the calibration point.
If the calculation is later expressed in another chart, the target is transported as a quadratic form, not recomputed by differentiating $V^{\mathrm{VV}}_r$ again in the new coordinates.
This convention is important because ordinary Hessians depend on the chosen chart, whereas the local quadratic target is the object the desk has specified.

Calibrating $C^k_{ij}$ means matching the covariant Hessian to the VV target. The matching condition is therefore
\begin{equation}
\label{eq:vv_matching_condition}
\widetilde H_{ij,r}=H^{\mathrm{VV}}_{ij,r}
\quad\Longleftrightarrow\quad
C^k_{ij}V^{\mathrm{BS}}_{k,r}
=
V^{\mathrm{BS}}_{ij,r}
-
H^{\mathrm{VV}}_{ij,r}.
\end{equation}
In the notation of Sec.~\ref{sec:framework}, this is the generic calibration equation with $H^\star=H^{\mathrm{VV}}$ and $b_{ij,r} = V^{\mathrm{BS}}_{ij,r}-H^{\mathrm{VV}}_{ij,r}$. The same connection coefficients are used across calibration instruments at the same state. The only VV-specific object in this calibration is the source of the target $\smash{H^{\mathrm{VV}}}$. The linear matching problem is the generic curvature-overlay problem of Sec.~\ref{sec:framework}.

In the reduced two-factor case with two calibration instruments whose baseline gradients are linearly independent, set
\[
D=\Delta_1\Vega_2-\Delta_2\Vega_1\neq0.
\]
Then, for each $(ij)\in\{SS,S\sigma,\sigma\sigma\}$,
\begin{equation}
\label{eq:vv_two_instrument_system}
\begin{bmatrix}
\Delta_1 & \Vega_1\\
\Delta_2 & \Vega_2
\end{bmatrix}
\begin{bmatrix}
C^S_{ij}\\
C^\sigma_{ij}
\end{bmatrix}
=
\begin{bmatrix}
b_{ij,1}\\[2pt]
b_{ij,2}
\end{bmatrix},
\end{equation}
and the closed-form solution is
\begin{equation}
\label{eq:vv_two_instrument_closed_form}
C^S_{ij}
=
\frac{b_{ij,1}\Vega_2-b_{ij,2}\Vega_1}{D},
\qquad
C^\sigma_{ij}
=
\frac{\Delta_1 b_{ij,2}-\Delta_2 b_{ij,1}}{D}.
\end{equation}

For a general calibration set, using the Greek design matrix $\mathcal G$ introduced in Sec.~\ref{sec:framework}, with weights $W=\mathrm{diag}(w_1,\dots,w_m)\succeq0$ and ridge parameter $\eta\ge0$, write $\|a\|_W^2=a^\top W a$.
This is a seminorm if some weights are zero.
We estimate
\begin{equation}
\label{eq:vv_ridge_minimizer}
u_{ij}
=
\operatorname*{\arg\min}_{u\in\mathbb{R}^d}
\left(
\|\mathcal G u-b_{ij}\|_W^2+\eta\|u\|_2^2
\right),
\qquad
u_{ij}
=
\begin{bmatrix}
C^1_{ij}\\
\vdots\\
C^d_{ij}
\end{bmatrix}.
\end{equation}
This is the standard weighted ridge least-squares problem, cf.~\cite{Bjorck1996LeastSquares}. For $\eta>0$, the normal matrix $\mathcal G^\top W\mathcal G+\eta I_d$ is positive definite, so the solution is unique and is given by
\begin{equation}
\label{eq:vv_ridge_solution}
u_{ij}
=
(\mathcal G^\top W \mathcal G+\eta I_d)^{-1}
\mathcal G^\top W b_{ij}.
\end{equation}
For $\eta=0$, the solution need not be unique if the calibration gradients do not span all included state directions. In that case, the fitted residual is still well defined, but one must choose a convention for the unidentified null-space component, such as the minimum-norm least-squares solution or a small ridge penalty.

The three usual VV systems for $(ij)=(SS),(S\sigma),(\sigma\sigma)$ share the same design matrix $\mathcal G$ but have different right-hand sides. In the two-factor case, the rank condition says that the calibration instruments' $(\Delta,\Vega)$ vectors cannot be nearly collinear. In an enlarged state, the corresponding condition is spanning of the relevant full-state Greek vectors, for example
\[
(\Delta_r,\Vega_r,RR01_r,BF01_r,\ldots).
\]
Only the directions spanned by the calibration instruments can be identified from the data; remaining components must be fixed by modelling choice or regularization.

As a diagnostic of target inconsistency, numerical noise or poor spanning by the calibration instruments, we monitor the fitted residual
\begin{equation}
\label{eq:vv_residual}
\rho_{ij}
=
\|\mathcal G u_{ij}-b_{ij}\|_W .
\end{equation}
When $\eta=0$, we also record the minimal residual
\begin{equation}
\label{eq:vv_minimal_residual}
\rho^{\min}_{ij}
=
\min_{u\in\mathbb{R}^d}\|\mathcal G u-b_{ij}\|_W .
\end{equation}
This is the distance, in the weighted calibration seminorm, from the target adjustment to the span of the calibration gradients. Thus exact matching on the positively weighted instruments is possible if and only if $\rho^{\min}_{ij}=0$. Equivalently, if $\mathcal G_+$ and $b_{ij,+}$ denote the rows with $w_r>0$, exact matching is possible if and only if $b_{ij,+}\in \operatorname{col}(\mathcal G_+)$.

If the smile target is locally identical to the baseline quadratic term, then $H^{\mathrm{VV}}_{ij,r}=V^{\mathrm{BS}}_{ij,r}$ for all calibration instruments and hence $b_{ij}=0$. Under full-rank identification the fitted correction is zero. More generally, the ridge solution with $\eta>0$, or the minimum-norm least-squares solution when $\eta=0$, also gives $u_{ij}=0$. This is the desired sanity check: the overlay is inactive when the target contains no additional local smile information. If $\eta=0$ and the system is underdetermined, an arbitrary least-squares solution may include a null-space component. In implementation this is fixed by the chosen regularization or minimum-norm convention.

The enlarged-state point is important but limited. A connection-only curvature overlay changes the scalar Hessian through the contraction $C^k_{ij}V_k$. Therefore a book that is first-order neutral with respect to every coordinate included in the state receives no correction from this channel. Delta- and ATM-vega-neutrality alone are not enough for this conclusion when the state also includes smile-shape coordinates and the book has nonzero exposure to them. Conversely, any residual curvature of a book neutral to the entire included state is not generated by a connection-only curvature overlay; it must be represented either by enlarging the state or by adding an independent residual quadratic term.

The VV calibration above does not require the fitted connection to be the Levi--Civita connection of a metric.
It is a locally calibrated affine connection, identified only through its contracted action on the relevant baseline Greeks.
Thus the connection should not be interpreted as a unique market-implied geometry. It is a device for representing the chosen local quadratic target.
When a metric interpretation is useful, for example to compare with the liquidity construction below, the corresponding Levi--Civita formulas and metric reconstruction equations are collected in Appendix~\ref{app:metric_reconstruction}.
In all cases, the displayed Vanna adjustment has the form
\begin{equation}
\widetilde H_{S\sigma}
=
V_{S\sigma}
-
C^k_{S\sigma}V_k,
\end{equation}
so the adjustment can act through any first-order exposure included in the state, not only through delta and ATM vega.


\section{Liquidity and Execution Costs}
\label{sec:liquidity}
\nosectionappendix

This section develops the cost/risk-overlay channel of the framework, with execution costs and liquidity as the worked example.
The starting point is simple: a small move in the risk factors leads to a hedge trade under the desk's rebalancing rule, and that hedge trade has a cost.
We approximate this cost locally by a quadratic form in hedge-trade space $\mathcal H$, calibrated from quoted widths, representative clips and, when relevant, cross-impact.
Composing this trade-cost model with the desk's hedge rule induces a quadratic penalty on factor moves, denoted by $g_\ell(x)$.
This penalty has two practical uses.
First, it measures the local execution cost of a factor move and can be used directly in liquidity-aware rebalancing rules.
Second, when it is smooth and non-degenerate on the relevant factor subspace, it can be treated as a local metric. Its Levi--Civita connection can then be used in the covariant Hessian to produce liquidity-adjusted second-order Greeks.
These two uses are related but distinct.

Within a local liquidity bucket, let $\delta q \in \mathcal H \simeq \mathbb{R}^m$ denote the vector of hedge trades, such as spot or vanilla hedges, expressed in each instrument's natural trade unit.
The bucket fixes the relevant liquidity regime while the cost coefficients may either be held fixed inside the bucket or allowed to vary smoothly with the state.
We use the local quadratic surrogate
\begin{equation}
  \label{eq:quadratic_cost_dq}
  L(\delta q)
  \approx
  \tfrac12\,\delta q^\top \Lambda(x)\,\delta q,
  \qquad
  \Lambda(x)\succ0,
\end{equation}
where $\Lambda$ is calibrated from quoted widths, representative clip sizes and, when relevant, cross-impact terms. This is not intended to replace a full execution model. It is a local approximation anchored to the sizes and market conditions relevant for the hedge bucket.

A desk rebalancing rule maps a factor move into hedge trades, e.g. a small factor move $\delta x$ leads to a hedge trade (by the desk rule) and therefore incurs an execution cost (loss, $L$).
Locally, we write this response as $\delta q \approx M(x)\,\delta x$.
If the rule is generated by a differentiable desired hedge position $q=q(x)$, then $M_{ri}(x)=\partial_{x^i}q_r(x)$.
More generally, $M(x)$ should be read as the local linearization of the desk's rebalancing policy.
Composing this linear hedge response with Eq.~\eqref{eq:quadratic_cost_dq} gives
\begin{equation}
  \label{eq:gLiq}
  L(\delta x)
  \approx
  \tfrac12\,\delta x^\top
  \underbrace{M(x)^\top\Lambda(x)M(x)}_{=\,g_\ell(x)}
  \delta x.
\end{equation}
Thus $g_\ell$ is the execution-cost penalty induced on factor moves by the hedge instruments, the cost matrix and the desk's rebalancing rule. In plain terms, $\Lambda$ says how expensive hedge instruments are to trade, $M$ says what the desk trades when factors move and $g_\ell$ says how expensive that factor move is after the hedge rule is applied.\footnote{Geometrically, $g_\ell$ is the pullback of the bilinear form $\Lambda$ through the map $\delta x\mapsto\delta q=M\delta x$. Units are inherited consistently: if $q_r$ has trade unit $u_r$, then $[\Lambda_{rs}]=\text{price}/(u_ru_s)$ and $[M_{ri}]=u_r/\text{unit}(x^i)$, so $[g_{\ell,ij}]=\text{price}/(\text{unit}(x^i)\text{unit}(x^j))$.}

The penalty $g_\ell$ charges only those factor moves that actually lead to hedge trades under the chosen rebalancing rule. If $M$ has full column rank on the factor directions being considered (equiv., the hedge rule reacts independently to moves in sufficient directions) then every nonzero move in those directions has positive local execution cost. If some direction is not hedged, or if hedge responses are nearly redundant, then $g_\ell$ is singular or close to singular: some factor moves appear cheap simply because the chosen hedge rule does not trade against them. In that case, cost comparisons can be restricted to the traded directions or a small baseline penalty can be added before computing distances or a Levi--Civita connection.

We record the basic positivity and invariance properties of the induced liquidity penalty, i.e. matrix $g_\ell$.
The matrix representation of $g_\ell$ changes under a change of factor coordinates but the scalar cost assigned to a given local move does not.

\begin{lemmarep}[Basic properties of the induced liquidity penalty]
\label{lem:pullback_metric}
Let $g_\ell(x)$ be as in Eq.~\eqref{eq:gLiq}. Then:

(i) Reparameterizing the risk factors does not change the scalar cost proxy $L(\delta x)$ for a given local move $\delta x$.

(ii) $g_\ell\succeq0$. It is positive definite on the factor subspace considered if and only if the hedge response $M$ has full column rank on that subspace.

(iii) Changing hedge trade units or the hedge basis does not change the induced factor cost, provided the trade-cost matrix is expressed in the corresponding units.
\end{lemmarep}

\ProofSeeAppendix{apx-lem:pullback_metric}

\begin{appendixproof}
  \phantomsection\label{apx-lem:pullback_metric}
  Throughout, fix a state $x$ and suppress the explicit $x$-dependence.

  (i) The object defined in Eq.~\eqref{eq:gLiq} is the quadratic form obtained by composing the trade cost with the linear response map $\delta x\mapsto\delta q$. To see the coordinate statement directly, write the same local move using two sets of factor coordinates, $x$ and $y=y(x)$. Their increments satisfy $\delta x = J\,\delta y$, with $J=\partial x/\partial y$. The hedge response is the same trade, so
  \[
  \delta q=M_x\,\delta x=M_y\,\delta y,
  \]
  which gives $M_y=M_xJ$ at the point. Hence
  \[
  g_{\ell}(y)
  =
  M_y^\top \Lambda M_y
  =
  J^\top(M_x^\top\Lambda M_x)J
  =
  J^\top g_{\ell}(x)J.
  \]
  Therefore the scalar cost proxy is unchanged:
  \[
  \delta y^\top g_{\ell}(y)\delta y
  =
  \delta x^\top g_{\ell}(x)\delta x.
  \]

  (ii) For any factor increment $w$,
  \[
  w^\top g_\ell w
  =
  w^\top M^\top\Lambda Mw
  =
  (Mw)^\top\Lambda(Mw)
  \ge 0,
  \]
  since $\Lambda\succ0$. Strict positivity for every nonzero $w$ in the factor subspace is equivalent to $Mw\neq0$ for every nonzero $w$ in that subspace, that is, to full column rank of $M$ on the subspace considered.

  (iii) Let $q'=Rq$ be an invertible linear reparameterization of hedge trades, including simple unit changes such as contracts versus vega notional. Then $\delta q'=R\delta q$, and the same scalar trade cost is written as
  \[
  \tfrac12\,\delta q^\top\Lambda\,\delta q
  =
  \tfrac12\,\delta q'^{\top}\Lambda'\,\delta q',
  \qquad
  \Lambda'=R^{-T}\Lambda R^{-1}.
  \]
  The corresponding hedge response is $M'=RM$, hence
  \[
  M'^\top\Lambda'M'
  =
  M^\top R^\top(R^{-T}\Lambda R^{-1})RM
  =
  M^\top\Lambda M
  =
  g_\ell.
  \]
  The induced factor cost is therefore unchanged.
\end{appendixproof}

The remaining modelling choice is the local hedge response $M(x)$.
When $M$ comes from a differentiable desired hedge position $q(x)$, it is the Jacobian of that position.
In this paper, however, $M(x)$ is best viewed as a policy input: it reflects the hedge instruments available to the desk and the exposures the desk chooses to control.
A convenient and transparent specification is the least-cost rebalancing rule. After a small factor move, choose the cheapest hedge trade, measured in the $\Lambda$ cost, that restores the target exposures:
\begin{equation}
  \label{eq:least_cost_problem}
  \min_{\delta q\in\mathbb{R}^m}
  \left(
  \tfrac12\,\delta q^\top\Lambda\,\delta q
  \right)
  \quad\text{subject to}\quad
  B\,\delta q=c.
\end{equation}
Here $B\in\mathbb{R}^{p\times m}$ collects the exposure changes delivered by each hedge instrument, one row per controlled exposure, and $c\in\mathbb{R}^p$ is the required change in the controlled exposure vector.

For example, if the desk controls $E(x)=(\Delta,\Vega)$, then after a small factor move $\delta E \approx J_E(x)\,\delta x$, where $J_E(x)$ is the Jacobian of the book exposures with respect to the factors $(S,\sigma)$. If the book is on target at the hedge time, the hedge trade must offset this drift, so $c \approx -\,J_E(x)\delta x$.

Assuming $\operatorname{rank}(B)=p$, the Karush--Kuhn--Tucker conditions give the minimum-cost hedge trade, cf.~\cite{BoydVandenberghe2004},
\begin{equation}
  \label{eq:min_cost_sol}
  \delta q
  =
  \Lambda^{-1} B^\top
  \big( B \, \Lambda^{-1} B^\top \big)^{-1} c.
\end{equation}
In the on-target case this has the form $\delta q=M(x)\delta x$, with
\[
M(x)
=
-\,\Lambda^{-1}B^\top
\big( B \, \Lambda^{-1} B^\top \big)^{-1} J_E(x),
\]
and therefore
\begin{equation}
  \label{eq:least_cost_solution}
  g_\ell(x)
  =
  M(x)^\top \Lambda \, M(x)
  =
  J_E(x)^\top \big( B \, \Lambda^{-1} B^\top \big)^{-1} J_E(x).
\end{equation}
Typical clip sizes $Q=(Q_1,\dots,Q_m)$ enter only upstream, when quoted widths are converted into the entries of $\Lambda$. Once $\Lambda$ is fixed for the bucket, the clip sizes do not appear in Eqs.~\eqref{eq:least_cost_problem} and~\eqref{eq:least_cost_solution}.

Equation~\eqref{eq:least_cost_solution} makes explicit what drives the liquidity penalty at book level. The matrix $\Lambda$ encodes instrument-level execution costs, $B$ specifies which exposures are controlled and which hedge instruments are used, and $J_E$ captures how the book's target exposures drift with the risk factors. The same expression applies whether the hedge set contains two instruments or a larger strip of liquid hedges.

To make this concrete, take spot and one at-the-money vanilla as hedges, with trade vector $\delta q=(\delta q_S,\delta q_C)^\top$ and $\Lambda=\operatorname{diag}(\lambda_S,\lambda_C)$. Suppose the desk restores delta and vega neutrality after a small move by requiring $B\,\delta q=-\delta E$, where $E=(\Delta_{\text{book}},\Vega_{\text{book}})^\top$, $\delta E\approx J_E(x) \, \delta x$,
\[
B(x)
=
\begin{pmatrix}
1 & \Delta_C(x)\\
0 & \Vega_C(x)
\end{pmatrix}
\qquad \text{and} \qquad
J_E(x)
=
\begin{pmatrix}
\Gamma_{\mathrm{book}} & \Vanna_{\mathrm{book}}\\
\Vanna_{\mathrm{book}} & \Volga_{\mathrm{book}}
\end{pmatrix}.
\]
The entries of $B$ are exposures per trade unit, consistent with the units used to calibrate $\Lambda$. The least-cost trade is obtained from Eq.~\eqref{eq:min_cost_sol}, $\delta q = \Lambda^{-1}B^\top(B\Lambda^{-1}B^\top)^{-1} \big(-J_E\,\delta x\big)$, and the induced penalty on $(S,\sigma)$ moves is given by Eq.~\eqref{eq:least_cost_solution}. Larger hedge sets and other neutrality constraints only change the matrices $B$ and $J_E$.

In implementation, poor conditioning of $B\Lambda^{-1}B^\top$ indicates that the chosen hedge instruments do not restore the target exposures robustly in the $\Lambda$ cost. One can then add another hedge, regularize $\Lambda$ or relax the neutrality constraints through a penalty on residual exposures, cf.~\cite{BoydVandenberghe2004}. If $\operatorname{rank}(B)<p$, a $\Lambda^{-1}$-weighted pseudoinverse gives the corresponding minimum-norm or least-squares hedge, cf.~\cite{GolubVanLoan2013}.

The induced matrix $g_\ell$ can be used directly to guide rebalancing. For a sequence of hedge updates at states
\[
x_0\to x_1\to\cdots\to x_N
\]
within a liquidity bucket, the leading-order execution cost implied by the local hedge rule is approximated by
\begin{equation}
  \label{eq:discrete_execution_energy}
  \mathcal E_N
  =
  \frac12
  \sum_{n=0}^{N-1} (x_{n+1}-x_n)^\top g_\ell(x_n) \, (x_{n+1}-x_n).
\end{equation}
The number of hedge updates is an operational input. This is important because a purely quadratic execution-cost model would otherwise mechanically favor splitting a fixed trade into arbitrarily many small trades. Thus Eq.~\eqref{eq:discrete_execution_energy} should be read as a local cost approximation for a given rebalancing frequency, not as a complete dynamic execution model.

The desk does not control the market path of the factors. The role of $g_\ell$ is to give a local cost scale for rebalancing decisions. A simple rule is a liquidity-cost trigger: starting from the last hedge state $x_{\mathrm{last}}$, rebalance when
\begin{equation}
  \label{eq:liquidity_distance_trigger}
  \frac12
  (x-x_{\mathrm{last}})^\top g_\ell(x_{\mathrm{last}})
  (x-x_{\mathrm{last}})
  \ge
  L_{\mathrm{trig}}.
\end{equation}
This gives one local quadratic band in factor space instead of separate coordinate-by-coordinate thresholds. Directions that are expensive to hedge receive a tighter effective band, while directions that are cheap to hedge receive a wider one.

For planned transitions, or as a local approximation to a multi-step hedge adjustment, $g_\ell$ also defines equal-cost intermediate hedge states. The continuous analogue of Eq.~\eqref{eq:discrete_execution_energy} is
\begin{equation}
  \label{eq:continuous_execution_energy}
  \mathcal E[\gamma]
  =
  \frac12
  \int_0^1
  \dot\gamma^i(t)\,
  g_{\ell,ij}(\gamma(t))\,
  \dot\gamma^j(t)\,dt,
\end{equation}
where $t$ indexes progress along the hedge schedule, not calendar time.
When $g_\ell$ is smooth and positive definite on the relevant subspace, a path minimizing this energy is the least-cost path induced by the local liquidity penalty.\footnote{In geometric terminology, this is a geodesic of the metric $g_\ell$, with the caveat that $g_\ell$ is a local execution-cost proxy rather than a full dynamic execution model.} It satisfies
\begin{equation}
  \label{eq:liquidity_geodesic}
  \ddot\gamma^a(t)
  +
  C^a_{ij}\!\left(g_\ell(\gamma(t))\right)
  \dot\gamma^i(t)\dot\gamma^j(t)
  =
  0.
\end{equation}
Thus geodesics provide a local decision rule for constructing cost-efficient intermediate hedge states or planned transitions. They do not prescribe the market path of the factors or the optimal number of hedge updates (those remain operational inputs). For the applications in this paper, Eq.~\eqref{eq:liquidity_geodesic} is mainly a conceptual guide. For short moves in a stable bucket, it is usually enough to freeze $g_\ell$ at the current state and work with the resulting local quadratic norm.

The constant-metric case gives an elementary benchmark.

\begin{propositionrep}[Equal-cost splitting under a frozen quadratic penalty]
\label{prop:equal_arc}
Fix an SPD matrix $g$ and a total factor move $\Delta x=x_N-x_0\in\mathbb{R}^d$. Among all decompositions $\Delta x = \sum_{n=0}^{N-1}\Delta x_n$, $\Delta x_n=x_{n+1}-x_n$, the quadratic cost $\mathcal E_N$ in Eq.~\eqref{eq:discrete_execution_energy} is minimized by the constant-step schedule $\Delta x_n = \Delta x / N$, $n = 0,\dots,N-1$. The minimum is
\begin{equation}
  \label{eq:minimal_constant_step}
  \mathcal E_N^{\min}
  =
  \frac{1}{2N}\Delta x^\top g\,\Delta x
  =
  \frac{1}{2N}\|x_N-x_0\|_g^2.
\end{equation}
\end{propositionrep}

\ProofSeeAppendix{apx-prop:equal_arc}

\begin{appendixproof}
  \phantomsection\label{apx-prop:equal_arc}
  Let $g^{1/2}$ be the SPD square root and set $y_n=g^{1/2}\Delta x_n$. Then
  \[
  \mathcal E_N
  =
  \frac12\sum_{n=0}^{N-1} y_n^\top y_n,
  \qquad
  \sum_{n=0}^{N-1}y_n=g^{1/2}\Delta x.
  \]
  By Cauchy--Schwarz or Jensen,
  \[
  \sum_{n=0}^{N-1} y_n^\top y_n
  \ge
  \frac1N
  \left(\sum_{n=0}^{N-1} y_n\right)^\top
  \left(\sum_{n=0}^{N-1} y_n\right),
  \]
  with equality if and only if all $y_n$ are equal. Since $g^{1/2}$ is invertible, this is equivalent to all $\Delta x_n$ being equal.
\end{appendixproof}

When $g_\ell$ is approximately diagonal and slowly varying near $x_0=(S_0,\sigma_0)$, say $g_\ell(x_0)\approx\mathrm{diag}(a,b)$, the scaled coordinates
\[
(\xi,\zeta)
=
\left(
\sqrt{a}\,(S-S_0),
\sqrt{b}\,(\sigma-\sigma_0)
\right)
\]
make local least-cost paths close to straight lines for small moves. A cost-balanced move satisfies $a(x_0)(\delta S)^2 \approx b(x_0)(\delta\sigma)^2$, equivalently $|\delta S/\delta\sigma| \approx \sqrt{b(x_0)/a(x_0)}$. More generally, for short steps in a stable bucket, one does not need to solve a geodesic ODE explicitly: Factorize $g_\ell(x_0)\approx R^\top R$, for example by Cholesky, work with local whitened displacements $u=R(x-x_0)$, take $N$ equal steps in $u$ and map back to $x$.

The same matrix $g_\ell$ can also be used to adjust second-order sensitivities. When $g_\ell$ is positive definite, or after regularization on the relevant subspace, let $\smash{C^k_{ij}}(g_\ell)$ denote its Levi--Civita connection. In this liquidity application we take the connection in the covariant Hessian to be $\smash{C^k_{ij}} = \smash{C^k_{ij}}(g_\ell)$, so that
\begin{equation}
  \label{eq:liquidity_adjusted_hessian}
  \widetilde H_{ij}
  =
  V_{ij}
  -
  C^k_{ij}(g_\ell)V_k.
\end{equation}
This gives liquidity-adjusted second-order Greeks. The construction should not be confused with the target-fitted connection in Sec.~\ref{sec:vv-connection}: there the connection is fitted to a smile target, while here it is induced by a cost penalty.

To compute $C^k_{ij}(g_\ell)$, one needs $g_\ell$ and its first derivatives. With $M\in\mathbb{R}^{m\times d}$ and $g_\ell = M^\top \Lambda \, M$,
\begin{equation}
  \label{eq:state_first}
  \partial_i g_\ell
  =
  (\partial_i M)^\top\Lambda \, M
  +
  M^\top (\partial_i\Lambda) \, M
  +
  M^\top \Lambda \, (\partial_i M).
\end{equation}
Component formulas are collected in Appendix~\ref{app:gl_components}. In practice, $\partial_i g_\ell$ is often obtained by finite differences of $g_\ell(x)$, with smoothing or bucketing, rather than by evaluating analytic third derivatives of option values. When $\Lambda$ is constant inside a bucket, the $\partial_i\Lambda$ terms in Eq.~\eqref{eq:state_first} drop. The next section records the resulting simplifications and scale properties.

A useful scale check follows immediately. Suppose that, within a bucket, execution costs are uniformly rescaled, $\Lambda\mapsto\alpha\Lambda$ with constant $\alpha>0$, and that the hedge rule $M$ is fixed. Then $g_\ell\mapsto\alpha g_\ell$. The Levi--Civita connection is unchanged under this constant rescaling so the liquidity-adjusted second-order Greeks in Eq.~\eqref{eq:liquidity_adjusted_hessian} are unchanged. Absolute costs and distance thresholds, however, are multiplied by $\alpha$. Thus uniform market-wide widening affect cost levels and trigger thresholds, while changes in relative liquidity across hedge instruments, cross-impact or state-dependence of $M$ and $\Lambda$ affect the induced liquidity geometry. When $M$ is generated by the least-cost neutrality rule in Eq.~\eqref{eq:least_cost_solution}, a uniform rescaling of $\Lambda$ also leaves $M$ unchanged. Non-uniform changes in relative costs generally do not.

Finally, the liquidity construction is used at book level in the sense of Sec.~\ref{sec:framework}: the hedge map, the induced penalty $g_{\ell,\pi}$ and, when used, the connection $C(g_{\ell,\pi})$ are built from the net portfolio rebalancing rule. Execution costs are therefore applied to the net hedge trade. Hence internal crossing and cost amplification are captured by the same quadratic cost rather than by separate add-ons at deal level.


\section{The Constant-$\Lambda$ Regime}
\label{sec:constLambda}
\nosectionappendix

This section studies a common implementation regime for the liquidity penalty introduced in Sec.~\ref{sec:liquidity}.
In many implementations, liquidity inputs are managed by bucket.
Dealers quote widths and firm sizes by product, venue, tenor and market regime, and these inputs are kept fixed until a time refresh, risk toggle or event-driven update.
We model this by taking the trade-space impact matrix $\Lambda$ to be constant in the state within a bucket, while allowing it to change from one bucket to the next.
Thus $\Lambda$ is piecewise constant across liquidity regimes.
All derivatives in this section are intra-bucket derivatives.
At a bucket boundary the desk refreshes $\Lambda$ and recomputes the local objects rather than differentiating through the jump.

A constant $\Lambda$ does not imply a constant liquidity penalty in factor space. It only means that the local cost of trading the hedge instruments is fixed within the bucket. The induced factor-space penalty $g_\ell(x) = M(x)^\top \Lambda \, M(x)$ may still vary with the state through the hedge response $M(x)$. Since $\partial_i\Lambda = 0$ inside the bucket, Eq.~\eqref{eq:state_first} reduces to
\begin{equation}
  \label{eq:state_first_constant}
  \partial_i g_\ell
  =
  (\partial_i M)^\top \Lambda \, M
  +
  M^\top \Lambda \, (\partial_i M),
  \qquad i=1,\ldots,d.
\end{equation}
Hence, in the constant-bucket case, the induced connection terms come from the state dependence of $M(x)$ and from the way this state dependence is weighted by $\Lambda$. Off-diagonal entries or relative weights in $\Lambda$ matter because they change the shape of the cost penalty, but they do not create a liquidity-induced Hessian correction by themselves if $g_\ell$ is locally constant.

A useful sanity check follows immediately.
If the components of $g_\ell$ are locally constant near a state $x_0$ in the chosen coordinates, then the Levi--Civita coefficients $C^k_{ij}(g_\ell)$ vanish at $x_0$, and the liquidity-adjusted Hessian coincides with the ordinary Hessian:
$\smash{\widetilde H_{ij}}(x_0)=V_{ij}(x_0)$.
Thus a nonzero mixed entry of $g_\ell$ is not by itself a Vanna-type correction or a mixed-Greek adjustment.
The correction comes from how the local cost penalty changes with the state, not merely from the presence of a cross term in the quadratic cost.

For numerics it is useful to factorize the impact matrix as $\Lambda=R^\top R$, for example by Cholesky, and write
\[
g_\ell = M^{\!\top} \Lambda \, M = (RM)^\top(RM).
\]
This Gram representation makes positive semidefiniteness explicit. Small eigenvalues of $g_\ell$ indicate that the hedge response does not charge one factor direction robustly, or that some factor responses are nearly redundant under the current hedge set and cost matrix. In those cases we either restrict cost comparisons to the traded directions or add a small positive definite baseline penalty before computing distances or Levi--Civita coefficients.

When a bucket changes, for example around a market event, we update $\Lambda$ and recompute $g_\ell$ and, when needed, the coefficients $C^k_{ij}(g_\ell)$. A uniform bucket-level rescaling $\Lambda\mapsto\alpha\Lambda$ with constant $\alpha>0$ rescales execution costs and distance thresholds but it does not change the Levi--Civita connection. What can change liquidity-adjusted Greeks is the shape of the penalty: relative costs across hedge instruments, cross-impact terms and the state dependence of $M(x)$. When $M$ is generated by the least-cost neutrality rule in Eq.~\eqref{eq:least_cost_solution}, a uniform rescaling of $\Lambda$ also leaves $M$ unchanged. Non-uniform changes in relative costs generally do not.

A two-factor, two-hedge example makes the mechanics transparent. Let the displayed factor block be $(S,\sigma)$, and let the hedge set consist of spot and one ATM vanilla option. To avoid overloading notation, denote the two hedges by $\mathrm{sp}$ and $\mathrm{van}$. Thus $\delta q = (\delta q_{\mathrm{sp}},\delta q_{\mathrm{van}})^\top$ and $\Lambda = \operatorname{diag}(\lambda_{\mathrm{sp}},\lambda_{\mathrm{van}})$. Write the local hedge response as
\[
\delta q=M(x)\delta x,
\qquad
M(x)=
\begin{pmatrix}
m_{\mathrm{sp},S}(x) & m_{\mathrm{sp},\sigma}(x)\\[2pt]
m_{\mathrm{van},S}(x) & m_{\mathrm{van},\sigma}(x)
\end{pmatrix}.
\]
Here the row label identifies the hedge instrument and the column label identifies the factor move. For example, $m_{\mathrm{sp},\sigma}$ is the spot trade generated by a unit volatility move under the desk's rebalancing rule (it is not a derivative of spot with respect to volatility). Then
\[
g_\ell(x)
=
M(x)^\top \Lambda M(x)
=
\begin{pmatrix}
g_{\ell,SS} & g_{\ell,S\sigma}\\[2pt]
g_{\ell,S\sigma} & g_{\ell,\sigma\sigma}
\end{pmatrix},
\]
with $g_{\ell,SS} = \lambda_{\mathrm{sp}}m_{\mathrm{sp},S}^2 + \lambda_{\mathrm{van}}m_{\mathrm{van},S}^2$, $g_{\ell,S\sigma} = \lambda_{\mathrm{sp}}m_{\mathrm{sp},S}m_{\mathrm{sp},\sigma} + \lambda_{\mathrm{van}}m_{\mathrm{van},S}m_{\mathrm{van},\sigma}$ and $g_{\ell,\sigma\sigma} = \lambda_{\mathrm{sp}}m_{\mathrm{sp},\sigma}^2 + \lambda_{\mathrm{van}}m_{\mathrm{van},\sigma}^2$. Even with diagonal execution costs in hedge space, the induced factor-space penalty is generally not diagonal. The reason is financial: the desk may use the same hedge instrument in response to more than one factor move. The mixed entry $g_{\ell,S\sigma}$ measures the cost-weighted interaction between the hedge trade generated by a spot move and the hedge trade generated by a volatility move. It should not be confused with the mixed connection coefficients in $C(g_\ell)$, which measure how the cost penalty changes across states.

We now recall how quoted widths and clips are converted into the diagonal entries of $\Lambda$. For a trade size $q$ in a given hedge instrument, crossing the spread gives the local slippage $\operatorname{slip}(q) = s^{\mathrm{price}}|q|$, where $s^{\mathrm{price}}$ is the per-unit half-spread in premium currency. We approximate this locally by the quadratic surrogate $\operatorname{slip}_{\mathrm{quad}}(q)=\tfrac12\lambda q^2$, cf.~\cite{AlmgrenChriss2000}, and anchor it at a representative clip size $Q > 0$: $\tfrac12\lambda Q^2=s^{\mathrm{price}}Q$. Thus
\begin{equation}
  \label{eq:lambda_sym}
  \lambda
  =
  \frac{2s^{\mathrm{price}}}{Q}.
\end{equation}
Equivalently, if the quoted half-spread is given for the full clip, so that $s^{\mathrm{price}}_{\mathrm{clip}} = s^{\mathrm{price}}_{\mathrm{per\,unit}}Q$, then
\begin{equation}
  \label{eq:lambda_clip}
  \lambda
  =
  \frac{2s^{\mathrm{price}}_{\mathrm{clip}}}{Q^2}
  =
  \frac{2s^{\mathrm{price}}_{\mathrm{per\,unit}}}{Q}.
\end{equation}
Thus $\lambda$ has units $\text{premium}/u^2$, where \emph{premium} is premium-currency amount and $u$ is the trade unit of $q$, for example notional, contracts or vega-risk units. Unit consistency is essential: $\Lambda$ and the hedge response $M$ must use the same trade units. For several independent hedge instruments, this gives $\Lambda = \operatorname{diag}(\lambda_1,\dots,\lambda_m)\succ0$. Cross-impact or common liquidity effects can be included through off-diagonal entries, provided the resulting matrix remains positive definite.

For vanilla options quoted in volatility points, quoted widths are first converted into premium widths using the desk's quote-vega convention. In the constant-$\Lambda$ regime this conversion is performed at the bucket reference state $x_b$ and then held fixed inside the bucket:
\begin{equation}
  \label{eq:vol_width_to_price_width}
  s^{\mathrm{price}}_{r,\mathrm{per\,unit}}
  =
  s^{\mathrm{vol}}_r\,\Vega^{\mathrm{quote}}_r(x_b,T).
\end{equation}
Here $\Vega^{\mathrm{quote}}_r$ is the positive quote-vega per trade unit used by the desk to translate a volatility width into a premium width. If the desk instead recomputes the vega conversion continuously as the state moves, then $\Lambda$ becomes state dependent and the full formula in Eq.~\eqref{eq:state_first}, including the $\partial_i\Lambda$ term, should be used. For spot, the price half-spread is used directly, e.g. in a non-\texttt{JPY} foreign-exchange pair, a pip is $10^{-4}$ in price units.

For asymmetric or tiered quotes, the same anchoring idea applies by fitting the local quadratic coefficient to the relevant executable clips.
The coefficient may be side-specific, tier-specific or smoothed across tiers, depending on how the desk records market depth.
If a single coefficient is selected for the bucket, the result remains in the constant-$\Lambda$ regime.
If the coefficient is allowed to depend on trade size, the object is better interpreted as a local execution-cost rule rather than a state-only metric. One may then freeze or localize it before using metric-based quantities such as Levi--Civita coefficients.
In the case studies below, we either use one bucket-level coefficient per hedge instrument or a smooth tiered version when market depth depends materially on trade size.

The quadratic cost is a local smooth surrogate for spread crossing.
A pure bid--ask cost is proportional to $|q|$, whereas the approximation
$\tfrac12\lambda q^2$ is anchored at the representative clip $Q$.
It therefore underestimates very small trades relative to a pure spread-crossing model and should be interpreted as a local penalty at the chosen execution scale.
If smaller trades are material, the desk should either choose a smaller calibration clip, impose a minimum executable clip or keep a separate proportional spread-crossing term.\footnote{A mixed proportional-plus-quadratic specification combines a spread-crossing term, proportional to $|q|$, with a quadratic impact term. The proportional component is common in option-hedging models with transaction costs while the quadratic component is standard in execution models, cf.~\cite{Leland1985,AlmgrenChriss2000}. The present paper does not develop the nonsmooth case.}
The quadratic part is the one that maps from hedge-trade space to factor space through
$g_\ell = M^\top \Lambda \, M$.

Operationally, a bucket implementation is straightforward.
The desk fixes the hedge universe and the rebalancing rule, converts quoted widths and clips into a trade-space matrix $\Lambda$, computes the local hedge response $M(x)$ and forms $g_\ell(x)=M(x)^\top \Lambda \, M(x)$.
The same object is then used either directly, to measure local execution costs and set liquidity-aware rebalancing thresholds, or, when a metric interpretation is needed, through its Levi--Civita connection $C(g_\ell)$ to compute liquidity-adjusted second-order Greeks.
If $g_\ell$ is poorly conditioned, a small positive definite baseline penalty can be added before computing distances or connection coefficients.


\section{Case Studies: FX Barrier Options}
\label{sec:example}
\nosectionappendix

This section illustrates the framework on two up--and--in call (UIC) foreign-exchange barrier options.
The first case is a one-year \texttt{EURUSD} option.
It is used mainly as a stability check for the curvature channel: the smile adjustment is expected to be small so a useful overlay should not introduce spurious variation into an already adequate local predictor.
We also use it as a high-liquidity benchmark for the cost channel.
The second case is a two-year \texttt{USDTRY} option and is used to illustrate the penalty side of the framework in a lower-liquidity market, where execution costs can become economically material.
All increments and costs are reported per unit notional in the premium currency unless otherwise stated.

The examples are diagnostic rather than a broad empirical validation exercise. Their purpose is to show how the two main channels of the framework are implemented. In the smile calculation, we match a Vanna--Volga (VV) target quadratic form and use the resulting adjusted second-order Greeks in a short-horizon Taylor predictor. In the liquidity calculation, we build a trade-space impact matrix $\Lambda$ from widths and clips, combine it with the hedge response $M(x)$ and obtain the induced factor-space penalty $g_\ell(x) = M(x)^\top \Lambda \, M(x)$. The initial market inputs are from 16 May 2022 for \texttt{EURUSD} and 2 May 2023 for \texttt{USDTRY}, including the relevant interest-rate curves and volatility surfaces.\footnote{All scripts and notebooks required to reproduce these experiments are publicly available as a GitHub repository at \href{https://github.com/Lu-Mengjue/curved-greeks-geometric-layer-option-pnl-adjustments}{\texttt{https://github.com/Lu-Mengjue/curved-greeks-geometric-layer-option-pnl-adjustments}}.}

For the reported P\&L comparison, we work on the displayed two-factor block and define the one-step model increment by
$\delta V_n = V(t_{n+1},S_{n+1},\sigma_{n+1}) - V(t_n,S_n,\sigma_n)$
on a reconstruction grid $(t_n)$.
We compare this increment with Taylor predictors that retain only spot and implied-volatility terms.
Therefore, the reported prediction errors also include omitted effects such as theta, rate exposure, smile-state moves and higher-order terms.
Time-series plots use weekly increments for readability, while distributional and error statistics are computed on a daily grid.

For the \texttt{EURUSD} smile case, we use a UIC barrier option with strike $K = 0.98$, barrier $B = 1.01$ and maturity $T = 1.0$. The benchmark valuation is the Reiner--Rubinstein BS barrier formula, cf.~\cite{ReinerRubinstein1991}. We compare four quantities: the closed-form Reiner--Rubinstein increment, the ordinary BS Taylor predictor, the full VV revaluation increment, cf.~\cite{Wystup2006Book,CastagnaMercurio2007}, and the connection-corrected Taylor predictor obtained by matching the VV quadratic target.

\begin{figure}[t]
  \centering
  \begin{subfigure}[b]{0.45\textwidth}
    \centering
    \includegraphics[width=\textwidth]{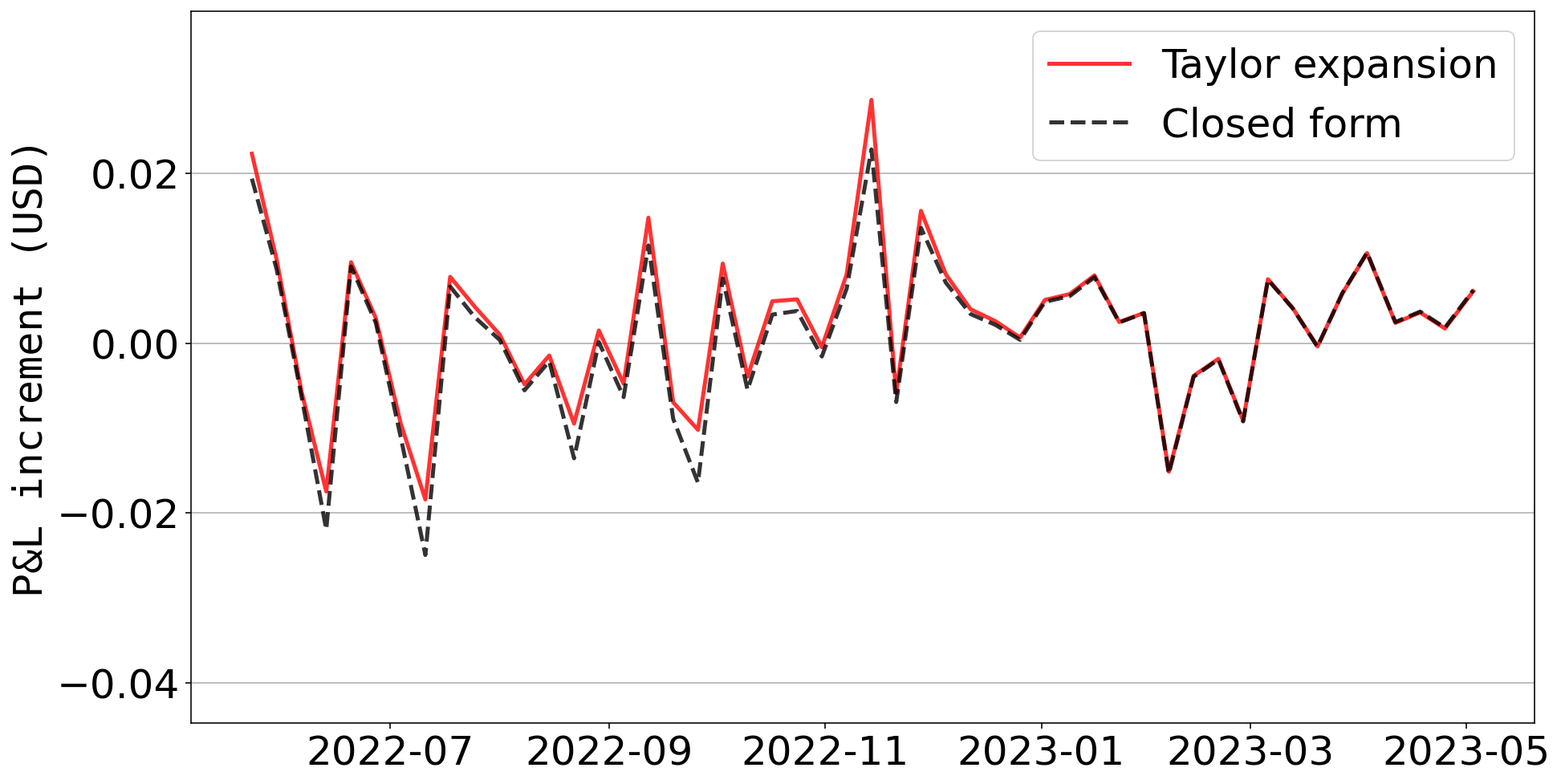}
    \caption{Reiner--Rubinstein vs. BS Taylor predictor.}
    \label{fig:fig1sub1}
  \end{subfigure}
  \hfill
  \begin{subfigure}[b]{0.45\textwidth}
    \centering
    \includegraphics[width=\textwidth]{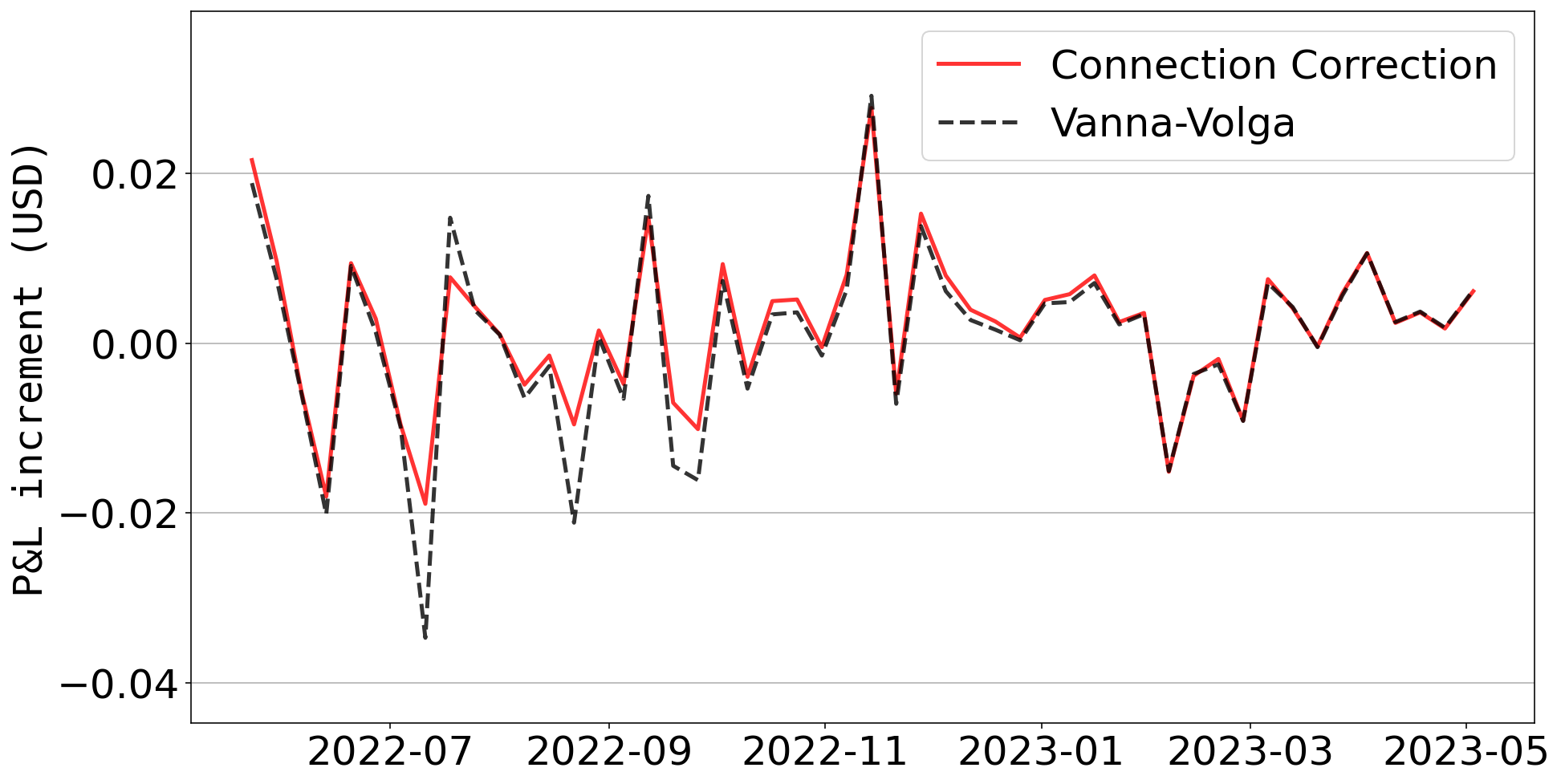}
    \caption{VV revaluation vs. connection correction.}
    \label{fig:fig1sub2}
  \end{subfigure}
  \caption{\texttt{EURUSD} UIC barrier weekly P\&L increments in \texttt{USD} per unit notional, with $K=0.98$, $T=1.0$, and $B=1.01$.}
  \label{fig:plandcorrection1}
\end{figure}

Figure~\ref{fig:plandcorrection1} shows the weekly increments. Panel~\ref{fig:fig1sub1} compares the closed-form Reiner--Rubinstein increment with the ordinary Taylor predictor. The largest deviations occur when omitted effects such as time decay, rate exposure and higher-order terms are more relevant. Panel~\ref{fig:fig1sub2} compares the VV revaluation increment with the connection-corrected predictor. In this case, the smile contribution is modest: the connection-corrected predictor remains close to the BS Taylor predictor and does not generate artificial variation when the target curvature is close to the baseline curvature.

\begin{figure}[t]
  \centering
  \begin{subfigure}[b]{0.45\textwidth}
    \centering
    \includegraphics[width=\textwidth]{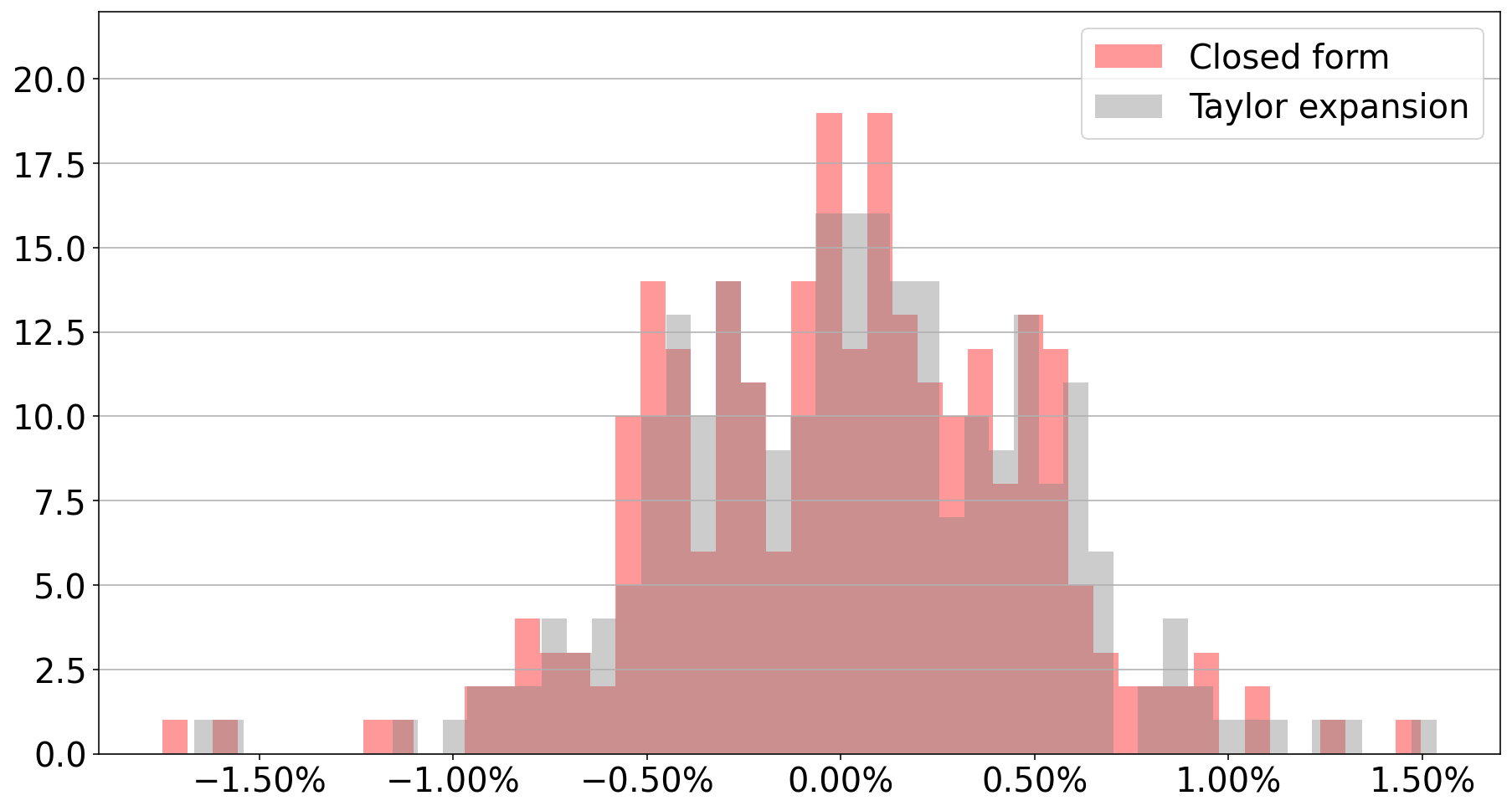}
    \caption{ReRu vs. Taylor Expansion.}
    \label{fig:fig2sub1}
  \end{subfigure}
  \hfill
  \begin{subfigure}[b]{0.45\textwidth}
    \centering
    \includegraphics[width=\textwidth]{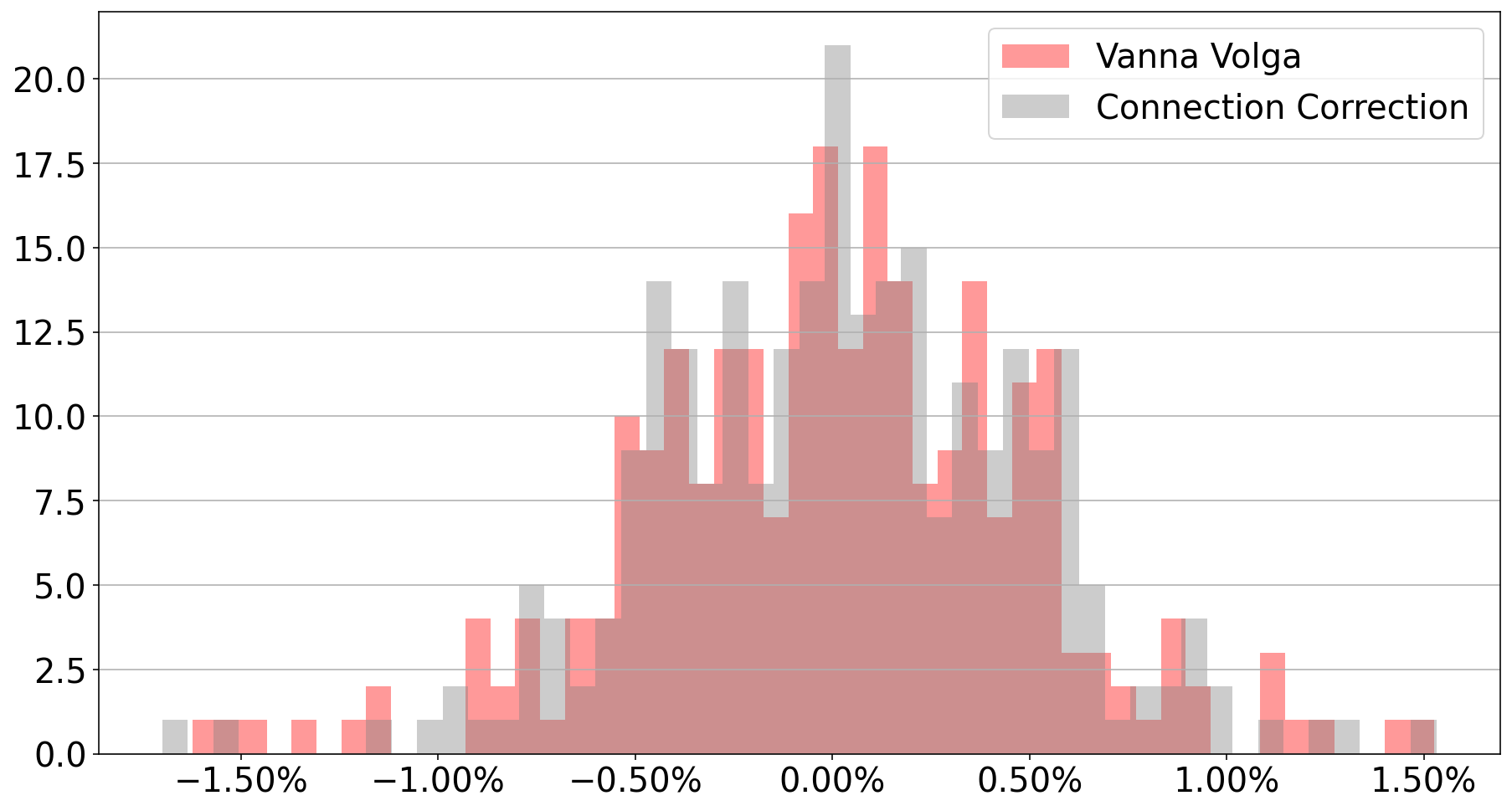}
    \caption{VV pricing vs. Connection Correction.}
    \label{fig:fig2sub2}
  \end{subfigure}
  \caption{Histogram of \texttt{EURUSD} UIC barrier P\&L increment with $K = 0.98$, $T = 1.0$, $B=1.01$ and daily reconstruction frequency.}
  \label{fig:plandcorrection2}
\end{figure}

Table~\ref{tab:eurusd_accuracy} reports daily one-step prediction errors relative to the Reiner--Rubinstein benchmark. The connection-corrected predictor has essentially the same accuracy as the ordinary BS Taylor predictor. This is the intended behavior in this setting. Since the benchmark is itself BS based, a smile correction is not expected to improve the error unless the target curvature differs materially from the baseline. The VV revaluation increment is more dispersed because it is a full revaluation under the VV construction and may also reflect changes in first-order sensitivities,\footnote{A full VV revaluation changes the local value function and therefore may also change first-order sensitivities. The connection-corrected predictor used here deliberately removes this effect: it uses VV only as a quadratic target and keeps the baseline first-order hedge convention. This reflects a common desk practice in which first-order hedges are governed by a stable production convention, while smile adjustments are introduced as overlays for marking, attribution or higher-order risk. If a desk instead hedges on VV first derivatives, the baseline gradient can be replaced by the VV gradient within the same local framework.} whereas the connection-corrected predictor preserves the baseline first-order hedge convention. In addition, we examine the distribution of daily one-step increments for each approach, depicted in Fig.~\ref{fig:fig2sub1}--\ref{fig:fig2sub2}.

\begin{table}[t]
\centering
\begin{tabular}{lccc}
\toprule
Predictor & MAE & RMSE & Corr. (Pearson) \\
\midrule
BS Taylor (second order) & 0.00023 & 0.00036 & 0.9982 \\
VV revaluation increment & 0.00034 & 0.00109 & 0.9761 \\
Connection correction    & 0.00022 & 0.00034 & 0.9984 \\
\bottomrule
\end{tabular}
\caption{Accuracy of daily one-step \texttt{EURUSD} UIC P\&L predictors, reported in \texttt{USD} per unit notional, relative to Reiner--Rubinstein closed-form increments.}
\label{tab:eurusd_accuracy}
\end{table}

The \texttt{EURUSD} case should therefore be read as a stability check rather than as evidence that smile adjustments improve a BS benchmark. When the target quadratic form is close to the baseline Hessian, the fitted term $C^k_{ij}V_k$ is small and the overlay becomes nearly inactive.

We next illustrate the liquidity calculation. 
The hedge universe is
\[
\mathcal I_{\mathrm{hedge}}
=
\{\,\text{spot},\ K_{\mathrm{ATM}}\text{ straddle}_{\tau},\ 25\Delta\text{ call}_{\tau}\,\},
\]
where $\tau$ denotes the hedge-option tenor used in the bucket. 
The reported liquidity calculation is two-factor: the state is $(S,\sigma)$ and the controlled exposure vector is 
\[
E(x)=(\Delta_{\mathrm{book}},\Vega_{\mathrm{book}}).
\]
Thus the exposure matrix $B$ in the least-cost hedge rule has dimension $2\times3$: the desk controls delta and ATM-vega using three available hedge instruments. 
The spot hedge primarily supplies delta, while the ATM straddle and the 25$\Delta$ call provide vanilla option hedges with different delta, vega and liquidity profiles.\footnote{We take the ATM-forward strike, $K_{\mathrm{ATM}}=F$, so the ATM straddle is centered at the forward.}
The 25$\Delta$ call is therefore included as an additional hedge instrument available to the least-cost rule rather than as a separate constraint on a smile-state exposure. 
Quoted widths and clips determine $\Lambda$ while the least-cost neutrality rule determines the hedge response and hence $g_\ell$, through Eqs.~\eqref{eq:least_cost_problem}--\eqref{eq:least_cost_solution}. 
Changing the controlled exposure vector, for example by adding an explicit wing or skew exposure, would change $B$, $J_E$, the hedge response $M$ and therefore the induced penalty $g_\ell$.

For \texttt{EURUSD}, we first use a high-liquidity bucket calculation. Table~\ref{tab:example2detail} reports indicative one-year liquidity inputs. The table reports bid--ask widths and reference clips. In constructing $\Lambda$ we use the corresponding half-widths. The impact matrix is diagonal, with entries calibrated from the quoted widths and clips as in Sec.~\ref{sec:constLambda}. For options, vol-point widths are converted into premium widths using the relevant quote-vega convention. Each weekly date is treated as a liquidity bucket: $\Lambda$ is fixed inside the one-step calculation and may be refreshed from week to week.

At the initial date, $\Lambda = \operatorname{diag} \left( 1.0\times10^{-12}, 5.0\times10^{-11}, 2.4\times10^{-10} \right)$ is the resulting diagonal impact matrix. The entries have units of premium currency divided by the square of the corresponding hedge trade unit. If all hedge trades are expressed in \texttt{EUR} notional this specializes to $\texttt{USD}/\texttt{EUR}^2$. If option trades are expressed in vega units or contracts, the corresponding entries use those units instead. Weekly liquidity inputs are allowed to vary by approximately 10--20\% around these indicative values to reflect ordinary variation in market depth. The entries are computed from Eq.~\eqref{eq:lambda_sym} after converting quoted widths into premium half-spreads per trade unit.

\begin{table}[t]
\centering
\begin{tabular}{lcccc}
\toprule
\multirow{2}{*}{Instrument} & Bid/Ask Vol & Bid/Ask Width & Clip Notional & Clip Vega \\
& (\%) & & (\texttt{EUR}) & (\texttt{USD} vega / vol pt.) \\
\midrule
Spot            & --          & 0.01 pip     & 10 MM & -- \\
ATM straddle    & 8.75 / 8.85 & 0.1 vol pts  & 10 MM & 50K \\
25$\Delta$ call & 8.40 / 8.60 & 0.2 vol pts  & 10 MM & 120K \\
\bottomrule
\end{tabular}
\caption{Indicative \texttt{EURUSD} one-year liquidity inputs.}
\label{tab:example2detail}
\end{table}

For the first weekly step, 23 May 2022, the estimated factor-space liquidity penalty is
\[
g_\ell
=
\begin{pmatrix}
4.54\times10^{-10} & -4.07\times10^{-10}\\[2pt]
-4.07\times10^{-10} & 3.65\times10^{-10}
\end{pmatrix}.
\]
The associated costs $\tfrac12\,\delta x^\top g_\ell\delta x$ are negligible relative to the option premium for typical weekly moves. This is consistent with the role of \texttt{EURUSD} as a high-liquidity benchmark.

The \texttt{USDTRY} case illustrates a different regime. For less liquid markets, quoted depth and spread often depend materially on trade size. To reflect this, we use a diagonal but trade-size dependent impact matrix $\Lambda(\delta q)$. This part of the example is used to compute execution costs under realistic tiered liquidity. It goes beyond the constant-\(\Lambda\) Riemannian setting used to define \(C(g_\ell)\), because the local cost now depends on trade size rather than only on the state.

\begin{figure}[t]
  \centering
  \includegraphics[width=0.8\textwidth]{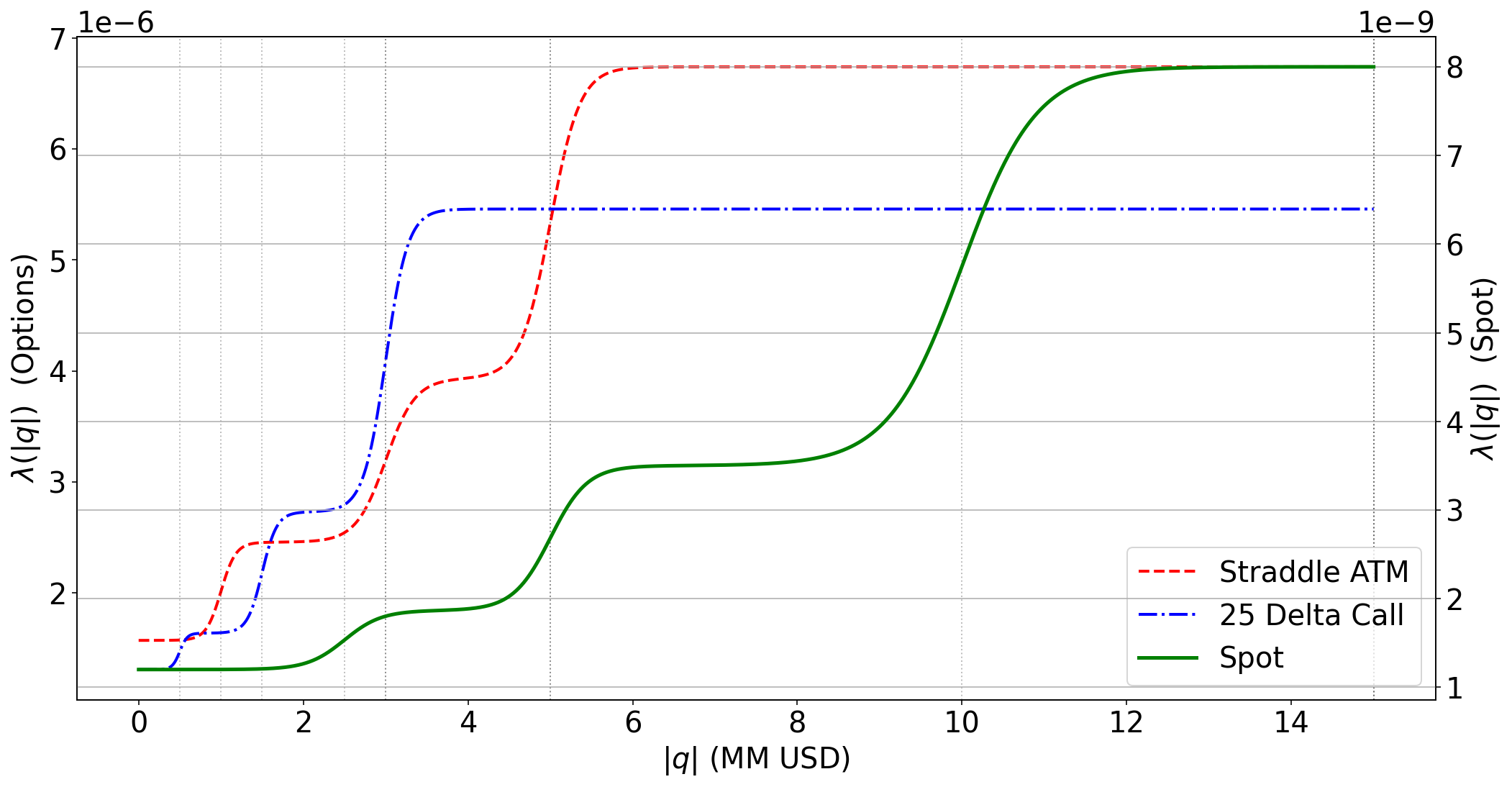}
  \caption{Smooth tiered impact functions for the \texttt{USDTRY} hedge set. The horizontal axis is trade size $|q|$ in \texttt{USD} millions; impact coefficients are in \(\texttt{TRY}/\texttt{USD}^2\).}
  \label{fig:impactfunc_usdtry}
\end{figure}

For each hedge instrument $i$, the diagonal impact coefficient is specified as a smooth tiered function of trade size. Let $j = 1,\ldots,J_i$ index the liquidity tiers, with plateau coefficients $\smash{\lambda_i^{(j)}}$. Transitions between adjacent tiers are smoothed with logistic functions. We use the effective coefficient
\begin{equation}
  \label{eq:usdtry_tiered_lambda}
  \Lambda_{ii}(q_i)
  =
  \lambda_i^{(1)}
  +
  \sum_{j=2}^{J_i}
  \left(\lambda_i^{(j)}-\lambda_i^{(j-1)}\right)
  s_{i,j}(|q_i|),
  \qquad
  s_{i,j}(u)
  =
  \frac{1}{1+\exp\!\left(-(u-c_{i,j})/\omega_{i,j}\right)}.
\end{equation}
Here $c_{i,j}$ is the transition level between tiers $j-1$ and $j$, $\omega_{i,j}$ is the smoothing width and $\lambda_i^{(j)}$ is the plateau impact level in tier $j$. The trade cost is evaluated as
\[
L_i(q_i)
=
\tfrac12\,\Lambda_{ii}(q_i)q_i^2.
\]
Each plateau is anchored to the corresponding quoted half-spread and representative clip using
\[
\lambda_i^{(j)}
=
\frac{2\,s^{\mathrm{price}}_{i,j}}{Q_{i,j}},
\]
with option vol widths first converted into premium widths. Trade sizes are measured in \texttt{USD} notional and impact coefficients in \(\texttt{TRY}/\texttt{USD}^2\). Figure~\ref{fig:impactfunc_usdtry} plots the resulting impact profiles and Table~\ref{tab:usdtry_tiers} reports the tier inputs.

\begin{table}[t]
\centering
\begin{tabular}{llcccc}
\toprule
Instrument & Quantity & Tier 1 & Tier 2 & Tier 3 & Tier 4 \\
\midrule
\multirow{3}{*}{Spot (\texttt{USDTRY})}
& Breakpoint upper bound (MM USD) & 1.0 & 5.0 & 10.0 & $\infty$ \\
& Clip size $Q$ (MM USD) & 1.0 & 1.5 & 2.0 & 2.0 \\
& Half-spread $s^{\mathrm{price}}$ (TRY/USD) & 0.0006 & 0.0014 & 0.0035 & 0.0080 \\
\midrule
\multirow{3}{*}{ATM straddle}
& Breakpoint upper bound (MM USD) & 1.0 & 3.0 & 5.0 & $\infty$ \\
& Clip size $Q$ (MM USD) & 0.25 & 0.40 & 0.60 & 0.70 \\
& Vol half-width $s^{\mathrm{vol}}$ (decimal vol) & 0.01 & 0.025 & 0.06 & 0.12 \\
\midrule
\multirow{3}{*}{25$\Delta$ call}
& Breakpoint upper bound (MM USD) & 0.50 & 1.50 & 3.0 & $\infty$ \\
& Clip size $Q$ (MM USD) & 0.10 & 0.20 & 0.30 & 0.30 \\
& Vol half-width $s^{\mathrm{vol}}$ (decimal vol) & 0.008 & 0.02 & 0.05 & 0.10 \\
\bottomrule
\end{tabular}
\caption{Tier inputs for the \texttt{USDTRY} liquidity illustration. Spot half-spreads are quoted in \texttt{TRY}/\texttt{USD}. Option widths are quoted in decimal volatility units and converted to premium widths before computing the impact coefficients.}
\label{tab:usdtry_tiers}
\end{table}

For numerical stability, trade sizes are rescaled from \texttt{USD} to million-\texttt{USD} units. The quadratic cost is unchanged when the impact coefficients are rescaled accordingly: if \(q_{\texttt{USD}}=10^6 q_{\mathrm{MM}}\), then \(\Lambda_{\mathrm{MM}}=10^{12}\Lambda_{\texttt{USD}}\).

\begin{figure}[t]
  \centering
  \begin{subfigure}[b]{0.32\textwidth}
    \centering
    \includegraphics[width=\textwidth]{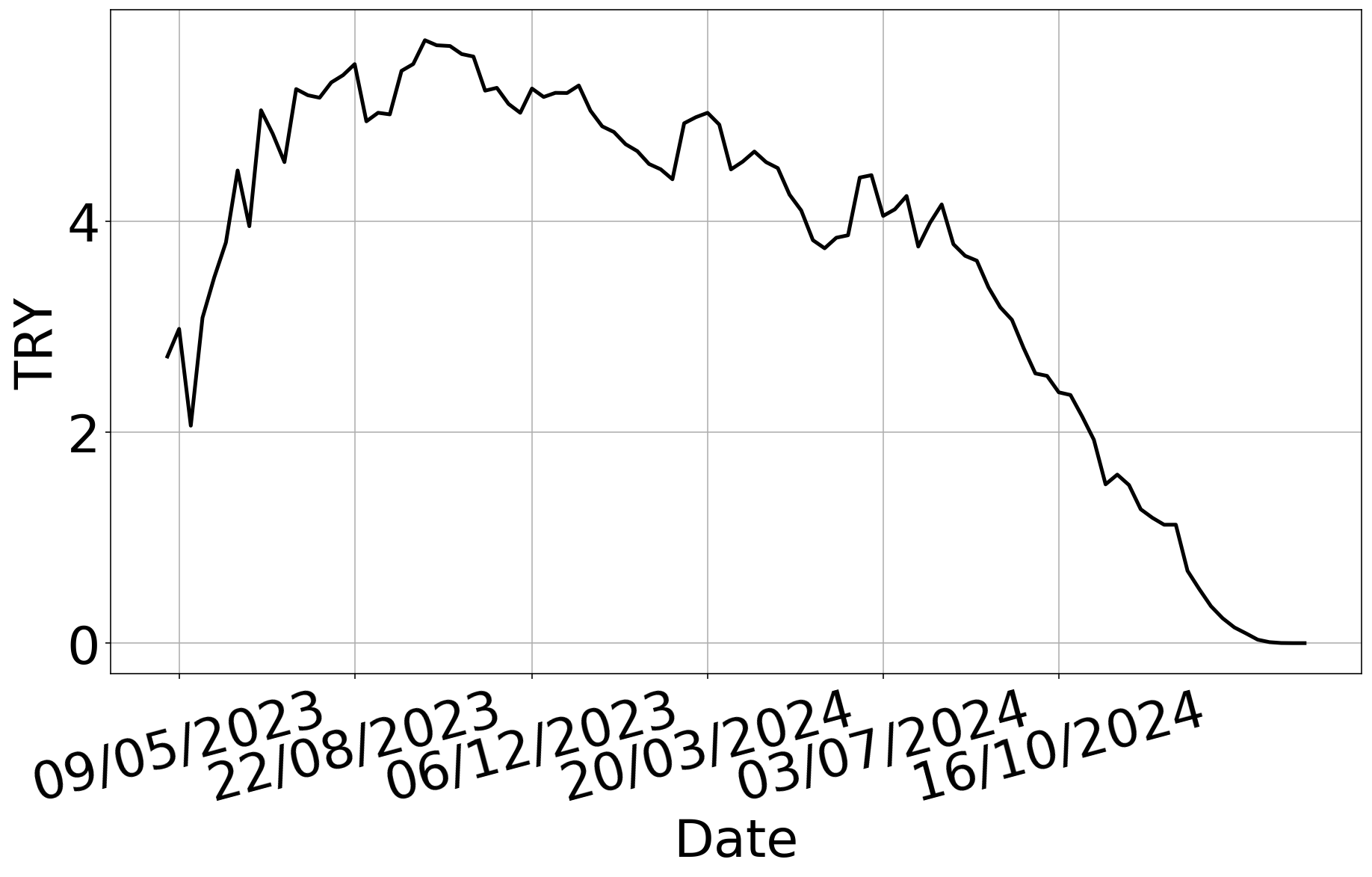}
    \caption{UIC option price.}
    \label{fig:uic_price_usdtry}
  \end{subfigure}
  \hfill
  \begin{subfigure}[b]{0.32\textwidth}
    \centering
    \includegraphics[width=\textwidth]{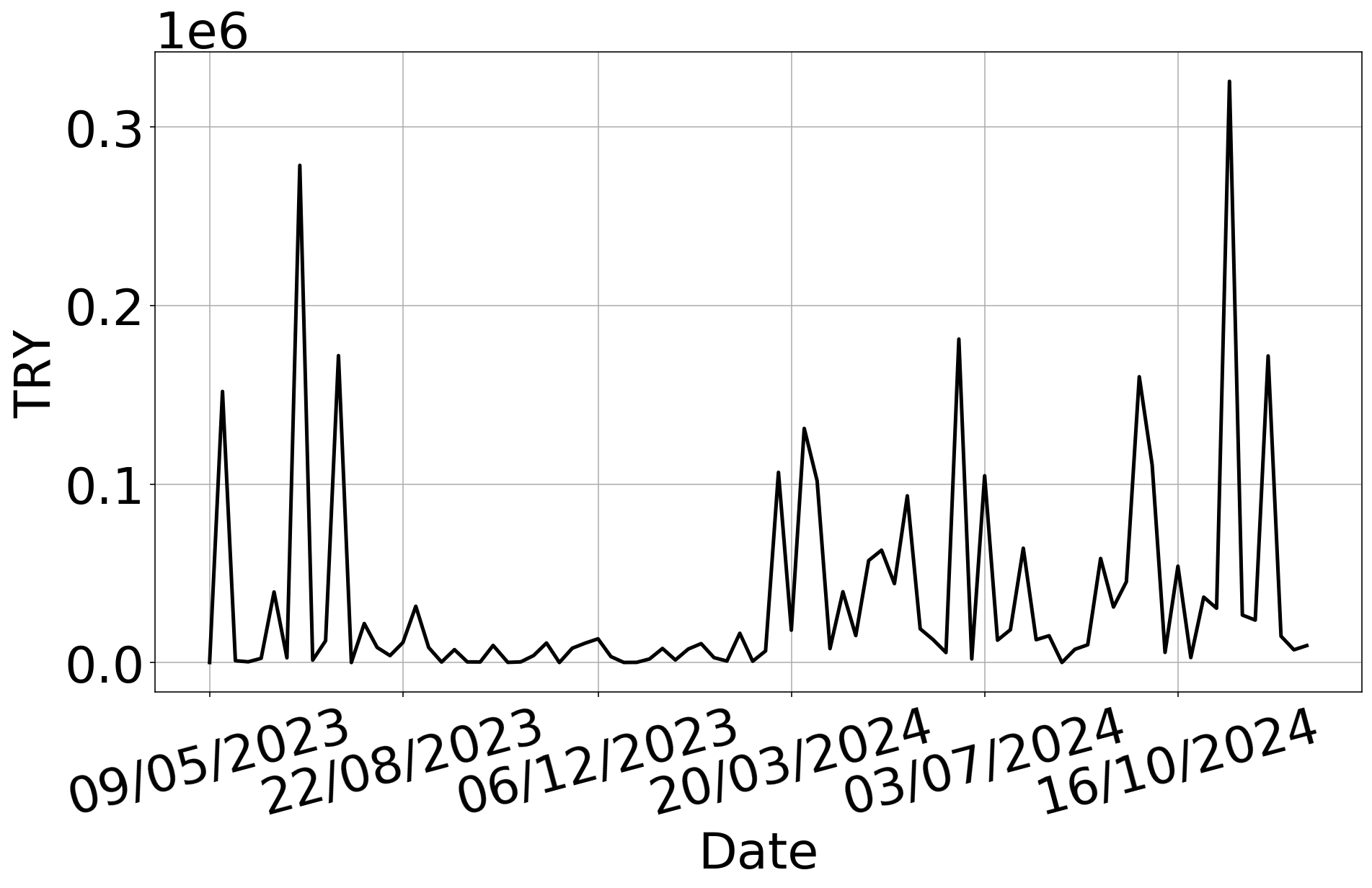}
    \caption{Weekly hedging cost.}
    \label{fig:impactfunc_usdtry_sub2}
  \end{subfigure}
  \hfill
  \begin{subfigure}[b]{0.32\textwidth}
    \centering
    \includegraphics[width=\textwidth]{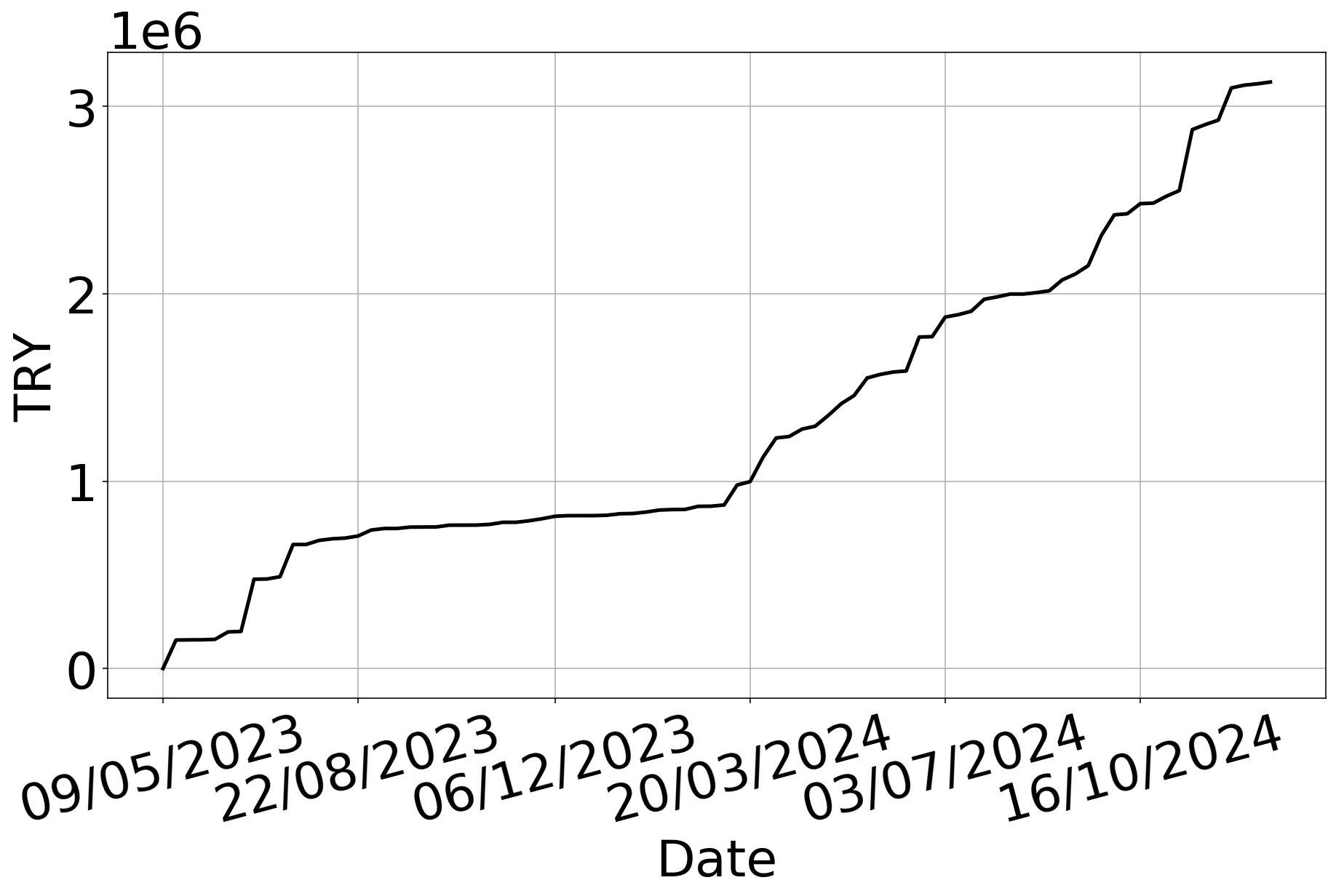}
    \caption{Cumulative hedging cost.}
    \label{fig:cumulative_usdtry_sub3}
  \end{subfigure}
  \caption{Weekly hedging of a \texttt{USDTRY} UIC barrier option with notional 10MM \texttt{USD}, $K=38.0$, $T=2.0$ and $B=40.0$.}
  \label{fig:USDTRYwithoutrehedge}
\end{figure}

Figure~\ref{fig:USDTRYwithoutrehedge} reports the weekly option value, weekly hedge cost and cumulative hedge cost for a \texttt{USDTRY} UIC option with notional 10MM \texttt{USD}, strike $K = 38.0$, maturity $T = 2.0$ and barrier $B = 40.0$. We use weekly rebalancing for comparability across dates. In the implementation, rebalancing is stopped on 25 December 2024 because the option value and Greeks have decayed to levels at which the computed hedge trades are dominated by numerical noise rather than meaningful risk, see Fig.~\ref{fig:uic_price_usdtry}.

\begin{figure}[t]
  \centering
  \begin{subfigure}[b]{0.3\textwidth}
    \centering
    \includegraphics[width=\textwidth]{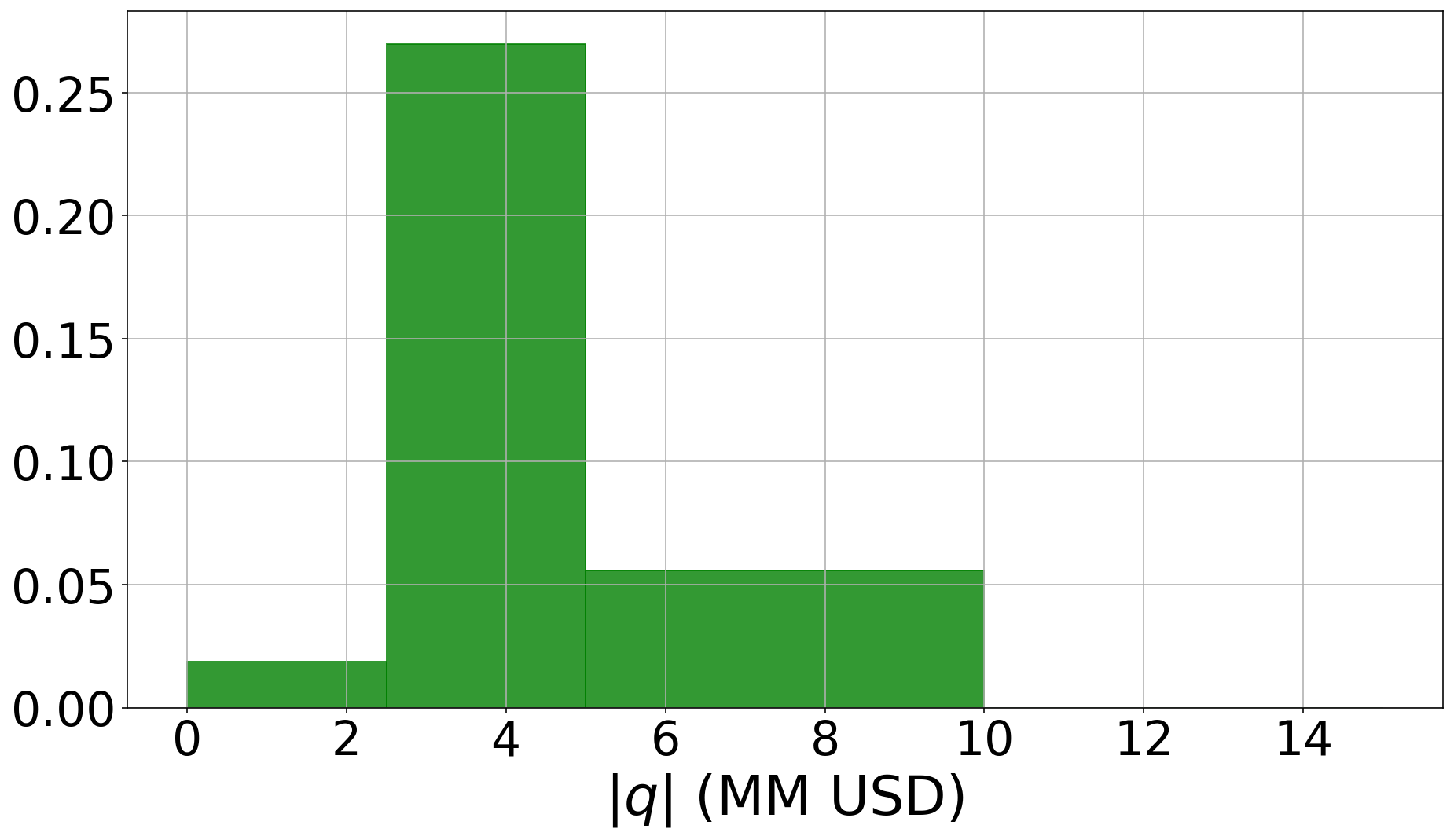}
    \caption{Spot.}
    \label{fig:q_spot}
  \end{subfigure}
  \hfill
  \begin{subfigure}[b]{0.3\textwidth}
    \centering
    \includegraphics[width=\textwidth]{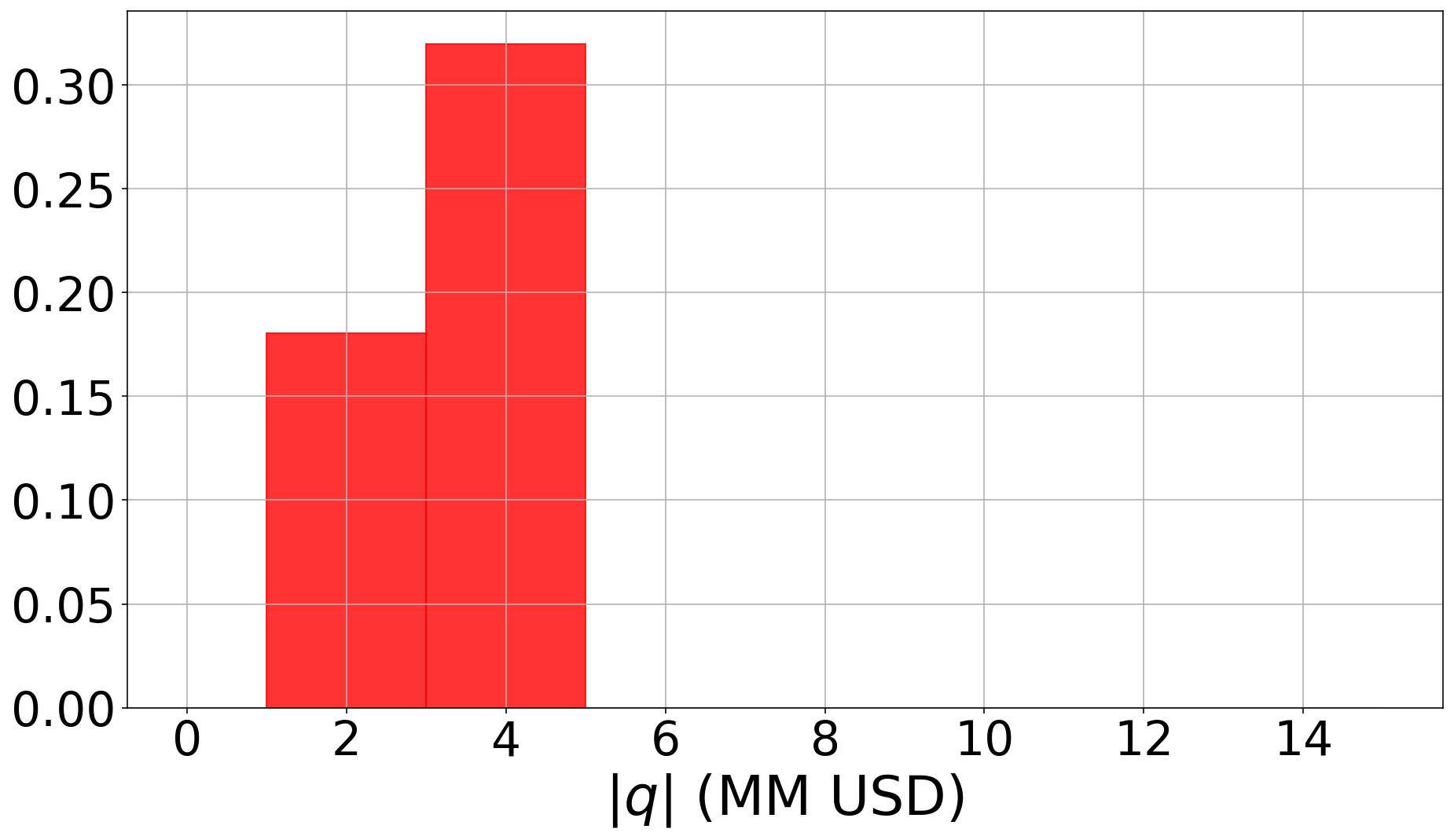}
    \caption{ATM straddle.}
    \label{fig:q_strad}
  \end{subfigure}
  \hfill
  \begin{subfigure}[b]{0.3\textwidth}
    \centering
    \includegraphics[width=\textwidth]{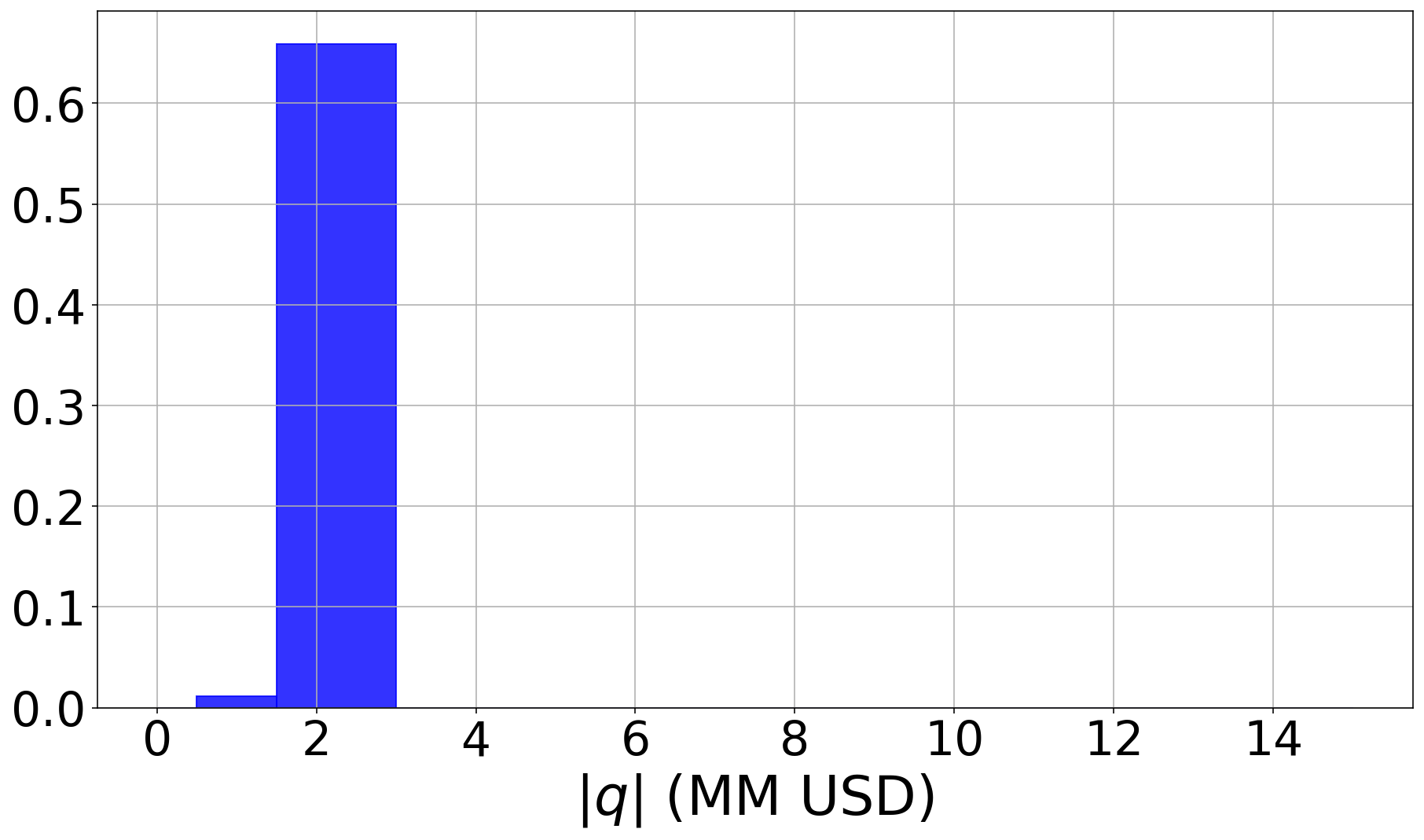}
    \caption{25$\Delta$ call.}
    \label{fig:q_25deltacall}
  \end{subfigure}
  \caption{Absolute hedge trades $|q|$ in \texttt{USD} millions at weekly rebalancing dates for the 10MM \texttt{USD} \texttt{USDTRY} UIC barrier option.}
  \label{fig:q_hedge}
\end{figure}

The cumulative liquidity cost reaches approximately 40\% of the initial option premium. This figure is not intended as a universal estimate for \texttt{USDTRY} but as the output of the stated hedge instruments, weekly rebalancing rule and tiered liquidity assumptions. Figure~\ref{fig:q_hedge} reports the distribution of absolute hedge trades by instrument. The liquidity cost is driven mainly by spot trades, while the 25$\Delta$ call occasionally requires larger trades that become disproportionately expensive under the tiered impact function. The point is that, in a lower-liquidity market, the penalty channel can be economically material and should be part of hedging design.

Taken together, the two cases illustrate the intended use of the framework. In a liquid market with mild smile curvature, the overlay is nearly inactive and does not distort the baseline predictor.
In a lower-liquidity market, the same framework turns quoted widths, clips and hedge rules into a local cost measure that can materially affect hedge design at portfolio scale.



\section{Further Applications and Extensions}
\label{sec:extensions}
\nosectionappendix

The previous sections develop two examples in detail: a curvature overlay, where a Vanna--Volga target is matched through a calibrated affine connection, and a cost/risk overlay, where execution costs induce a local quadratic penalty on factor moves. The same interface can be used for other desk features but only when they admit a reliable local quadratic approximation over the hedging horizon. The extensions in this section should therefore be read as implementation templates not as empirical validations beyond the case studies already reported.

There are two basic routes.
In the curvature route, the desk specifies a target quadratic form $H^\star$ for the second-order P\&L term while the baseline scalar value and first-order hedge convention may be kept fixed as in the main construction.
In some applications the desk may instead choose a different scalar value to expand, for example an xVA-adjusted value or an internal stress value (in that case the first-order convention should be stated explicitly).
In the cost/risk route, one adds a positive semidefinite quadratic penalty to the factor move.
The two routes serve different purposes.
A target quadratic form modifies the local P\&L predictor.
A penalty measures the cost, risk or undesirability of moving in a particular factor direction, and can be used either directly for rebalancing rules or, when non-degenerate or regularized, through its Levi--Civita connection in the covariant Hessian.

A natural curvature extension is to estimate the target quadratic form directly from realized short-horizon P\&L. After subtracting the first-order hedge term, and after removing deterministic carry or theta terms when they are material, one may fit locally
\begin{equation}
  \label{eq:empirical_pnl_target}
  \delta V_n - V_i\,\delta x_n^i
  \approx
  \tfrac12\,H^{\mathrm{emp}}_{ij}\,\delta x_n^i\delta x_n^j
  +
  \varepsilon_n,
\end{equation}
using observations from a bucket, regime or nearby state window. Equivalently, one may include an intercept or other nuisance terms in the regression and retain only the fitted symmetric quadratic coefficient. The fitted matrix $H^{\mathrm{emp}}$ can then be used as the target $H^\star$ in the same matching equations used for Vanna--Volga. This route is useful when the desk has enough realized P\&L observations to estimate a stable local quadratic term. If the data are sparse, the quadratic design is poorly conditioned or regime changes are frequent, the fitted target should be regularized or replaced by a model-implied, market-implied or desk-specified target.

Other curvature overlays arise from model transfer and marking conventions. A desk may keep a simple production model for prices and first-order hedges but import the local Hessian from a richer model, a scenario engine or an internal stress model. Similarly, sticky-strike, sticky-delta and sticky-moneyness conventions specify how volatility quotes are carried when the state moves. They affect the local quadratic P\&L layer but do not define a cost or distance. In rates, curve roll-down and tenor-bucket transport play a similar role: they specify how sensitivities are carried across nearby curve states or tenor grids. These examples are naturally connection-type overlays when the object being specified is a local curvature rule rather than a penalty.

A simple cost/risk extension is risk normalization. In multi-factor settings, or even in a low-dimensional display when the relative scale of the coordinates is unstable, it is often useful to measure factor moves in risk units rather than raw units. Let
\begin{equation}
  \label{eq:state_covariance}
  \Xi(x)
  \approx
  \operatorname{Cov} \, [\delta x \mid x]
\end{equation}
be an empirical covariance matrix of factor moves over the hedging horizon, estimated with a rolling, exponentially weighted or regime-filtered window. After shrinkage, eigenvalue clipping or another regularization step if needed, the inverse covariance $g_{\mathrm{cov}}(x) = \Xi(x)^{-1}$ defines the local Mahalanobis penalty
\begin{equation}
  \label{eq:mahalanobis_penalty}
  \tfrac12 \, \delta x^\top g_{\mathrm{cov}}(x) \, \delta x.
\end{equation}
This term is not an execution-cost model and does not by itself change the mark-to-market value. It measures how large a move is relative to the local scale and correlation of the factors.

Several penalty terms can be combined when they are expressed in compatible units. For example,
\begin{equation}
  \label{eq:geff}
  g_{\mathrm{eff}}(x)
  =
  g_\ell(x)
  +
  \alpha\,g_{\mathrm{cov}}(x),
  \qquad
  \alpha>0,
\end{equation}
combines execution costs with risk normalization. Since $\delta x^\top \! g_\ell \, \delta x$ is measured in premium-currency units while $\delta x^\top g_{\mathrm{cov}} \, \delta x$ is dimensionless or measured in statistical risk units, $\alpha$ carries the units needed to put both terms on the same desk-objective scale. It may be chosen by matching the two penalties on a reference move, by imposing a risk budget or by tuning an out-of-sample objective when enough data are available.

Once $g_{\mathrm{eff}}$ is fixed, it can be used in the same way as $g_\ell$: to define penalty-weighted triggers, equal-cost or equal-risk rebalancing rules, or a Levi--Civita connection $C(g_{\mathrm{eff}})$ for adjusted second-order sensitivities. For short moves, freezing $g_{\mathrm{eff}}$ at the current state gives a local quadratic norm and hence simple equal-penalty steps in whitened coordinates. For larger planned transitions, when $g_{\mathrm{eff}}$ is treated as a smooth positive definite metric on the relevant subspace, the corresponding least-penalty paths are geodesics of $g_{\mathrm{eff}}$. In this paper we use this primarily as an organizing principle for rebalancing rules rather than as a separately validated dynamic hedging strategy.

Scenario and gap terms can also be summarized locally by quadratic penalties when the desk wants a smooth surrogate for stress-direction exposure. Choose representative stress moves\footnote{For example, a $2\%$ spot gap with a $30$bp volatility jump, a joint spot-volatility shift calibrated to historical event windows, or a desk stress scenario.} $\{\delta x^{(s)}\}_{s=1}^N$, with nonnegative weights $w_s$, and choose a reference metric $g_0$ that puts the factors in comparable units. Setting $\ell^{(s)} = g_0\,\delta x^{(s)}$, define
\begin{equation}
  \label{eq:ggap_projection}
  g_{\mathrm{gap}}
  =
  \sum_{s=1}^N
  w_s\,\ell^{(s)}\ell^{(s)\top}.
\end{equation}
Then
\begin{equation}
  \label{eq:gap_penalty}
  \delta x^\top g_{\mathrm{gap}} \, \delta x
  =
  \sum_{s=1}^N
  w_s \big( \ell^{(s)\top} \delta x \big)^2,
\end{equation}
so moves aligned with the chosen stress directions receive a larger penalty. If only the direction of a stress move is meant to matter, the covectors $\ell^{(s)}$ can be normalized before forming $g_{\mathrm{gap}}$. This construction does not turn jumps into a local diffusion model. It only provides a local quadratic surrogate for scenario exposure over the hedging horizon. The effective penalty may then be enlarged, for example,
\begin{equation}
  \label{eq:geff_gap}
  g_{\mathrm{eff}}(x)
  =
  g_\ell(x)
  +
  \alpha\,g_{\mathrm{cov}}(x)
  +
  \beta\,g_{\mathrm{gap}}(x),
\end{equation}
with $\beta$ chosen in units consistent with the desk objective.

The same idea applies to event regimes. Around central-bank meetings, fixings, earnings announcements or other scheduled events, the inputs defining the local penalty can be bucketed in time:
\begin{equation}
  \label{eq:geff_time}
  g_{\mathrm{eff}}(x,t)
  =
  g_\ell(x,t)
  +
  \alpha(t)\,g_{\mathrm{cov}}(x,t)
  +
  \beta(t)\,g_{\mathrm{gap}}(x,t).
\end{equation}
The weights and inputs can be kept fixed inside a bucket and recomputed when the regime changes. This formalizes a common desk practice: the same hedge rule may become effectively tighter, looser or more scenario-aware as an event approaches, without changing the underlying pricing model but by changing the quadratic penalty used to measure and penalize moves.

Higher-dimensional smile states fit naturally into the same framework. Instead of $x=(S,\sigma)$, one may work with a quoting state such as $x = \big( \log F,\, \sigma_{\mathrm{ATM}},\, \mathrm{RR}_{\Delta},\, \mathrm{BF}_{\Delta} \big)$ or with a parametric smile state $x=(\log F,\theta_1,\dots,\theta_p)$. This is useful in FX and equity index options, where a two-factor display $(S,\sigma)$ may be sufficient for reporting a local P\&L predictor, but portfolios often have first-order exposure to additional smile coordinates such as skew and curvature. The curvature construction then matches a target quadratic form in the enlarged state, while the cost/risk construction estimates either a covariance penalty $g_{\mathrm{cov}}$, an execution-cost penalty $g_\ell=M^\top \Lambda \, M$ or a combination of both. Operationally, the workflow is unchanged: compute sensitivities with respect to the chosen quoting variables, map factor moves into hedge trades through the desk rule and apply the same local quadratic formulas. As in the Vanna--Volga calibration, only the directions spanned by available sensitivities and calibration instruments are identified. The remaining directions require regularization or a modelling convention.

Internal xVA, margin and capital objectives can also enter as quadratic penalties. Suppose $Y\!(x)$ is a vector of quantities whose short-horizon variation the desk wants to control, such as selected xVA sensitivities, margin add-ons or capital-relevant exposures. Locally, $\delta Y(x) \approx J_Y(x)\,\delta x$. Given a positive semidefinite weight matrix $W_Y$, the penalty
\begin{equation}
  \label{eq:gY}
  g_Y(x) = J_Y(x)^\top W_Y \, J_Y(x)
\end{equation}
can be added to $g_{\mathrm{eff}}$. The weights in $W_Y$ convert changes in $Y$ into the desk's chosen objective units, such as cost, capital usage or limit consumption. This does not require a new pricing model. It requires local sensitivities of the controlled quantity and a desk choice of weights or limits. If the controlled quantity is nonsmooth, for example near a margin threshold, the same construction should be interpreted bucketwise or after local smoothing. If the desk instead wants to forecast the P\&L of an xVA-adjusted value itself, then the curvature route applies: use the adjusted scalar value as $V$, with its first-order convention stated explicitly, or fit a target quadratic form for its local P\&L while preserving the chosen baseline gradient.

Finally, real execution costs may be asymmetric or nonlinear. Bid--ask imbalance, inventory skew and depth-dependent impact can be handled locally by using different bucketed penalties by side, piecewise quadratic approximations or state-dependent and trade-size-dependent cost matrices. These extensions go beyond the symmetric quadratic examples developed in the paper, but they preserve the same organizing principle: once a desk feature is represented locally, possibly bucket by bucket or side by side, as either a target quadratic form or a quadratic penalty, it can be added to the existing model through the same overlay.

The purpose of this section is not to claim that all such extensions are equally easy or equally well identified.
Some require reliable realized P\&L data, some require stable covariance estimates and some require desk choices of scenario weights or internal cost units.
Some effects are not local quadratic objects at all: full smile dynamics, jump timing, path dependence, multi-period execution and no-arbitrage evolution of the state variables require additional modelling beyond the overlay.
The point is narrower: the framework provides a common local interface for adding desk features that can be represented, over the hedging horizon, as either curvature targets or quadratic penalties, without rebuilding the base pricing model and, in the model-preserving mode used throughout the paper, without changing the first-order hedge convention.


\section{Local Overlay and No-Arbitrage}
\label{sec:local_geometry_dynamics}
\nosectionappendix

The framework developed in the paper is local by design.
It starts from the value function, risk factors and hedge convention already used by the desk, and modifies or supplements only the short-horizon quadratic layer.
This is why the construction can add smile, liquidity and other desk-level features without replacing the underlying pricing model.
It is useful, however, to clarify what changes if the same geometric correction is interpreted dynamically.

There are two readings. In the first, used throughout the paper, the connection is a local P\&L and hedging overlay. The base pricing model remains unchanged. In the second, the connection is promoted to part of an effective pricing dynamics. This second reading is mathematically natural but more restrictive. Once the correction enters a generator or pricing PDE, the resulting dynamics must satisfy the martingale restrictions associated with the chosen numeraire and traded assets. Moreover, the connection must then be a property of the state dynamics, applied consistently across traded claims, not a deal-specific or book-specific calibration object.

To see the point, suppose the desk state $X_t$ has, under a pricing measure, local generator
\begin{equation}
  \label{eq:base_local_generator}
  \mathcal L V
  =
  b^i\partial_i V
  +
  \frac12 a^{ij}\partial_{ij}V,
\end{equation}
where $a=\Sigma\Sigma^\top$ is the local covariance matrix. Replacing the ordinary Hessian by the covariant Hessian gives the formal operator
\begin{equation}
  \label{eq:local_generator}
  \widetilde{\mathcal L} V
  =
  b^i\partial_i V
  +
  \frac12 a^{ij}
  \left(
  \partial_{ij}V
  -
  C^k_{ij}\partial_k V
  \right)
  =
  \left(
  b^k
  -
  \frac12 a^{ij}C^k_{ij}
  \right)
  \partial_k V
  +
  \frac12 a^{ij}\partial_{ij}V.
\end{equation}
Thus, at generator level, the connection term is a drift adjustment. This should not be confused with the local use of the paper. Here, $C^k_{ij}$ is introduced to adjust short-horizon quadratic P\&L. If it is instead treated as part of a pricing dynamics, the adjusted drift must be compatible with no-arbitrage restrictions.

For example, if $S$ is a tradable equity spot with dividend yield $q$, then under the usual risk-neutral measure the adjusted generator must satisfy
\begin{equation}
  \widetilde{\mathcal L} S = (r-q) S.
\end{equation}
For an FX spot under the domestic risk-neutral measure, the corresponding condition is
\begin{equation}
  \label{eq:local_generator_eq_fx}
  \widetilde{\mathcal L}S=(r_d-r_f)S.
\end{equation}
These conditions do not prevent the use of the geometric overlay. They only say that drift changes cannot be imposed freely in directions corresponding to tradable prices. The connection-induced drift must therefore either be offset by a compensating drift choice, be restricted to state components that are not themselves tradable prices, or be interpreted as a local P\&L and hedging correction rather than as a new risk-neutral dynamics.

This distinction is especially relevant for liquidity.
In the present contribution, liquidity is not introduced by changing the marginal price process of the underlying asset or hedge instruments.
The primitive is the executable cost of changing hedge positions.
For the smooth local quadratic layer used in this paper,\footnote{A pure bid--ask cost has a proportional component and is not differentiable at zero. The expansion here refers to the smooth quadratic surrogate used in the paper or to a local smoothing/anchoring of the executable-cost function around the relevant hedge size.} write $\mathcal C_t(\delta q)$ for the incremental cash required to execute a signed net hedge trade $\delta q$ and expand around the current execution point as
\begin{equation}
  \label{eq:execution_cost_expansion}
  \mathcal C_t(\delta q)
  =
  P_t\cdot\delta q
  +
  \frac12\,\delta q^\top\Lambda_t\,\delta q
  +
  o(\|\delta q\|^2),
\end{equation}
where $P_t$ is the marginal executable hedge-price vector and $\Lambda_t$ is the local liquidity-cost matrix.
The execution-cost penalty used in the paper is the excess over marginal execution,
\begin{equation}
  \label{eq:excess_execution_cost}
  \kappa_t(\delta q)
  =
  \mathcal C_t(\delta q)-P_t\cdot\delta q
  \approx
  \frac12\,\delta q^\top\Lambda_t\,\delta q.
\end{equation}
Pulling this quadratic term back through the hedge map gives $g_\ell(t,x)=M(t,x)^\top\Lambda_t \, M(t,x)$.

No-arbitrage should be imposed on the marginal executable prices $P_t$, when those prices are modelled as tradable marginal prices. The quadratic term has a different role: it is the additional cost of trading size and is paid when the hedge position changes. This separation is economically important. If the marginal price admitted an arbitrage, a trader could scale trades down so that the quadratic execution cost became negligible relative to the first-order arbitrage gain. By contrast, the quadratic term represents the local cost of trading size. It is not itself a marginal price process and need not be a martingale. Its local discipline is different: it should be nonnegative and should not reward instantaneous round trips.\footnote{By a round trip we mean a sequence of hedge trades that starts and ends with the same position, typically zero net position. In a local execution-cost model, buying and then reversing the trade at the same marginal prices should not generate a profit purely from the cost term.} Positive semidefiniteness of $\Lambda_t$ is the corresponding local condition in the quadratic approximation. This is not a complete theory of no-arbitrage with transaction costs or transient impact but it is the appropriate local discipline for the execution-cost layer used here.

The same separation explains why the liquidity adjustment is naturally book-level. The marginal valuation can remain based on the desk's arbitrage-free pricing model while the liquidity contribution depends on the net hedge trade generated by the whole portfolio. Let $\delta q_\pi$ be the hedge trade required by the existing portfolio $\pi$ and $\delta q_{d_0}$ the additional hedge trade generated by adding a candidate deal $d_0$. Under the local quadratic execution-cost penalty $\kappa_t(\delta q)$, the incremental liquidity charge of adding $d_0$ to the book is
\begin{equation}
\label{eq:incremental_liquidity_charge}
\delta\kappa_t(d_0\mid\pi)
=
\kappa_t(\delta q_\pi+\delta q_{d_0})
-
\kappa_t(\delta q_\pi)
=
\delta q_{d_0}^{\top}\Lambda_t\,\delta q_\pi
+
\tfrac12\,\delta q_{d_0}^{\top}\Lambda_t\,\delta q_{d_0}.
\end{equation}
Thus the liquidity contribution of a deal is not a stand-alone number independent of the existing book. If the candidate deal offsets existing hedge trades, the cross term is negative and the incremental liquidity charge is reduced. It may even be negative relative to the current book if the new deal materially reduces the existing rebalancing need.\footnote{A negative incremental charge does not mean that execution costs are negative in absolute terms. It means that the new deal reduces the execution-cost penalty of the combined book relative to the existing book.} If it reinforces existing trades, the charge increases. This is precisely the netting mechanism captured by applying the quadratic cost to the book hedge trade.

In discrete time, suppressing cash-account accrual or working in discounted units, the same idea can be written schematically as a self-financing equation with execution costs. Let $q_n$ be the hedge position held over the interval $[t_n,t_{n+1}]$, and let $P_n$ be the marginal hedge-price vector. After the price move, the hedge is rebalanced from $q_n$ to $q_{n+1}$, paying the excess execution cost $\kappa_{n+1}(q_{n+1}-q_n)$. The marked-to-market wealth then satisfies
\begin{equation}
  \label{eq:self_finance}
  Y_{n+1}
  =
  Y_n
  +
  q_n\cdot(P_{n+1}-P_n)
  -
  \kappa_{n+1}(q_{n+1}-q_n).
\end{equation}
The marginal cost $P_{n+1}\cdot(q_{n+1}-q_n)$ of the rebalance is financed by the cash account and cancels in marked-to-market wealth. The remaining loss is the excess cost of execution. In the local quadratic approximation, $\kappa_{n+1}(\delta q) \approx \frac12\,\delta q^\top\Lambda_{n+1} \, \delta q$. The first term in Eq.~\eqref{eq:self_finance} is the usual mark-to-market gain on the hedge. The last term is the cost of changing the hedge using the available liquidity. This is the applied sense in which the paper extends the frictionless local P\&L picture.

The construction is not tied to BS. A different base model changes the state variables, the value function, the hedge response and the interpretation of the trade units, but not the local form of the overlay:
\[
\delta V
\approx
\partial_iV\,\delta x^i
+
\frac12
\left(
\partial_{ij}V
-
C^k_{ij}\partial_kV
\right)
\delta x^i\delta x^j,
\qquad
g_\ell(x)=M(x)^\top\Lambda M(x).
\]
For interest-rate books, for example, the state may contain curve and volatility factors, the hedge instruments may be swaps, futures, bonds, caps, floors or swaptions, and the trade units may be DV01, PV01, notional or vega. The same local mapping from hedge-space costs to factor-space penalties applies once units are made consistent. If, however, the geometric layer is promoted to a full pricing dynamics, then the usual term-structure drift restrictions must also be imposed.

A full dynamic extension would choose hedge trades over many dates, balancing residual exposure and execution costs. That leads naturally to stochastic control, nonlinear PDEs or backward stochastic equations in an incomplete market. Such a development is beyond the scope of this paper. The local version is more modest and is the one used here: it provides a way to adjust short-horizon P\&L, compare hedge choices, set cost-aware rebalancing thresholds and measure the incremental liquidity cost of adding risk to an existing book.

The conclusion is that the geometric layer is model-agnostic but not arbitrage-agnostic. As a local overlay, it can sit on top of BS, stochastic-volatility, term-structure or curve-based models without changing their marginal pricing dynamics. If it is promoted to an SDE or PDE, then the adjusted generator must satisfy the martingale restrictions of the chosen pricing measure, and the connection must be a legitimate state-dynamics object rather than a payoff-specific or book-specific overlay. This separation is useful in practice: the desk model is preserved where it should be preserved, while smile adjustments, liquidity costs and book-level netting can be represented through a disciplined local quadratic layer.


\section{Conclusion and Future Research}
\label{sec:conclusion}
\nosectionappendix

This paper introduces a local geometric overlay for short-horizon option P\&L and hedging. 
The purpose of the overlay is not to replace the pricing model used by the desk. 
Instead, it provides a systematic way to modify or supplement the local quadratic layer with features such as smile corrections, execution costs, risk normalization or scenario penalties, while preserving the baseline marginal pricing model and the first-order hedge convention in the model-preserving mode studied here. 
This separation is useful in practice: the desk can keep the model it already uses for pricing and first-order hedging, and enrich the quadratic layer where many important desk adjustments enter.

The framework has two complementary channels. 
The first is the curvature-overlay channel. 
A target quadratic form is matched through a covariant Hessian,
\[
\widetilde H_{ij}
=
\partial_{ij}V
-
C^k_{ij}\partial_k V.
\]
This allows the quadratic term in the local P\&L predictor to reproduce a desk target, such as a Vanna--Volga smile adjustment, an alternative-model Hessian or, when data are available, a locally estimated realized P\&L target. 
The first-order term is left unchanged. 
Calibration is local and reduces to small linear systems whose conditioning has a clear financial interpretation: the calibration instruments must span the relevant first-order state directions.

The second channel is the cost/risk-overlay channel. 
Execution costs are represented by a quadratic cost in hedge-trade space,
\[
L(\delta q)
\approx
\tfrac12\,\delta q^\top\Lambda\,\delta q.
\]
Combined with a local hedge response \(\delta q\approx M(x)\delta x\), this induces a factor-space penalty
\[
g_\ell(x)
=
M(x)^\top \Lambda \, M(x).
\]
This object measures the local cost of factor moves and supports cost-aware rebalancing rules. 
When it is smooth and non-degenerate on the relevant subspace, or after regularization, it can also be treated as a metric whose Levi--Civita connection gives liquidity-adjusted sensitivities (second-order). 
This metric interpretation is optional: the primary object on the cost side is the penalty itself. 
The construction is naturally book-level: execution costs apply to the net hedge trade. 
Therefore, internal crossing, netting and cost amplification arise from the same quadratic cost rather than from add-ons per deal.

A useful consequence of the geometric formulation is coordinate awareness. 
This does not mean that the numerical entries of the adjusted Hessian are the same in every coordinate system. 
They are not: an adjusted Gamma with respect to spot, forward or log-forward has different units and therefore different numerical values. 
The invariant object is the scalar local predictor. 
Once the connection, target and increments are transformed consistently, the same local market move receives the same adjusted quadratic P\&L contribution, regardless of whether it is expressed in spot, forward, log-forward or another smooth coordinate system. 
This removes a source of artificial variation in second-order P\&L attribution and makes comparisons across systems and risk-factor parameterizations more stable. 
In the paper, however, this invariance is a property of the overlay rather than its main objective.

The two foreign-exchange barrier case studies illustrate the developed channels. 
The \texttt{EURUSD} case acts mainly as a stability check: when the smile target is close to the baseline curvature, the fitted correction is small and the overlay does not distort an already accurate local predictor. 
The same high-liquidity setting also provides a benchmark for the cost channel. 
The \texttt{USDTRY} case illustrates the penalty channel in a lower-liquidity market, where execution costs generated by quoted widths, clips, hedge rules and tiered market depth can become economically material. 
The examples are not intended as broad empirical validation, but to show how the framework is implemented from desk observables.

The construction is not specific to foreign exchange. 
The same trade-space to factor-space mapping applies in other asset classes as long as units are handled consistently. 
In equities or commodities, ticks and contract multipliers replace pips; in rates, DV01/PV01 and swaption vega replace FX vega. 
What matters is that \(\Lambda_{rs}\) is expressed in premium currency per product of hedge-trade units \(u_ru_s\), and that \(M\) is computed using the same trade units.

The framework is deliberately local. 
It is designed for short-horizon P\&L, hedging and attribution, not as a full global pricing model. 
This distinction matters for no-arbitrage. 
As a local overlay, the method can sit on top of an arbitrage-free base model without changing its marginal pricing dynamics. 
If the overlay is embedded in an SDE or PDE generator, then the induced drift adjustment must satisfy the martingale restrictions of the chosen pricing measure, and the connection must be a legitimate state-dynamics object applied consistently across traded claims rather than a payoff-specific or book-specific overlay. 
Similarly, the liquidity layer should be interpreted as an execution-cost layer on hedge trades, not as a change in the marginal arbitrage-free price process.

Several limitations follow from this local perspective. 
The approximation can degrade near payoff kinks, barrier events or genuinely nonlocal jumps. 
Second derivatives and empirical local regressions can be noisy. 
Calibration becomes unstable when hedge instruments do not span the relevant state directions. 
A connection-only curvature overlay acts through the first-order exposure vector, so a book that is neutral to every coordinate included in the state receives no correction from this channel. 
Remaining curvature must then be represented by enlarging the state or by adding an independent residual quadratic term. 
Cost/risk penalties may be poorly conditioned if the hedge rule does not trade against all factor moves. 
Finally, a connection fitted directly to a target quadratic form need not be the Levi--Civita connection of a smooth metric. 
The metric interpretation is appropriate when the second-order structure is induced by costs, risks or constraints.

Future work can proceed in several directions. 
The most immediate is empirical: estimating target quadratic forms from realized short-horizon P\&L and testing whether the overlay improves hedging error or risk attribution out of sample. 
A second direction is robust calibration, including regularization, regime detection and uncertainty estimates for the fitted connection or penalty matrix. 
A third direction is dynamic: embedding the local hedge optimizer in a multi-period control problem with execution costs, while preserving the no-arbitrage restrictions of the base model. 
Further applications include higher-dimensional implementations with smile coordinates, curve factors, marking conventions, tenor transport, xVA sensitivities or capital constraints. 
These would test how far the same local quadratic interface can be pushed in realistic portfolio settings.

Overall, the paper provides a model-preserving way to organize short-horizon P\&L adjustments (second-order). 
By separating curvature targets from cost and risk penalties, and by expressing both through local geometric objects, the framework gives desks a disciplined language for incorporating smile adjustments, liquidity costs and other quadratic features at book level without rebuilding the underlying pricing model.



\newpage

\bibliographystyle{apacite}
\bibliography{Bibliography}


\clearpage
\appendix

\section{Levi--Civita interpretation and metric reconstruction}
\label{app:metric_reconstruction}

This appendix records the optional Levi--Civita interpretation of a selected affine connection and the associated metric-reconstruction formulas used for interpretation and normalization.
The Vanna--Volga (VV) calibration in Sec.~\ref{sec:vv-connection} does not require any metric assumption and, in general, identifies only the contracted action of the connection on the relevant baseline Greeks.
Metric reconstruction is therefore meaningful only after a particular representative connection has been selected, for example through a normalization, regularization or modelling convention.
In the liquidity construction of Sec.~\ref{sec:liquidity}, by contrast, the starting point is already a quadratic penalty induced by execution costs, and this penalty may be treated as a metric when it is smooth and non-degenerate on the relevant subspace.

We first give the explicit two-dimensional formulas for the displayed $(S,\sigma)$ block. If one chooses to interpret the recovered connection as the Levi--Civita connection of a $C^2$ Riemannian metric on this block, write
\[
g(x)
=
\begin{pmatrix}
a(S,\sigma) & c(S,\sigma) \\[2pt]
c(S,\sigma) & b(S,\sigma)
\end{pmatrix},
\qquad
D=ab-c^2>0,
\qquad
g^{-1}
=
\frac{1}{D}
\begin{pmatrix}
b & -c \\
-c & a
\end{pmatrix}.
\]
The torsion-free, metric-compatible Levi--Civita coefficients are
cf.~\cite{do2016differential,Lee2018Riemannian}
\begin{align}
C^{S}_{SS}
&=
\frac{b\,a_S-2c\,c_S+c\,a_\sigma}{2D},
&
C^{S}_{S\sigma}
&=
\frac{b\,a_\sigma-c\,b_S}{2D},
&
C^{S}_{\sigma\sigma}
&=
\frac{b(2c_\sigma-b_S)-c\,b_\sigma}{2D},
\label{eq:genGammaS}
\\
C^{\sigma}_{SS}
&=
\frac{a(2c_S-a_\sigma)-c\,a_S}{2D},
&
C^{\sigma}_{S\sigma}
&=
\frac{a\,b_S-c\,a_\sigma}{2D},
&
C^{\sigma}_{\sigma\sigma}
&=
\frac{a\,b_\sigma-c(2c_\sigma-b_S)}{2D}.
\label{eq:genGammas}
\end{align}
The mixed adjusted coefficient is
\[
\widetilde H_{S\sigma}
=
\overline H_{S\sigma}
-
C^{S}_{S\sigma}\Delta
-
C^{\sigma}_{S\sigma}\Vega.
\]
Therefore matching a VV target $H^{\mathrm{VV}}_{S\sigma}$ on the displayed
block amounts to
\begin{equation}
\label{eq:vannaProjection}
\widetilde H_{S\sigma}
=
H^{\mathrm{VV}}_{S\sigma}
\quad\Longleftrightarrow\quad
C^{S}_{S\sigma}\Delta
+
C^{\sigma}_{S\sigma}\Vega
=
\overline H_{S\sigma}
-
H^{\mathrm{VV}}_{S\sigma}.
\end{equation}

Under this two-dimensional Levi--Civita interpretation, a nonzero Vanna correction $\widetilde{\Vanna}\neq\Vanna$ requires the contracted mixed connection term $C^{S}_{S\sigma}\Delta + C^{\sigma}_{S\sigma}\Vega$ to be nonzero.
Thus at least one mixed connection coefficient must act through a nonzero first-order exposure, and the two contributions must not cancel.
From Eqs.~\eqref{eq:genGammaS}--\eqref{eq:genGammas}, the mixed coefficients $C^{S}_{S\sigma}$ and $C^{\sigma}_{S\sigma}$ depend on the cross-variable derivatives $a_\sigma$ and $b_S$ of the diagonal metric coefficients.
A constant off-diagonal term $c$ alone does not generate mixed coefficients if $a_\sigma=b_S=0$. In the diagonal case $g=\operatorname{diag}(a,b)$, i.e. $c\equiv0$, one has
\[
C^{S}_{S\sigma}
=
\frac{a_\sigma}{2a}
=
\tfrac12\,\partial_\sigma(\log a),
\qquad
C^{\sigma}_{S\sigma}
=
\frac{b_S}{2b}
=
\tfrac12\,\partial_S(\log b).
\]
Thus the mixed coefficients vanish for a separable diagonal metric with $a=a(S)$ and $b=b(\sigma)$, and no Vanna correction is generated by the Levi--Civita term on the two-coordinate block.

We next describe metric reconstruction. This step is optional and is used only for interpretation, normalization or comparison with metric-induced constructions such as liquidity. If a fitted torsion-free connection is the Levi--Civita connection of some metric $g$, then
\begin{equation}
  \label{eq:LC}
  C^{k}_{ij}
  =
  \tfrac12\,g^{k\ell}
  \big(
  \partial_i g_{j\ell}
  +
  \partial_j g_{i\ell}
  -
  \partial_\ell g_{ij}
  \big),
  \qquad
  \nabla g\equiv0.
\end{equation}
Equivalently, the metric satisfies the linear first-order system
\begin{equation}
  \label{eq:metricPDE}
  \partial_k g_{ij}
  =
  C^{\ell}_{ki}\,g_{\ell j}
  +
  C^{\ell}_{kj}\,g_{i\ell}.
\end{equation}
In the two-dimensional formulas above, $k\in\{S,\sigma\}$. In a higher-dimensional state the same equation applies with $i,j,k$ running over the chosen state coordinates.

The exact reconstruction problem is overdetermined in general. Not every torsion-free affine connection is metrizable, and a path-independent solution of Eq.~\eqref{eq:metricPDE} exists only when the corresponding integrability conditions are satisfied.\footnote{Equivalently, the initial bilinear form must be compatible with the holonomy of the connection. If the connection is known to be Levi--Civita of a metric, then the compatible metric is determined locally up to the usual freedom allowed by the connection, often only an overall constant scale in irreducible cases. Scaling the metric by a positive constant leaves the Levi--Civita connection unchanged.} In empirical calibration, the recovered coefficients may also contain finite-difference noise or small inconsistencies. For this reason, metric reconstruction should be interpreted as an optional diagnostic or projection step rather than as part of the VV calibration itself.

In practice one may solve Eq.~\eqref{eq:metricPDE} on a grid in a least-squares sense and, if needed, project the resulting matrices pointwise onto the SPD cone, for example by eigenvalue clipping.
This projection is a numerical diagnostic or regularization step.
After projection, the metric need not reproduce the fitted connection exactly.
Several normalizations or anchors can be used to fix scale or provide an initial metric:
\begin{itemize}\itemsep2pt
\item Shock covariance scale. Let $\Xi=\operatorname{Cov}[\delta x]$ over the forecast horizon, restricted if desired to the displayed block such as $(\delta S,\delta\sigma)$, and use $g\propto\Xi^{-1}$ as a reference metric at a chosen point.
  This is the Mahalanobis normalization, cf.~\cite{Mahalanobis1936}.
  When a likelihood or statistical model is available, this is also related to the information-geometric viewpoint, cf.~\cite{Amari2016}.
  Correlations affect angles and distances, while mixed Levi--Civita coefficients arise from state dependence of the metric coefficients, i.e. from $\partial g\neq0$.

\item Market-sensitivity scale. For a vector of liquid hedge prices $P=(P_1,\dots,P_m)$ with sensitivity matrix $A_{ri}=\partial_{x^i}P_r$, take $g = A^\top W \, A$, where $W$ is an inverse-variance, liquidity or bid--ask weight. This gives an SPD normalization when $W\succ0$ and $A$ has full column rank. Otherwise it is positive semidefinite and may require regularization. This construction is analogous to the liquidity pullback in Sec.~\ref{sec:liquidity}, where a quadratic penalty, and when non-degenerate a metric, is built from a trade map $\delta q=M(x)\delta x$ and a trade-space impact matrix $\Lambda$.
  
\item Anchor normalization. At a reference state $(S_0,\sigma_0)$, impose for example $g_{SS}(S_0,\sigma_0)=1/S_0^2$ and $g_{\sigma\sigma}(S_0,\sigma_0)(\Delta\sigma_{\mathrm{ref}})^2 = 1$, or simply $\det g(S_0,\sigma_0) = 1$. Here $\Delta\sigma_{\mathrm{ref}}$ should be expressed in the volatility units used in the paper.
\end{itemize}

To fix the numerical scale of a reconstructed metric, suppose that $\widehat g$ is an SPD solution or approximation, $x_0$ is a reference state and $v_0$ is a reference displacement. Set
\[
\alpha
=
\frac{L_0^2}{v_0^\top\widehat g(x_0)v_0},
\qquad
g=\alpha\,\widehat g,
\]
so that $\|v_0\|_g^2=L_0^2$. Typical anchors are $v_0=(\varepsilon_S S_0,0)$ with $\varepsilon_S=1\%$, or $v_0=(0,\varepsilon_\sigma)$ with $\varepsilon_\sigma$ equal to a reference volatility move expressed in the chosen volatility units. This final scaling does not affect the Levi--Civita coefficients or the corresponding geodesic paths because a constant rescaling of a metric leaves its Levi--Civita connection unchanged.
It does, however, rescale distances, penalties and any threshold expressed in metric units.


\section{Coordinate changes and connection coefficients}
\label{app:connection_transform}

This appendix records the coordinate-change identities used in the paper and a simple example of the coordinate issue discussed in Sec.~\ref{sec:framework}. The main text relies only on the fact that the adjusted quadratic layer is intrinsic. The formulas below are included for implementation and for the common FX change of variables from forward level $F$ to log-forward $z=\log F$.

\subsection{General coordinate identities}

Let $x=(x^i)$ and $y=(y^\alpha)$ be two coordinate systems on the state space, related by a smooth invertible map $y=y(x)$ with inverse $x=x(y)$. For a scalar value function $V$, first derivatives transform as
\begin{equation}
\label{eq:first_derivative_transform}
V_\alpha
=
\frac{\partial x^i}{\partial y^\alpha}V_i.
\end{equation}
Ordinary second derivatives transform as
\begin{equation}
\label{eq:ordinary_hessian_transform}
V_{\alpha\beta}
=
\frac{\partial x^i}{\partial y^\alpha}
\frac{\partial x^j}{\partial y^\beta}
V_{ij}
+
\frac{\partial^2 x^k}{\partial y^\alpha\partial y^\beta}
V_k.
\end{equation}
The second term in Eq.~\eqref{eq:ordinary_hessian_transform} is the reason the ordinary Hessian is not a coordinate-free quadratic object. By contrast, once a desk target $H^\star$ has been specified as a local quadratic form, it transforms tensorially:
\begin{equation}
\label{eq:target_tensor_transform_y}
H^\star_{\alpha\beta}(y)
=
\frac{\partial x^i}{\partial y^\alpha}
\frac{\partial x^j}{\partial y^\beta}
H^\star_{ij}(x),
\qquad
H^\star_{ij}(x)
=
\frac{\partial y^\alpha}{\partial x^i}
\frac{\partial y^\beta}{\partial x^j}
H^\star_{\alpha\beta}(y).
\end{equation}
This is the rule used when a desk target, such as a Vanna--Volga quadratic adjustment, is specified in one quoting chart and evaluated in another. The target is transported as a quadratic form, not recomputed by taking ordinary second derivatives in the new chart.

Let $\nabla$ be an affine connection, with coefficients $C^m_{ij}(x)$ in the $x$ chart and $\widehat C^\alpha_{\beta\gamma}(y)$ in the $y$ chart. The coefficients transform as
\begin{equation}
\label{eq:connection_transform_rule}
\widehat C^{\alpha}_{\beta\gamma}(y)
=
\frac{\partial y^\alpha}{\partial x^m}
\frac{\partial x^i}{\partial y^\beta}
\frac{\partial x^j}{\partial y^\gamma}
C^{m}_{ij}(x)
+
\frac{\partial y^\alpha}{\partial x^m}
\frac{\partial^2 x^m}{\partial y^\beta\,\partial y^\gamma}.
\end{equation}
Combining Eqs.~\eqref{eq:ordinary_hessian_transform} and~\eqref{eq:connection_transform_rule} gives the tensorial transformation of the covariant Hessian:
\begin{equation}
\label{eq:covariant_hessian_transform_appendix}
\widetilde H_{\alpha\beta}
=
V_{\alpha\beta}
-
\widehat C^\gamma_{\alpha\beta}V_\gamma
=
\frac{\partial x^i}{\partial y^\alpha}
\frac{\partial x^j}{\partial y^\beta}
\left(
V_{ij}-C^k_{ij}V_k
\right)
=
\frac{\partial x^i}{\partial y^\alpha}
\frac{\partial x^j}{\partial y^\beta}
\widetilde H_{ij}.
\end{equation}
Therefore, if $\delta x$ and $\delta y$ denote the components of the same tangent displacement at the base state, so that $\delta y^\alpha = \frac{\partial y^\alpha}{\partial x^i}\delta x^i$, then
\begin{equation}
\label{eq:quadratic_predictor_coordinate_invariance_appendix}
V_i\delta x^i
+
\tfrac12\widetilde H_{ij}\delta x^i\delta x^j
=
V_\alpha\delta y^\alpha
+
\tfrac12\widetilde H_{\alpha\beta}\delta y^\alpha\delta y^\beta.
\end{equation}
This is the coordinate-invariance statement used in the paper.
It is a local statement about tangent increments.
It does not mean that the numerical entries of the adjusted Hessian are the same in every coordinate system.
They are not: an adjusted Gamma with respect to spot, forward or log-forward has different units and therefore different numerical values.
The invariant object is the quadratic P\&L contribution assigned to a given local market move, provided the connection, target and increments are transformed consistently.
Thus the desk may report different adjusted Greeks in different coordinates while the predicted second-order P\&L for the same tangent move is unchanged.
For finite observed coordinate differences, such as $F_{n+1}-F_n$ versus $\log(F_{n+1}/F_n)$, one must either use the corresponding tangent approximation at the base point or re-expand consistently in the chosen chart.
In particular, the relation $\delta z=F^{-1}\delta F$ is a tangent relation at the base state, not an exact identity for finite moves.

\subsection{A one-dimensional forward/log-forward example}
\label{app:forward_logforward_example}

We illustrate the distinction between coefficient values and quadratic P\&L contributions with a one-dimensional example.
Consider a state written either as the forward level $F$ or as the log-forward
\[
z=\log F.
\]
At the base state, a tangent move satisfies
\[
\delta z=\frac{1}{F}\,\delta F.
\]
Let $V=V(F)$ be a scalar value function.
Ordinary derivatives transform as
\[
V_z = F V_F,
\qquad
V_{zz}=F^2V_{FF}+F V_F.
\]
The extra term $F V_F$ is the familiar non-tensorial term in the ordinary Hessian.
It appears because $z=\log F$ is a nonlinear coordinate change.

Now let $C^F_{FF}$ be the connection coefficient in the $F$ coordinate.
The adjusted second-order coefficient is
\[
\widetilde H_{FF}=V_{FF}-C^F_{FF}V_F.
\]
Under the coordinate change $z=\log F$, the connection coefficient becomes
\[
\widehat C^z_{zz}=F C^F_{FF}+1,
\]
where the hat denotes the coefficient in the $z$ coordinate.
Therefore
\[
\begin{aligned}
\widetilde H_{zz}
&=
V_{zz}
-
\widehat C^z_{zz}V_z  \\
&=
\left(F^2V_{FF}+F V_F\right)
-
\left(F C^F_{FF}+1\right)F V_F  \\
&=
F^2\left(V_{FF}-C^F_{FF}V_F\right)  \\
&=
F^2\widetilde H_{FF}.
\end{aligned}
\]
Thus the adjusted second-order coefficient is not numerically invariant:
\[
\widetilde H_{zz}\neq \widetilde H_{FF}
\]
in general.
This is expected, since the two quantities have different units.
However, the quadratic P\&L term for the same tangent move is invariant:
\[
\frac12\,\widetilde H_{zz}(\delta z)^2
=
\frac12\,F^2\widetilde H_{FF}
\left(\frac{\delta F}{F}\right)^2
=
\frac12\,\widetilde H_{FF}(\delta F)^2.
\]
This is the sense in which the geometric quadratic layer is coordinate-aware.
A desk may report different adjusted second-order Greeks in forward and log-forward coordinates, but the local quadratic P\&L contribution assigned to the same move is unchanged.

By contrast, the ordinary Hessian coefficient does not transform as a quadratic form:
\[
V_{zz}=F^2V_{FF}+F V_F.
\]
The extra first-order term means that the ordinary Hessian, viewed by itself, is tied to the coordinate convention.
The full Taylor expansion in a chosen coordinate system remains meaningful, but the ordinary second-order coefficient alone is not an intrinsic quadratic P\&L object.
The covariant Hessian removes the coordinate-convention term through the connection, so that the adjusted quadratic term transforms as a genuine quadratic form.

\subsection{The two-dimensional log-forward chart}
\label{app:log_forward_chart}

We now specialize these formulas to the log-forward chart.
Let $x=(F,\sigma)$ and $y=(z,\sigma)$, with $z=\log F$.
All quantities below are evaluated at the base state.
\[
\frac{\partial z}{\partial F}=\frac{1}{F},
\qquad
\frac{\partial F}{\partial z}=F,
\qquad
\frac{\partial^2F}{\partial z^2}=F,
\qquad
\frac{\partial^2F}{\partial z\,\partial\sigma}=0,
\qquad
\frac{\partial^2\sigma}{\partial(\cdot)\partial(\cdot)}=0.
\]
For ordinary derivatives of a scalar value function,
\begin{equation}
\label{eq:log_forward_first_derivatives}
V_z=F V_F,
\qquad
V_\sigma \ \text{unchanged},
\end{equation}
and
\begin{equation}
\label{eq:log_forward_second_derivatives}
V_{zz}=F^2V_{FF}+F V_F,
\qquad
V_{z\sigma}=F V_{F\sigma},
\qquad
V_{\sigma\sigma}\ \text{unchanged}.
\end{equation}
The additional term $F V_F$ in $V_{zz}$ is precisely the non-tensorial term in the ordinary Hessian.

For a target quadratic form $H^\star$, there is no such additional term. The components in the log-forward chart are
\begin{equation}
\label{eq:target_log_forward_transform}
H^\star_{zz}=F^2 H^\star_{FF},
\qquad
H^\star_{z\sigma}=F H^\star_{F\sigma},
\qquad
H^\star_{\sigma\sigma}\ \text{unchanged}.
\end{equation}
Conversely,
\begin{equation}
\label{eq:target_forward_transform}
H^\star_{FF}=F^{-2}H^\star_{zz},
\qquad
H^\star_{F\sigma}=F^{-1}H^\star_{z\sigma},
\qquad
H^\star_{\sigma\sigma}\ \text{unchanged}.
\end{equation}

Applying Eq.~\eqref{eq:connection_transform_rule} gives the connection coefficients in the log-forward chart:
\begin{equation}
\label{eq:connection_log_forward_z_components}
\widehat C^z_{zz}
=
F C^F_{FF}+1,
\qquad
\widehat C^z_{z\sigma}
=
C^F_{F\sigma},
\qquad
\widehat C^z_{\sigma\sigma}
=
\frac{1}{F}C^F_{\sigma\sigma},
\end{equation}
and
\begin{equation}
\label{eq:connection_log_forward_sigma_components}
\widehat C^\sigma_{zz}
=
F^2 C^\sigma_{FF},
\qquad
\widehat C^\sigma_{z\sigma}
=
F C^\sigma_{F\sigma},
\qquad
\widehat C^\sigma_{\sigma\sigma}
=
C^\sigma_{\sigma\sigma}.
\end{equation}
Here hats denote coefficients in the $(z,\sigma)$ chart, while the unhatted coefficients on the right-hand side are in the $(F,\sigma)$ chart.

As a useful check, the covariant Hessian transforms tensorially:
\begin{equation}
\label{eq:cov_hess_log_forward_transform}
\widetilde H_{zz}=F^2\widetilde H_{FF},
\qquad
\widetilde H_{z\sigma}=F\widetilde H_{F\sigma},
\qquad
\widetilde H_{\sigma\sigma}\ \text{unchanged}.
\end{equation}
Indeed,
\[
\begin{aligned}
\widetilde H_{zz}
&=
V_{zz}
-
\widehat C^z_{zz}V_z
-
\widehat C^\sigma_{zz}V_\sigma \\
&=
(F^2V_{FF}+FV_F)
-
(F C^F_{FF}+1)(F V_F)
-
F^2C^\sigma_{FF}V_\sigma \\
&=
F^2
\left(
V_{FF}
-
C^F_{FF}V_F
-
C^\sigma_{FF}V_\sigma
\right)
=
F^2\widetilde H_{FF}.
\end{aligned}
\]
The same calculation gives the remaining identities in Eq.~\eqref{eq:cov_hess_log_forward_transform}.

Starting from a spot-based chart $(S,\sigma)$, one may first pass to $(F,\sigma)$ under a frozen-carry approximation over the short horizon and then apply the log-forward formulas above.
If rates, dividends or carry are material over the horizon, they should be included in the state rather than treated as fixed constants in the coordinate change.


\section{Component formulas for $g_\ell$ and its derivatives}
\label{app:gl_components}

This appendix records component expressions for $g_\ell(x) = M(x)^\top \Lambda(x) \, M(x)$, the liquidity cost matrix, and its first derivatives. The formulas are useful for implementation checks and for special cases such as diagonal $\Lambda$, bucketed $\Lambda$ or simple hedging rules. They are not needed for the main conceptual arguments. We write the formulas for the displayed two-factor block $(S,\sigma)$. In a larger state the same identities hold with the corresponding factor indices.

Let $M\in\mathbb{R}^{m\times 2}$ denote the hedge response restricted to the $(S,\sigma)$ block and write
\[
g_\ell(x)
=
\begin{pmatrix}
a(x) & c(x)\\
c(x) & b(x)
\end{pmatrix},
\qquad
a=g_{\ell,SS},\quad
b=g_{\ell,\sigma\sigma},\quad
c=g_{\ell,S\sigma}.
\]
Let $r,s$ run over hedge instruments, while $S$ and $\sigma$ label the two factor columns of $M$. From $g_\ell=M^\top\Lambda M$ one has
\begin{equation}
  \label{eq:app_gl_components}
  a
  =
  \sum_{r,s}\Lambda_{rs}\,M_{r,S}\,M_{s,S},
  \qquad
  b
  =
  \sum_{r,s}\Lambda_{rs}\,M_{r,\sigma}\,M_{s,\sigma},
  \qquad
  c
  =
  \sum_{r,s}\Lambda_{rs}\,M_{r,S}\,M_{s,\sigma}.
\end{equation}
If $\Lambda$ depends on the state, derivatives pick up both $M$ and $\Lambda$ terms. For $i\in\{S,\sigma\}$,
\begin{align}
  \label{eq:app_d_a}
  \partial_i a
  &=
  \sum_{r,s}
  \Big[
  (\partial_i\Lambda_{rs})\,M_{r,S}M_{s,S}
  +
  \Lambda_{rs}
  \big(
  (\partial_i M_{r,S})M_{s,S}
  +
  M_{r,S}(\partial_i M_{s,S})
  \big)
  \Big],
  \\
  \label{eq:app_d_b}
  \partial_i b
  &=
  \sum_{r,s}
  \Big[
  (\partial_i\Lambda_{rs})\,M_{r,\sigma}M_{s,\sigma}
  +
  \Lambda_{rs}
  \big(
  (\partial_i M_{r,\sigma})M_{s,\sigma}
  +
  M_{r,\sigma}(\partial_i M_{s,\sigma})
  \big)
  \Big],
  \\
  \label{eq:app_d_c}
  \partial_i c
  &=
  \sum_{r,s}
  \Big[
  (\partial_i\Lambda_{rs})\,M_{r,S}M_{s,\sigma}
  +
  \Lambda_{rs}
  \big(
  (\partial_i M_{r,S})M_{s,\sigma}
  +
  M_{r,S}(\partial_i M_{s,\sigma})
  \big)
  \Big].
\end{align}
In matrix shorthand,
\begin{equation}
  \label{eq:app_state_first}
  \partial_i g_\ell
  =
  (\partial_i M)^\top \Lambda \, M
  +
  M^\top (\partial_i\Lambda) \, M
  +
  M^\top \Lambda \, (\partial_i M),
  \qquad
  i\in\{S,\sigma\}.
\end{equation}
Here $\partial_iM$ denotes the total local derivative of the hedge-response matrix with respect to the state coordinate $x^i$. If $M$ is generated by a least-cost neutrality rule, this derivative includes the state dependence of the book sensitivities, hedge sensitivities and, when applicable, liquidity inputs entering the hedge rule.

Equation~\eqref{eq:app_state_first} is Eq.~\eqref{eq:state_first} written on the displayed two-factor block. If $\Lambda$ is bucketed-constant, all $\partial_i\Lambda$ terms drop. If $\Lambda$ is trade-size dependent rather than state dependent, as in a tiered impact specification, these formulas apply after freezing or evaluating the effective local $\Lambda$ used for the state calculation. Without such a localization, the cost is a trade-size dependent execution rule rather than a standard state-only metric, and the Levi--Civita formulas do not apply directly.

As a special case, if $\Lambda = \operatorname{diag}(\lambda_1,\dots,\lambda_m)$, then
\[
a=\sum_r \lambda_r M_{r,S}^2,
\qquad
b=\sum_r \lambda_r M_{r,\sigma}^2,
\qquad
c=\sum_r \lambda_r M_{r,S}M_{r,\sigma},
\]
and the derivative formulas simplify accordingly:
\[
\partial_i a
=
\sum_r
\left[
(\partial_i\lambda_r)M_{r,S}^2
+
2\lambda_r M_{r,S}\partial_i M_{r,S}
\right],
\]
\[
\partial_i b
=
\sum_r
\left[
(\partial_i\lambda_r)M_{r,\sigma}^2
+
2\lambda_r M_{r,\sigma}\partial_i M_{r,\sigma}
\right],
\]
and
\[
\partial_i c
=
\sum_r
\left[
(\partial_i\lambda_r)M_{r,S}M_{r,\sigma}
+
\lambda_r
\big(
(\partial_i M_{r,S})M_{r,\sigma}
+
M_{r,S}(\partial_i M_{r,\sigma})
\big)
\right].
\]
In the bucketed-constant diagonal case, the $\partial_i\lambda_r$ terms also
drop.

Given $a,b,c$ and their derivatives, one has
\[
\det(g_\ell)=ab-c^2,
\qquad
g_\ell^{-1}
=
\frac{1}{\det(g_\ell)}
\begin{pmatrix}
b & -c\\
-c & a
\end{pmatrix},
\]
whenever $g_\ell$ is nonsingular on the displayed block. Since $g_\ell$ is positive semidefinite by construction, this means positive definite on that block. The Levi--Civita coefficients $C^k_{ij}(g_\ell)$ can then be obtained by substituting $g_\ell$, $\smash{g_\ell^{-1}}$ and $\partial_i g_\ell$ into the standard Levi--Civita formula, cf.~Eq.~\eqref{eq:LC} in App.~\ref{app:metric_reconstruction}. These coefficients enter liquidity-adjusted second-order Greeks through
\[
\widetilde H_{ij}
=
V_{ij}
-
C^k_{ij}(g_\ell)V_k.
\]

In practice, under neutrality rules, $M(x)$ depends on book and hedge sensitivities, so $\partial_iM$ may involve higher derivatives of option values. For this reason it is often simpler numerically to compute $\partial_i g_\ell$ by finite differences of $g_\ell(x)$, possibly with smoothing or bucketing, rather than evaluating analytic third derivatives.

The positivity and rank conditions for $g_\ell$ are discussed in Lemma~\ref{lem:pullback_metric} and in Sec.~\ref{sec:constLambda}. In this appendix, the inverse matrix and the Levi--Civita coefficients are used only when $g_\ell$ is positive definite on the displayed block, or after a numerical
regularization of the form
\[
g_\ell\leftarrow g_\ell+\varepsilon_g g_0,
\]
where $\varepsilon_g>0$ and $g_0$ is a fixed SPD baseline penalty in consistent units. The regularization is only needed when $g_\ell$ is used as a genuine metric. It is not required for computing the scalar quadratic execution cost.

\end{document}